\documentclass[12pt]{article}
\usepackage{latexsym}
\renewcommand{\theequation}{\thesection.\arabic{equation}}
\newlength{\extraspace}
\newlength{\extraspaces}
\newcommand{\be}{\begin{equation}
\addtolength{\abovedisplayskip}{\extraspaces}
\addtolength{\belowdisplayskip}{\extraspaces}
\addtolength{\abovedisplayshortskip}{\extraspace}
\addtolength{\belowdisplayshortskip}{\extraspace}}
\newcommand{\ee}{\end{equation}}
 
\newcommand{\ba}{\begin{eqnarray}
\addtolength{\abovedisplayskip}{\extraspaces}
\addtolength{\belowdisplayskip}{\extraspaces}
\addtolength{\abovedisplayshortskip}{\extraspace}
\addtolength{\belowdisplayshortskip}{\extraspace}}
\newcommand{\ea}{\end{eqnarray}}
 
\newcommand{\bas}{\begin{eqnarray*}
\addtolength{\abovedisplayskip}{\extraspaces}
\addtolength{\belowdisplayskip}{\extraspaces}
\addtolength{\abovedisplayshortskip}{\extraspace}
\addtolength{\belowdisplayshortskip}{\extraspace}}
\newcommand{\eas}{\end{eqnarray*}}
 
\newcounter{subequation}[equation]
\makeatletter
\expandafter\let\expandafter\reset@font\csname reset@font\endcsname
\def\subeqnarray{\arraycolsep1pt
    \def\@eqnnum\stepcounter##1{\stepcounter{subequation}
        {\reset@font\rm(\theequation\alph{subequation})}}\eqnarray}

\makeatother
 
\newenvironment{theorem}[1]
{\vspace{3mm}\noindent {\bf #1 :} }{\vspace{2mm}}
\newcommand{\bt}[1]{\begin{theorem}{#1}}
\newcommand{\et}{\end{theorem}}
 
\newcommand{\newsection}[1]{
\vspace{12mm}
\pagebreak[3]
\addtocounter{section}{1}
\setcounter{equation}{0}
\setcounter{subsection}{0}
\addcontentsline{toc}{section}
{\protect\numberline{\arabic{section}}{#1}}
 
\begin{flushleft}
{\large\bf \thesection. #1}
\end{flushleft}
\nopagebreak
\medskip
\nopagebreak}
 
\newcommand{\newsubsection}[1]{
\vspace{1cm}
\pagebreak[3]
 
\addtocounter{subsection}{1}
\addcontentsline{toc}{subsection}{\protect
\numberline{\arabic{section}.\arabic{subsection}}{#1}}
\noindent{ \bf \thesubsection. #1}
\nopagebreak
\vspace{2mm}
\nopagebreak}

\newcommand{\newsubsubsection}[1]{
\vspace{0.8 cm}
\pagebreak[3]
 
\addtocounter{subsubsection}{1}
\addcontentsline{toc}{subsubsection}
{\protect\numberline{\arabic{section}.\arabic{subsection}.%
\arabic{subsubsection}}{#1}}
\noindent{ \sc \thesubsubsection. #1}
\nopagebreak
\vspace{2mm}
\nopagebreak}

\newcommand{\NP}[1]{Nucl.\ Phys.\ {\bf #1}}
\newcommand{\PL}[1]{Phys.\ Lett.\ {\bf #1}}
\newcommand{\CMP}[1]{Comm.\ Math.\ Phys.\ {\bf #1}}
\newcommand{\PR}[1]{Phys.\ Rev.\ {\bf #1}}
\newcommand{\PRL}[1]{Phys.\ Rev.\ Lett.\ {\bf #1}}

\newcommand{\FP}[1]{Fortschr.\ Phys.\ {\bf #1}}

\newcommand{\is}{\! & \! = \! & \!}

\newcommand{\inv}{^{-1}}

\newcommand{\ps}{p\llap{/}} 
\newcommand{\ds}{\partial\llap{/}} 
\newcommand{\Ds}{D\llap{/}}
\begin{document}
%
\newcommand{\Cal}{{\cal C}}
\renewcommand{\l}{\lambda}
\renewcommand{\a}{\alpha}
\renewcommand{\b}{\beta}
\renewcommand{\d}{\delta}
\renewcommand{\k}{\kappa}
\newcommand{\ld}{\buildrel \leftarrow \over \d}
\newcommand{\rd}{\buildrel \rightarrow \over \d}
\newcommand{\e}{\eta}
\renewcommand{\o}{\omega}
\newcommand{\dem}{\d_\o}
\newcommand{\p}{\partial}
\newcommand{\pmu}{\p_\mu}
\newcommand{\pmo}{\p^\mu}
\newcommand{\pnu}{\p_\nu}
\newcommand{\s}{\sigma}
\renewcommand{\r}{\rho}
\newcommand{\bpsi}{\bar\psi}
\newcommand{\dslash}{\p\llap{/}}
\newcommand{\pslash}{p\llap{/}}
\newcommand{\ve}{\varepsilon}
\newcommand{\uvi}{\underline{\varphi}}
\newcommand{\vi}{\varphi}
\newcommand{\ue}{u^{(1)}}
\newcommand{\Am}{A_\mu}
\newcommand{\An}{A_\nu}
\newcommand{\Fmnu}{F_{\mu\nu}}
\newcommand{\Fmno}{F^{\mu\nu}}
\newcommand{\Ga}{\Gamma}
\newcommand{\Gao}{\Gamma^{(o)}}
\newcommand{\Gae}{\Gamma^{(1)}}
\newcommand{\Gacl}{\Gamma_{cl}}
\newcommand{\Gagf}{\Gamma_{{\rm g.f.}}}
\newcommand{\Gainv}{\Gamma_{\rm inv}}
\newcommand{\pvi}{\partial\varphi}
\newcommand{\om}{{\bf w}}
\newcommand{\mn}{\mu\nu}
\newcommand{\Tmn}{T_{\mn}}
\newcommand{\Gmn}{\Ga_{{\mn}}}
\newcommand{\hT}{\hat T}
\newcommand{\emt}{energy-mo\-men\-tum ten\-sor}
\newcommand{\eit}{Ener\-gie-Im\-puls-Ten\-sor}
\newcommand{\Tc}{T^{(c)}}
\newcommand{\Tcr}{\Tc_{\rho\sigma}(y)}
\newcommand{\ha}{{1\over 2}}
\newcommand{\dalam}{{\hbox{\frame{6pt}{6pt}{0pt}}\,}}
\newcommand{\wtm}{{\bf W}^T_\mu}
\newcommand{\nnp}{\nu'\nu'}
\newcommand{\mnp}{\mu'\nu'}
\newcommand{\emn}{\eta_{\mn}}
\newcommand{\mpm}{\mu'\mu'}
\newcommand{\tbw}{\tilde{\bf w}}
\newcommand{\tw}{\tilde w}
\newcommand{\hw}{\hat{\bf w}}
\newcommand{\bw}{{\bf w}}
\newcommand{\bW}{{\bf W}}
\newcommand{\ubW}{\underline{\bW}}
\newcommand{\hmn}{h^{\mn}}
\newcommand{\gmn}{g^{\mn}}
\newcommand{\ga}{\gamma}
\newcommand{\Gf}{\Gamma_{\hbox{\hskip-2pt{\it eff}\hskip2pt}}}
\newcommand{\T}{\buildrel o \over T}
\newcommand{\Lf}{{\cal L}_{\hbox{\it eff}\hskip2pt}}
\newcommand{\np}{\not\!\p}
\newcommand{\lp}{\partial\llap{/}}
\newcommand{\ah}{{\hat a}}
\newcommand{\han}{{\hat a}^{(n)}}
\newcommand{\hak}{{\hat a}^{(k)}}
\newcommand{\dv}{{\d\over\d\vi}}
\newcommand{\zze}{\sqrt{{z_2\over z_1}}}
\newcommand{\zez}{\sqrt{{z_1\over z_2}}}
\newcommand{\Hmn}{H_{\(\mn\)}}
\newcommand{\hfrac}[2]{\hbox{${#1\over #2}$}}  
\newcommand{\smdm }{\underline m \p _{\underline m}}
 \newcommand{\tsmdm }{\underline m \tilde \p _{\underline m}} 
 \newcommand{ \Wh }{{\hat {\bf W}}^K}
\newcommand{ \CS }{Callan-Symanzik}
\newcommand{ \G}{\Gamma}
\newcommand{ \bl }{\b _ \l}
\newcommand{ \kdk }{\k \p _\k}
\newcommand{\mdm }{m \p _m} 
\newcommand{ \te }{\tau_{\scriptscriptstyle 1}}
\newcommand{ \mhi }{m_H}
\newcommand{ \mf }{m_f}
\newcommand{\bare}{^o}
\newcommand{\Pol}{{\mathbf P}}
\newcommand{\brs}{{\mathrm s}}
\newcommand{\cw}{\cos \theta_W}
\newcommand{\cws}{\cos^2 \theta_W}
\newcommand{\sw}{\sin \theta_W}
\newcommand{\sws}{\sin^2 \theta_W}
\newcommand{\cg}{\cos \theta_G}
\newcommand{\sg}{\sin \theta_G}
\newcommand{\cwg}{\cos (\theta_W- \theta_G)}
\newcommand{\swg}{\sin (\theta_W- \theta_G)}
\newcommand{\cv}{\cos \theta_V}
\newcommand{\sv}{\sin \theta_V}
\newcommand{\cvg}{\cos (\theta_V- \theta_G)}
\newcommand{\svg}{\sin (\theta_V- \theta_G)}
\newcommand{\fsc}{{e^2 \over 16 \pi ^2}}
\newcommand{\cvt}{\cos \Theta^V_3}
\newcommand{\svt}{\sin \Theta^V_3}
\newcommand{\cvf}{\cos \Theta^V_4}
\newcommand{\svf}{\sin \Theta^V_4}
\newcommand{\cgt}{\cos \Theta^g_3}
\newcommand{\sgt}{\sin \Theta^g_3}



\begin{titlepage}
%
\renewcommand{\thefootnote}{\fnsymbol{footnote}}
\begin{flushright}
June 1997
\end{flushright}
\vspace{1cm}
 
\begin{center}
{\Large {\bf Renormalization of the
Electroweak  Standard Model to All Orders}}
\footnote{to be published in 
Ann.~Phys.~(NY) 1997}
 \\[4mm]
{\makebox[1cm]{  }       \\[1.5cm]
{\bf Elisabeth Kraus} \footnote{Supported by 
Deutsche Forschungsgemeinschaft}\\ [3mm]
{\small\sl Physikalisches Institut, Universit\"at Bonn} \\
{\small\sl Nu\ss allee 12, D-53115 Bonn, Germany}} 
\vspace{1.5cm}
 
{\bf Abstract}
\end{center}
\begin{quote}
We give the renormalization of the standard model of electroweak
interactions to all orders of perturbation theory by using the
method of algebraic renormalization, which is based on general properties
of renormalized perturbation theory and not on a specific regularization 
scheme.  The Green functions
of the standard model are uniquely constructed to all orders, if
one defines the model by the Slavnov-Taylor identity, Ward-identities
of rigid symmetry and  a specific form of the abelian local gauge
Ward-identity, which continues the Gell-Mann Nishijima relation to
higher orders. Special attention is directed to the mass diagonalization
of massless and massive neutral vectors and ghosts.  For
obtaining off-shell infrared finite expressions it is required to
take into account higher order corrections into the functional symmetry
operators. It is shown, that the normalization conditions
of the on-shell schemes are in agreement with the 
most general symmetry transformations allowed by the algebraic constraints.
\end{quote}
\vfill
\renewcommand{\thefootnote}{\arabic{footnote}}
\setcounter{footnote}{0}
\end{titlepage}
\tableofcontents
\newpage
\newsection{Introduction}

The standard model of electroweak interaction has been tested 
in the last few years with  precision experiments of remarkable accuracy
\cite{datasum}. Theoretical predictions are based on the consistent
perturbative formulation of the standard model of electroweak
interaction as a renormalizable and unitary quantum field theory, which
allows the derivation of unambiguous results  for 
physical scattering processes order by order in perturbation theory.
In order to match the level of accuracy  given by experiments,
it is also necessary to take into account also higher order quantum corrections
to the different processes considered. 
Conversely, present experiments enable  the
standard model to be tested beyond tree approximation. 
Higher order corrections
to the electroweak processes have been computed and evaluated quite
systematically by several groups. (For a review see \cite{YeRep}
and references listed below.)
The agreement between experimental
results and theoretical predictions is quite impressive 
 and by now there
is  no evidence --- either from theoretical arguments or from
experiments --- for physics beyond the standard model.

The evaluation of higher order corrections in the standard model
is quite an involved task. First, one has to remove the divergencies
which appear in the naive perturbative expansion of Green functions
in the course of renormalization and
 one has to establish the defining symmetries of the theory.
 At the same time, the independent parameters of the standard
model have to be specified and fixed by normalization conditions
in such a way that the remaining undetermined constants, such
as masses and the coupling strength, can be taken as input parameters
from experiment.
Finally, there remains the explicit evaluation of  higher order
loop diagrams. Nearly all calculations have been carried out in
the framework of dimensional regularization. There
the one-loop order has been studied quite systematically
(see \cite{YeRep})
and the computations have reached a high field-theoretic standard.
 However, there is no abstract approach which
  analyses the
renormalization 
of the electroweak standard model to higher orders. With the present
article we fill this gap, with special attention being paid to
the symmetries, normalization conditions and infrared-finiteness
of off-shell Green functions. In particular, the analysis  does
not refer to invariance properties of a scheme, but is based on
properties of finite renormalized perturbation theory.
(For a review of algebraic renormalization see \cite{PISO95}.)
The purpose of   a systematic analysis is twofold.
First, it is evident that such an analysis will support explicit
calculations by allowing  symmetries to be established and 
possible breakings of the symmetries to be characterised
 quite systematically. In particular, if one
wants to take into account  higher order corrections in
theoretical predictions, either by
summing up one loop induced higher order corrections or by explicit
evaluation of some higher order diagrams, it has to be ensured that
the defining symmetries are not violated at any stage of the calculations
and that Green functions exist to higher orders, once they are
specified in 1-loop order. Higher order existence of Green functions
can be destroyed in the standard model due to off-shell infrared-divergencies,
 whenever a photon mass counterterm
is enforced by symmetries and by lower order normalization conditions. In fact
it appears that infrared finiteness and the establishment of  symmetries 
cannot be considered as separate from each other.
Apart from these practical reasons the analysis is also important
in its own right. Since the standard model has been so successful by now,
we are convinced that  electroweak interaction can only be embedded into
a more complete theory of fundamental interactions, once one
understands  their structure 
in its quantized version as prescribed by the standard model.

From the algebraic point of view,
 the abstract approach to the quantized version
of the standard model is similar
 to the construction of the classical Lagrangian
\cite{GLA61, WEI67, SAL68}. If one
takes the charged currents of weak interactions as given in the
lepton sector, it is seen that the algebra is closed, when one
includes the weak and electromagnetic currents into the group structure.
Coupling these currents to the gauge bosons of weak interactions and
to the photon, and requiring local $SU(2)\times U(1)$ gauge
invariance, the algebra, and at the same time the classical action, is
uniquely determined and  the transformation of all
further fields is restricted \cite{GLA61}.
  Introducing a complex scalar doublet
with one physical Higgs field, one generates
all masses  by the mechanism of spontaneous symmetry breaking
and the final standard model
Lagrangian is invariant under spontaneously broken $SU(2)
\times U(1) $ symmetry, which is a natural algebraic generalization
of unbroken symmetry. The implementation
of symmetries is
also the main ingredient of abstract renormalization.

At  an early stage it was observed in the framework
of dimensional regularization
that gauge theories
 are renormalizable \cite{HO71}, in the sense
that divergencies  can be absorbed into a redefinition of coupling
constants, mass parameters and fields. If one
uses the renormalizable `t Hooft gauges, the divergence structure of a
spontaneously broken theory is seen to be no worse than that of unbroken
theories \cite{SY72, lee}.

The main advances in the systematic definition of renormalizable gauge
theories occurred, when it was observed that gauge theories, 
including the gauge fixing and Faddeev-Popov part \cite{FAD67}, are
invariant under nonlinear symmetry transformations, the 
Becchi-Rouet-Stora (BRS) trans\-for\-mations
\cite{BRS74, TYU75}.
It is then possible to derive and postulate the Slavnov-Taylor
identities,
which are the functional version of BRS-symmetry, as expressing the defining
symmetries of gauge theories in the quantized version. In particular,
the program of algebraic renormalization has been applied 
to the abelian Higgs-Kibble model \cite{BRS75}
and spontaneously broken gauge
theories with semisimple gauge groups \cite{BRS76}.
With the help of the action principle in its quantized version
\cite{LOW71, LAM73}, and algebraic consistency
 it was shown that the Green functions are
completely characterized by normalization conditions on the mass and
coupling constants and the Slavnov-Taylor identity. 
If the Adler-Bardeen anomalies \cite{ADL69, BEJA69, BAR69} are absent,  
the Slavnov-Taylor
identity can be established to all orders for off-shell Green functions.
Then one is finally able to
prove unitarity of the physical S-matrix, i.e.~compensation of unphysical
fields in physical scattering processes and gauge parameter independence
of the physical S-matrix \cite{BRS75, BRS76, KUOJ78}.
  Algebraic renormalization
yields  finite Green functions by requiring
invariance under symmetry transformations instead of defining them
by an invariant scheme. In  gauge theories without parity violation
a specification by an invariant scheme
 is quite satisfactory. However, if anomalies are
not forbidden for reasons of symmetry, then 
 the algebraic renormalization becomes important, if one wants to formulate 
the theory consistently to all orders of perturbation theory.

In the early papers only gauge theories with a semisimple gauge
structure were considered. Later on the renormalization procedure
 was extended to
non-semisimple groups with several abelian factors \cite{BABE78}.
 But the analysis, as
it is carried out there, is not immediately
 applicable to the standard model, due to the
restriction to massive fields and due to the fact that 
the Green functions are not
specified for on-shell fields.
However, we shall use some technical components of this paper
such as  the form of the Slavnov-Taylor identity
and the use of the Callan-Symanzik operator for solving the cohomology.

In the remainder of this introduction we shall outline the procedure
of renormalizing the standard model, as it is presented in the 
 paper. As the first step we have to specify  all the
symmetry transformations which characterize the tree approximation
and higher order Green functions. 
 It is important
to note that the weak hypercharges 
are determined by  requiring  electromagnetic current
conservation according to the Gell-Mann Nishijima
formula. In the procedure of quantization 
electromagnetic gauge invariance is replaced by BRS-symmetry and
the Slavnov-Taylor identity. For deriving the analog of the
Gell-Mann Nishijima relation, we have to establish
the local $U(1)$ Ward identity in addition to
the Slavnov-Taylor identity. 
 For specifying the abelian
factor, however, it is necessary to have  invariance under
rigid $SU(2)\times  U(1) $ Ward identities. This constraint 
restricts order by order the independent parameters of the gauge fixing
functions, but rigid invariance is immediately established on the matter and
Yang-Mills parts of the action. Only if one includes
all these symmetry transformations, are the finite Green functions 
uniquely specified as being those of weak and electromagnetic interactions.

The symmetry invariants  are free parameters  and
have to be fixed by normalization conditions. Here,
the abstract analysis  benefits from the fact, that different
parameterizations have been considered for one-loop calculations
and have been discussed quite extensively in the past (for a review see
\cite{hollik}),
 since their definition
also enters the theoretical predictions of higher orders. It has been
pointed out that those schemes are adequate, which allow the computation of 
different processes without switching to different parameter sets
\cite{BO90}. 
On-shell schemes 
which specify the mass parameters as physical masses on the 2-point
functions    \cite{PAVE79, CO79, SIMA80, FLJE81, BACH80, AOHI82, COLO83, 
BOHO86, HO90} 
 are certainly a safe
choice, because  all S-matrix elements are computed without adjusting
further parameters when taking the LSZ-limit. 
  Throughout  this paper, we adopt
an on-shell definition for the masses and in particular require
mass diagonalization for massive/massless particles on-shell. In
the abstract approach, such on-shell conditions are crucial,
not only 
for physical particles but also for unphysical fields, when one
finally wants  to prove unitarity of the physical
S-matrix \cite{BRS76, KUOJ78}.
 As far as the residua are concerned, we remain quite general in the
construction, and do not specify special conditions. We finally see
 that some of the normalization conditions of residua can
be eliminated by requiring a simple form of rigid Ward identities,
but this is not essential at any stage of the procedure. The critical
point in the analysis is the observation that  on-shell
conditions indeed fix more parameters than there are naive  invariants.
Requiring symmetries in their explicit tree form,  one is
unable simultaneously to adjust the $W$-mass and to diagonalize 
 the neutral mass matrix at the mass of the $Z$-boson
and at $p^2 = 0$. 
    These normalization conditions are also deeply connected with
 off-shell infrared divergencies to higher orders.
It has been pointed out already
 in \cite{AOHI82} that complete on-shell schemes
are compatible with the Slavnov-Taylor identity, 
and there are scattered remarks in the literature  that
on-shell schemes are in agreement with the symmetries if the
transformations are themselves subject to renormalization 
(see e.g.~\cite{JB1}). But neither the Slavnov-Taylor identity nor
rigid or local Ward-identities have been given in an explicit 
 form
valid for the Green functions of the standard model.
The Slavnov-Taylor identity in its homogeneous form as given 
in \cite{AOHI82} 
is not quite an adequate choice for the 
$SU(2) \times U(1)$-symmetry of the standard model, since
one has to split off the abelian factor explicitly as done
in \cite{BABE78}. In terms of on-shell fields, 
all symmetry transformations depend  on
the weak mixing angle in the tree approximation,
and it is
seen, that due to off-shell infrared divergencies, the symmetry
operators have to be modified order by order in perturbation theory.
For this reason we start the analysis by characterizing the
symmetry transformations by algebra and field content, and
find in this way
all higher order deformations which are compatible with
the symmetries. These general symmetry operators
 finally allow us  to construct unique Green functions 
in the on-shell schemes, without introducing off-shell infrared
divergencies. 

For the present paper, we  restrict ourselves to a diagonal quark
mass matrix, because we are mainly interested in the
renormalization of the vector sector. Apart from this we
stay quite general and proceed as far as possible along the lines
of concrete calculations. In particular, we use the general $R_\xi$-gauges,
although in a modified form with an auxiliary field
which couples to the gauge fixing functions.
Particular
 attention is paid to the solution of the classical approximation,
which gives the local four-dimensional invariants of symmetry
transformations. In the higher order construction of finite Green functions
we use the BPHZL scheme
 \cite{ZIM69, LOW76}. In this scheme, massless
particles are treated quite systematically by establishing
those normalization conditions in the scheme which are necessary
for the computation of finite Green functions to all orders. 
These normalization conditions are essentially the conditions
for mass diagonalization of massless/massive field at $p^2 = 0$
(i.e.~for the $Z$-boson and photon and the respective Faddeev Popov fields)
and are established in the above-cited on-shell schemes by adjustment
of counterterms.

The plan of this paper is as follows:
In section 2, we give the classical action  in
renormalizable gauges compatible with rigid symmetry
and local $U(1)$-gauge symmetry. We also present
the symmetry transformations of the tree approximation in
a functional form. These are the Slavnov-Taylor identity,
rigid Ward-identities and the local $U(1)$ Ward identity. 
In section 3, we outline the method of algebraic
renormalization. In section 4 we
solve the algebra of symmetry operators and obtain
the general consistent symmetry operators of the standard model.
 Section 5 is
devoted to solving the symmetries
for  the local  four-dimensional field polynomials.
This analysis allows us to give the free parameters
of the model and also to list the invariant
counterterms of higher orders. In  section 5.4 
a complete
treatment of the ghost equations is also included.
 In section 6, we derive the
Callan-Symanzik equation of 1-loop order.
By means of symmetric differential operators, it is
possible to characterize symmetric nonlocal contributions of
higher orders and in particular
to 
 determine the  independent parameters of the theory in a
scheme-independent
way.  In section 7, we proceed 
 to higher orders and prove that Green functions can be constructed
in agreement with the infrared normalization conditions
to all orders, if one takes into account the modifications
of the symmetry operators to higher orders as suggested by the tree
approximation. 
\newpage

\newsection{The tree approximation of the standard model}
\newsubsection{The gauge invariant part of the action}

The standard model of electroweak interactions is a non-abelian
gauge theory with the non-semisimple gauge group 
$SU(2) \times U(1)$.  The gauge structure is essentially
determined in the matter sector: It is seen, that the
matter currents of weak interactions, the charged current $J _{CC}^\mu$
and the neutral current $J_{NC}^\mu $, together with
the electromagnetic current $j_{em}^\mu$  form a closed representation 
with respect to $SU(2) \times U(1)$
\cite{GLA61}. In order to embed these currents
into a gauge theory, one groups the fermions into left-handed doublets,
which transform under  
the fundamental representation of $SU(2)\times U(1)$,
and right handed singlets, which only transform with respect
to the abelian subgroup. The decomposition of the Dirac spinors
into left and right handed fields is defined by the following projections:
\begin{eqnarray}
f^L = \hfrac 12 (1 -\ga _5) f  & \qquad  & f^R = \hfrac 12 (1 +\ga _5) f 
\nonumber \\
\overline {f^L} = 
\bar f \hfrac 12 (1 + \ga _5) 
& \qquad & \overline {f^R} = \bar f \hfrac 12 (1 -\ga _5) 
\end{eqnarray}

The fermions appear in families: Each family consists of a neutrino
$\nu_i$, a charged lepton $e_i$ with electric charge $Q_e = -1$,
and the up and down-type quarks $u_i$ and $d_i$ with charge
$Q_u = \frac 23 $ and $Q_d = - \frac 13 $. For simplicity we suppress
the colour index of the quarks throughout the paper.
 The lepton
  doublets $F^L_{l_i}$ and quark doublets $F^L_{q_i}, i = 1,2,3$,
are given by
\begin{eqnarray}
\label{lepd}
F^{L}_{l_i} = \left( \begin{array}{c} \nu^L _i \\
                                    e^L _i \end{array} \right)
 & = & \left( \begin{array}{c} \nu^L _e \\
                                    e^L  \end{array} \right) \,
\left( \begin{array}{c} \nu^L _{\mu} \\
                                    \mu^L \end{array} \right) \,
\left( \begin{array}{c} \nu^L _{\tau} \\
                                    \tau^L  \end{array} \right)
\\
F^{L}_{q_i}  = \left( \begin{array}{c} u^L _i \\
                                    d^L _i \end{array} \right)
& = & \left( \begin{array}{c} u^L  \\
                                    d^L  \end{array} \right) \,
\left( \begin{array}{c} c^L  \\
                                    s^L \end{array} \right) \,
\left( \begin{array}{c} t^L  \\
                                    b^L  \end{array} \right)
\end{eqnarray}
 The singlets only comprise the charged fermions:
\begin{equation}
f^R_i = e_i^R, u_i^R , d_i^R
\end{equation}

The $SU(2)$ and $U(1)$ gauge transformations $(\a = +,-,3)$:
\begin{eqnarray}\label{gtferm}
\epsilon_\a (x)  \delta_\alpha  F^L_{\delta_i} 
 =   i \epsilon_\a(x)  {\tau^T_\alpha  \over 2}  F^L_{\delta_i}
& \qquad &  \epsilon_\a(x) \delta _\a f_i^R = 0 \nonumber \\
\epsilon_4 (x) \delta _4 F^L_{\delta_i}  = -
i \epsilon_4(x)  \frac {Y_W^{\delta}}2 F^L_{\delta_i} 
& \qquad &  \epsilon_4(x) \delta_4 f_i^R = -i Q_f f_i^R
\end{eqnarray}
give rise to the matter currents of  electroweak interactions
\begin{eqnarray}\label{current}
J^\mu_+ = -\frac 12 \sum_{\delta_i} \overline {F^L_{\delta_i}} \ga ^\mu 
\tau _- F^L_{\delta_i} & \qquad &
J^\mu_- = -
\frac 12 \sum_{\delta_i}
 \overline {F^L_{\delta_i}} \ga ^\mu \tau _+ F^L_{\delta_i} \nonumber \\
J^\mu_3 = -\frac 12 
\sum_i \overline {F^L_{\delta_i}} \ga^\mu \tau _3 F^L_{\delta_i}
& \qquad &
J^\mu
_4 =  \frac 12 \sum_{\delta_i} Y^{\delta}_W
  \overline {F^L_{\delta_i}} \ga^\mu F^L_{\delta_i
}
      + \sum_{f_i} Q _f\overline {f^R_i} \ga^\mu f^R_{i} 
\end{eqnarray}
with $\delta = l, q $.
If one
identifies  out of the neutral currents the electromagnetic
current
\begin{equation}
j^\mu_{em} = \sum_{i,f} Q_f \bar f_i \gamma ^\mu f_i =  J^\mu_4 -J^\mu _3
\end{equation}
the weak hypercharge and the electric charge are related according
to the Gell-Mann  Nishijima formula:
\begin{equation}
\label{gmn}
\frac 12 (\tau_3 + Y _W ) = Q
\end{equation}
which means explicitly
\begin{equation}
Y_W ^{l} = -1 \quad \hbox{and} \quad Y_W^{q} = \frac 13 
\end{equation}

In (\ref{gtferm}) and (\ref{current}) $\tau _\alpha, \alpha = +,- 3$,
denote the generators of the charged 
fundamental representation of $SU(2)$. They are defined by
\begin{equation}
\label{pauli}
\tau_+ = \left( \begin{array} {cc} 0& \sqrt 2 \\ 
                                  0 & 0 \end{array}\right) \quad 
\tau_- = \left( \begin{array} {cc} 0& 0 \\ 
                                  \sqrt 2 & 0 \end{array}\right) \quad 
\tau_3 = \left( \begin{array} {cc} 1& 0 \\ 
                                  0 & -1 \end{array}\right) 
\end{equation}
and satisfy the following commutation relations:
\begin{equation}
\bigl[ \tau_\alpha , \tau _\beta \bigr]  =  2 i \epsilon _{\alpha
\beta \gamma} \tau_\gamma^T
\end{equation}
The structure constants $\epsilon _{\alpha
\beta \gamma} $ are imaginary and completely antisymmetric in
all three indices:
\begin{equation}
\label{eabc}
\epsilon _{+-3} = -i
\end{equation}
According to the Noether construction of gauge theories
 the matter action consists of the
kinetic terms and the currents coupled to a $SU(2)$-triplet of
vector fields $W^\mu_\alpha , \alpha = +,-, 3$, and an abelian vector
field $W^\mu_4$
\begin{eqnarray}
\label{gamat}
\Gamma_{matter}  &=& \sum_{i=1}^{N_F} \int \Bigl(  i \overline
 {F^L_{l_i}}  \dslash  F^L_{l_i}
       +  i \overline {F^L_{q_i}}  \dslash
        F^L_{q_i} + 
 i \overline {f^R_i}  \dslash f^R_{i}\\
                && \phantom{\sum \int }
     - g_2 (W^\mu_+ J_{\mu -} + W^\mu_- J_{\mu +} + W^\mu_3 J_{\mu 3})
     - g_1 W^\mu_4 J_{\mu4}  \Bigr)     \nonumber \\
& \equiv & \sum_{i=1}^{N_F} \int \Bigl( \overline
 {F^L_{l_i}} i \Ds F^L_{l_i}
       + \overline {F^L_{q_i}} i \Ds F^L_{q_i} + \overline
 {f^R_i} i \Ds f^R_{i} \Bigr) \nonumber
\end{eqnarray} 
The covariant derivatives are therefore given by:
\begin{eqnarray}
  D^{\mu} F^L_{\delta_i} &=& \partial^\mu F^L_{\delta_i} - i 
g_2
 \frac{{\tau}_\a}{2} F^L_{\delta_i} W^{\mu }_\a + i g_1 \frac {Y_W^\delta} 2
 F^L_{\delta_i} W^{\mu}_ 4\qquad \delta= l,q\\
  D^{\mu} f_i^R &=& \partial^\mu f_i^R  +
 i g_1 Q_f  f_i^R W^{\mu}_4 \nonumber
\end{eqnarray}
The gauge transformations of the vectors are uniquely determined from
gauge invariance of the matter action:
\begin{equation}
\label{gtvec}
\begin{array}{ccl}
\epsilon_\a (x)  \delta_\alpha W_\beta^\mu  & =&  \bigl( \tilde I_{\alpha\beta}
\frac 1 {g_2} \partial^\mu +
W^\mu _\gamma \tilde I _{\gamma \gamma'}\epsilon_{
\gamma' \beta \alpha} \bigr)
\epsilon_\alpha (x)  \\
\epsilon_\alpha (x)  \delta_\alpha W^\mu_4  & = & 0 \end{array}
 \qquad 
\begin{array}{ccl} \epsilon_4(x)  \delta_4 W^\mu_\alpha  & =& 0 \\
\epsilon_4 (x)  \delta_4 W^\mu_4 & = &
\frac 1 {g_1} \partial^\mu \epsilon _4 (x) \end{array}
\end{equation} 
The matrix $\tilde I_{\a \b}$ is the charge conjugation matrix:
\begin{equation}
\label{tildei}
{\tilde I} = 
\left(\begin{array}{cccc}
0&1&0&0\\ 1&0&0&0\\ 0&0&1&0 \\ 0&0&0&1 \end{array}\right) \quad\quad
\begin{array}{l}
\tilde { I} _{+-} = \tilde { I} _{-+} =\tilde { I} _{33}=
\tilde { I} _{44} = 1 \\
\tilde { I} _{\a\b} = 0,\: \hbox{ else}   \end{array}
\end{equation}
From (\ref{gtvec}) the Yang-Mills part which involves the kinetic terms of
the vectors is determined
\begin{eqnarray}
\label{gaym}
  \Gamma_{YM} & = &
  -\frac 1 4 \int \Bigl( G_\alpha^{\mu\nu}\tilde{I}_{\a \a'}G_{\mu\nu \a'}
                          + F^{\mu\nu} F_{\mu \nu} \Bigr)
\end{eqnarray}
with
\begin{eqnarray}
  G^{\mu\nu}_\a & = & \partial ^\mu W_\a^{\nu} - \partial^{\nu} W_\a^\mu
  +  g_2 \tilde{I}_{\a\a'} 
  {\epsilon}_{\a'\b \gamma} W^{\mu}_\beta W^{\nu}_\gamma\\
  F^{\mu\nu} & = & \partial ^\mu W_4^\nu - \partial^{\nu} W_4^\mu
\end{eqnarray}

The bosons of weak interactions as well as the charged fermions are
massive. The mass terms break chiral gauge invariance and have
to be generated by the spontaneous breaking of the gauge symmetry.
In the standard model all the masses are generated by introducing
one complex scalar doublet $\Phi$ and its complex conjugate $\tilde \Phi$:
\begin{equation}
 \Phi \equiv\left(
    \begin{array}{c}
      \phi^+(x)\\
    \frac  1{\sqrt 2}(H(x) + i\chi(x))
    \end{array}
  \right) 
\qquad \tilde \Phi \equiv i\tau_2 \Phi^* = \left(
    \begin{array}{c}
            \frac 1{\sqrt 2}(H(x) - i\chi(x)) \\
            -\phi^-(x) 
    \end{array}
  \right) 
\end{equation}
$\phi^\pm$ are charged, $H$ and $\chi$ neutral  scalar fields.
The doublet transforms under the fundamental representation and includes
in its transformation a constant shift $\mathrm v$ into the
direction of the neutral component of the scalar doublet:
\begin{eqnarray}
\label{gtscal}
\epsilon_\a (x)  \delta_\alpha  \Phi
& = &  i \epsilon_\a (x)  {\tau^T_\alpha  \over 2}  ( \Phi + \hbox{v})
\nonumber \\
\epsilon_4 (x) \delta _4 \Phi  & = &
-i \epsilon_4(x)  \frac {Y_W^{s}}2 ( \Phi + \hbox{v})
\end{eqnarray}
with
\begin{equation}
\label{shift}
{ \hbox{v}} = \left( \begin{array}{c} 0 \\
                                 \hbox{$\frac 1{\sqrt 2}$}
   v \end{array} \right)
\end{equation}
The weak hypercharge is determined from (\ref{gmn})
\begin{equation}
Y_W^{s} = 1
\end{equation}
As a response to the transformation (\ref{gtscal}), the 
gauge invariant parts of the action $\Gamma_{scalar}$
and $\Gamma_{Yuk}$, depend on the shift:
\begin{eqnarray}
\label{gascal}
\Gamma _{scalar}    &=&\int\Bigl( (D^\mu\Phi 
 )^\dagger D_{\mu}\Phi -\frac 1 2\frac{m_H^2} {v^2}
(\Phi^\dagger\Phi +   {\hbox{v}} 
 ^\dagger \Phi
+ \Phi^\dagger   {\hbox{v}})  ^2 \Bigr)
\\
\label{gayuk}
\Gamma_{Yuk}  & = &- \sum_i^{N_F} \int \frac {\sqrt 2} v\bigl( 
   m_{e_i}  
\overline
 {F^L_{l_i}} (\Phi +   {\hbox{v}}) e^R_i \nonumber \\
        & & \phantom{\sum \int \frac{sqrt 2} v}
         +  m_{u_i}  \overline
 {F^L_{q_i}} (\Phi +   {\hbox{v}}) u^R_i 
                 + m_{d_i}  \overline
 {F^L_{q_i
}} (\tilde \Phi + \tilde{  {\hbox{v}}}) d^R_i + \hbox{h.c.} \bigr) 
\end{eqnarray}

The Yukawa interaction contains via the shift all mass terms of 
the fermions $m_{f_i}$. We have chosen the couplings of the
Yukawa interactions in such a form, that the mass terms are parametrized
by the mass of the respective fermions.
 For the purpose of this paper we  forbid 
mixing between different
families  and especially assume CP-invariance throughout the
paper. 

The scalar part consists of the kinetic terms of the scalars
and the scalar potential, which includes the mass of the
Higgs field $m_H^2$. In order to have a proper particle 
interpretation, we have arranged the terms such that the 
contributions linear in $H(x)$ drop out.
Via the covariant derivative 
\begin{equation}
  D_{\mu}\Phi = \partial_{\mu}\Phi - i  ( g_2
 \frac{{\tau}_\a}{2} W_{\mu \a}  - g_1 \frac {Y_W ^{s}}2 
 W_{\mu 4} ) ( \Phi + \mathrm{v})
\end{equation}
the masses   of the gauge fields are generated by eating  up
 the massless Goldstone bosons $\phi_+ , \phi_-$ and $\chi $:
\begin{equation}
\frac 12 {g_2 ^2 v^2 \over 4} 
(2 W^\mu_+ W_{\mu -} + W^\mu_3 W_{\mu 3})
+  {g_2 g_1 v^2 \over 4} W^\mu _3 W _{\mu4} 
+\frac 12 {g_1 ^2 v^2 \over 4} W^\mu_4  W _{\mu 4} 
\end{equation}
Physical fields are constructed by diagonalizing the mass matrix
with an orthogonal matrix:
\begin{equation}
\label{othetaw}
W^\mu_\alpha = O_{\a a} (\theta_W) V^\mu_a \qquad
O_{\a a} (\theta_W) = 
\left(\begin{array}{cccc}
1&0&0&0\\ 0&1&0&0\\ 0&0&\cw&-\sw\\ 0&0&\sw & \cw \end{array}\right) 
\end{equation}
The fields which are generated by the rotation are the physical
on-shell fields $V^\mu_a = (W^\mu_+, W^\mu_- , Z^\mu, A^\mu)$.
Throughout the paper roman indices $a,b,c$ are reserved  to on-shell
indices $a,b,c = +,-,Z,A$, whereas Greek indices $\a , \beta, \gamma
$ denote group indices of $SU(2) $ and $U(1)$  $\a , \beta, \gamma =
+,-,3,4$.
In the tree approximation 
one calculates the following relations between the ratio of the
gauge parameters and
 the weak mixing angle $\theta_W$:
\begin{equation}
\label{coupthetawrel}
{g_1 \over g_2} = \tan \theta _W
\end{equation}
$W^\mu_\pm$ are the charged bosons of weak interactions with mass $M_W^2$,
$Z^\mu$ is the neutral boson with mass $M_Z^2$, $A^\mu$ the massless photon:
\begin{equation}
M_W^2 = {g^2_2 v^2 \over 4} \qquad M_Z^2 =  {g^2_2 
v^2 \over 4 \cos^2 \theta_W}
\end{equation}
If one eliminates the parameters $\theta_W$ and $ v$ in favour of
the masses $M_W $ and $M_Z$, one arrives at the on-shell parameter set
\begin{equation}
\label{onshellpar}
M_W,\, M_Z, \, m_{f_i}, \, m_H 
\end{equation}
which specifies 
the  particles   by their masses and electric charge.
The weak mixing angle is then defined by the mass ratio of the
W- and Z-mass
\begin{equation}
\label{weakmix}
\cos \theta _W \equiv {M_W \over M_Z} \, .
\end{equation}
If one chooses the on-shell set for parametrizing the free parameters
of the standard model, then one remains with one coupling
constant, which in the QED-like parametrizations
is taken to be the coupling of  the electromagnetic current to the photon:
\begin{equation}
\label{emcoup}
e = g_2 \sin \theta_W
\end{equation}

The gauge invariant  part of the classical action $\Gamma_{GSW} $ 
\cite{GLA61, WEI67, SAL68}
is given by the sum
of the gauge invariant parts (\ref{gamat}) (\ref{gaym}) 
(\ref{gascal}) 
and (\ref{gayuk}):
\begin{equation}
\label{gagsw}
\Gamma_{GSW} =
\Gamma_{YM} + \Gamma_{scalar} + \Gamma_{matter} + \Gamma_{Yuk}
\end{equation}
It is completely specified by the gauge transformations, the masses
of the interacting particles,  their electric charge and the
electromagnetic coupling.

We want to summarize the gauge transformations of the on-shell fields
within the QED-like on-shell parameter set 
(\ref{onshellpar}) and (\ref{emcoup}). In the spirit of the subsequent
considerations we express the gauge transformations thereby
in a functional operator acting on
$\Gamma _{GSW}$:
\begin{eqnarray}
\label{gaugeward}
 \Bigl( - {\mathbf w}_+ -  
\frac \sw e \partial ^\mu {\delta \over \delta W^\mu _-} \Bigr)
\Gamma_{GSW}  & = & 0 \\
 \Bigl( - {\mathbf w}_- -  \frac \sw  e 
\partial ^\mu {\delta \over \delta W^\mu_+} \Bigr)
\Gamma_{GSW}  & = &  0 \\
 \Bigl(- {\mathbf w}_3 -  \frac \sw e \partial ^\mu \bigl( \cw 
{\delta \over \delta Z  ^\mu} - \sw {\delta \over \delta A ^\mu} \bigr) \Bigr)
\Gamma_{GSW}  & =  & 0 \\
 \Bigl(\phantom{-}{\mathbf w}_4 -  \frac  \cw  e\partial ^\mu \bigl( \sw 
{\delta \over \delta Z  ^\mu} + \cw {\delta \over \delta A ^\mu} \bigr) \Bigr)
\Gamma_{GSW} & =  & 0
\end{eqnarray}
 The  functional operators of $SU(2)$-transformations
are given by ($\a = +,-,3$)
\begin{eqnarray}
\label{wardna1} 
{\mathbf w}_\alpha & =  \tilde I_{\alpha\alpha'} & \Biggl(
  V^\mu _b O^T_{b\beta}(\theta_W) \hat \ve_{\beta\gamma\alpha}
           O_{\gamma c}(\theta_W)  \tilde I _{cc'}
\frac{\delta}{\delta V^\mu _{c'}}  \nonumber \\
&  & +   i (\Phi + {\mathrm v}) ^\dagger
\frac { \tau _{\alpha'} } 2  \frac{\overrightarrow 
\delta}{\delta \Phi^\dagger}
 -  i \frac{\overleftarrow
\delta}{\delta \Phi} \frac { \tau _{\alpha'} } 2  (\Phi +{\mathrm v})  
\nonumber \\
& &  + \sum_{i=1}^{N_F} \sum_{\delta = l,q}
 \Bigl( i \overline{F ^L_{\delta_i}} \frac {\tau _{\alpha'} } 2  
\frac{\overrightarrow \delta}{\delta \overline{ F^L_{\delta_i}} }
 -  i \frac{\overleftarrow
\delta}{\delta F^L_{\delta_i}} \frac { \tau _{\alpha'} } 
  2  F^L_{\delta_i}  \Bigr) \Biggr)
\end{eqnarray}
The transformation of the on-shell vectors depends 
on the weak mixing angle:
\begin{equation}
\label{hateabc}
O^T_{b\beta}(\theta_W) \ve _{\b \ga \a}
O_{\ga c} (\theta_W) \equiv \hat \varepsilon_{bc,\alpha} = \left\{
    \begin{array} {ccc}
       {\hat \ve}_{Z+,-} &=& -i \cos\theta_W\\
       {\hat \ve}_{A+,-} &=& i \sin\theta_W \\
        \hat \ve_{+-,3}  &=& -i 
    \end{array}\right.
\end{equation}
The abelian Ward operator is given by
\begin{eqnarray}
\label{wardab1}
{\mathbf w}_4 & = & 
\frac i2 (\Phi +{\mathrm v} ) ^\dagger   \frac{\overrightarrow 
\delta}{\delta \Phi^\dagger
}
 -  \frac i2 \frac{\overleftarrow
\delta}{\delta \Phi}  ( \Phi + {\mathrm v} )   \nonumber \\
& & + \sum^{N_F}_{i = 1} \Biggl( \sum_{\delta = l,q}
Y_W^{\delta} \Bigl( \frac i2 
 \overline{F _{\delta_i}^L}   \frac{\overrightarrow 
\delta}{\delta \overline{ F_{\delta_i}^L} }
 -  \frac i2 \frac{\overleftarrow 
\delta}{\delta F_{\delta_i}^L}    F_{\delta_i}^L 
\Bigr)   \nonumber \\
& & \phantom{+ \sum^{N_F}_{i=1} \Biggl(} - \sum_{f^R}
  Q_f \Bigl( i \overline{f_i ^R}   \frac{\overrightarrow 
\delta}{\delta \overline{ f_i^R} }
 -  i \frac{\overleftarrow
\delta}{\delta f_i^R}   f_i^R  
 \Bigr) \Biggr) 
\end{eqnarray}
In the notation we understand summation over all fermion singlets and
doublets.
The Ward operators satisfy the local $SU(2) \times U(1)$-algebra:
\begin{eqnarray}
\label{localg}
\bigl[ {\mathbf w}_\a (x), {\mathbf w}_\beta (y) \bigr] & = &
\delta (x-y)
\ve_{\a\beta \gamma} \tilde I_{\ga \ga'} {\mathbf w}_{\ga'} (x)
\\ 
\bigl[ {\mathbf w}_\a (x), {\mathbf w}_4 (y) \bigr] & = &
0 \nonumber
\end{eqnarray}

It is obvious that $\Gamma_{GSW}$  is also invariant with respect 
to rigid transformations which are obtained by taking the infinitesimal
parameters $\epsilon_\a $ as constants or, equivalently, by integrating
the local Ward operators $(\alpha = +,-,3,4)$:
\begin{equation}
\label{wardrig1}
{\cal W_\alpha} \Gamma_{GSW}  = 0 \qquad \mbox{and} \quad
{\cal W_\alpha} = \int {\mathbf w}_\alpha 
\end{equation}
Rigid symmetries can be established for off-shell Green functions
to all orders of perturbation theory in a modified form and turn out
together with the abelian Ward identity to be important ingredients
for defining the standard model in its quantized version.

\newsubsection{The gauge fixing and rigid transformations}

 The perturbative construction of Green functions  and finally the S-matrix
starts with
 the specification of the free fields and their respective propagators.
In the standard model the scalars $\phi_\pm$ and $\chi$ are unphysical
fields being absorbed into the longitudinal polarization of the
massive vectors $W_\pm$ and $Z$. Eliminating them by a gauge transformation,
however, leads to propagators with a bad ultraviolet behaviour,
and renormalizability by power counting is not evident anymore.
For a systematic treatment of higher orders
one better uses the renormalizable gauges as the $R_\xi$-gauges.
If  one constructs the off-shell Green functions 
in the renormalizable gauges,
one is able to refer to power counting properties of renormalized
perturbation theory and, especially, to the quantum action principle.
In the end one has then to prove unitarity of the physical S-matrix, i.e.\
it has to be shown that the unphysical fields, as
the scalar component of the vectors and the Goldstone bosons, do not
appear in physical scattering processes.

The free field propagators are calculated from the bilinear
parts of the gauge invariant action $\Gamma_{GSW}$
and the gauge fixing part $\Gamma_{g.f.}$:
\begin{equation}
\label{gabil}
\Gamma^{(bil)} = \Ga^{(bil)} _{GSW} + \Ga_{g.f.}
\end{equation}
The gauge fixing  in the $R_\xi$-gauges is given
by:
\begin{eqnarray}
\Gamma_{ {g.f.}}&= 
& \int - \frac 1 { \xi _W} F_+ F_- -\frac 1 {2\xi_Z } F_Z F_Z
- \frac 1 {2 \xi_A} F_A F_A
\end{eqnarray}
with
\begin{eqnarray}
F_{\pm}&\equiv&\partial_{\mu}W^{\mu}_{\pm} \mp iM_{W}\zeta_{W}\phi_{\pm}
\nonumber{}\\
F_Z&\equiv& \partial_{\mu}Z^{\mu}-M_{Z}\zeta_{Z}{{\chi}}\\
F_A&\equiv & \partial_{\mu}A^{\mu}  \nonumber 
\end{eqnarray}
The free field propagators are seen to have a good UV-behaviour
which guarantees renormalizability by power counting:
\begin{equation}
G_{\varphi_k \varphi_l} \longrightarrow p ^{-2 (2-d_{\varphi_k})}
\quad \mbox{if} \quad p^2 \to \infty
\end{equation}
where $d_{\varphi_\alpha}$ is
 the mass dimension of the field $\varphi_k$. They also have
 good infrared behaviour, i.e.~they diverge for the massless particles
not stronger than $ p^{-2}$ as for the photon field. $p^{-4}$ 
infrared divergent terms are removed by introducing mass terms for
the would-be Goldstone fields into the gauge fixing functions.

Adding such a gauge fixing with arbitrary gauge parameters to the
action one does not keep any knowledge about the $SU(2) \times U(1)$
structure of the standard model in the free field propagators,
but treats the bilinear action as if it were composed of several 
$U(1)$-factors.
But as a consequence of the gauge construction $\Ga ^{(bil)}_{GSW} $ 
{\it has}
definite transformation properties under rigid unbroken 
$SU(2) \times U(1)$: 4-dimensional terms are invariants; 
the 3-dimensional terms,
which are the fermion mass terms and the mixed scalar-vector 
terms,  together with their 
variations  transform as a vector under unbroken $SU(2) \times U(1)$.
The mass terms of the vectors are composed with their variations
and second variations to a second rank tensor. In order not to spoil
these transformation properties by the gauge fixing part, one has
to choose:
\begin{equation}
\label{riggauge}
\xi \equiv
\xi_W = \xi _Z  = \xi_A \qquad \mbox{and} \qquad 
\zeta \equiv \zeta_Z = \zeta _W
\end{equation}
Instead of requiring the complicated transformation behaviour
of the mass terms
one can introduce an external scalar field 
$\hat\Phi $ and its complex conjugate, which couples
to the masses and their variations (see also \cite{DEWEI}):
\begin{equation}
\hat \Phi = {\hat \phi^ + \choose \frac 1 {\sqrt 2 }
( \hat H + i \hat \chi )} \quad
\hat \Phi^\dagger = {\hat \phi^ - \choose \frac 1 {\sqrt 2 }
( \hat H - i \hat \chi )} 
\end{equation} 
Under rigid transformations it transforms in the same way 
as  the scalar doublet $\Phi$, 
but includes a different shift parameterized by $\zeta {\mathrm v}$
into the transformation $(\epsilon _\a, \epsilon_4 = \,$ const.):
\begin{eqnarray}
\label{gtextscal}
\epsilon_\a  \delta_\alpha  \hat \Phi
 & =  & i \epsilon _\a  {\tau^T_\alpha  \over 2}  ( \hat \Phi + 
\zeta \hbox{v})
\nonumber \\
\epsilon_4  \delta _4\hat \Phi   & = & -
i \epsilon_4  \frac {Y_W^{s}}2 ( \hat \Phi + \zeta \hbox{v})
\end{eqnarray} 
Algebraically  this  is the same procedure as one carries out
 if one introduces the scalar doublet and the Higgs mechanism
for generating the masses of the fermions and vectors,  but the
external field is required to be
non-propagating and does not have physical meaning.
The gauge fixing functions can be enlarged by the external field
in such a way that they
transform  as a vector under the
adjoint representation:
\begin{equation}
\label{gfrig}
F_a \to {\cal F}_a = \partial_{\mu}V_a^{\mu}- 
i \frac e{\sin \theta_W}
\Bigl((\hat \Phi + \zeta \mbox{v} )^\dagger \frac {\tau^T_{a}} 2
( \Phi +  \mbox{v} ) -
( \Phi + \mbox{v} )^\dagger \frac {\tau^T_{a}} 2
( \hat \Phi + \zeta \mbox{v} ) \Bigr) 
\end{equation}
 The gauge fixing part is then invariant under the rigid transformations,
if one includes the transformations of the external fields:
\begin{equation}
\label{gaugefix}
\Gamma_{g.f.} = \int -\frac 1 {2\xi} {\cal F}_a \tilde I _{ab}
{\cal F} _b  \qquad \epsilon_\a \delta_\a \Gamma_{g.f.} = 0
\end{equation}
$V_a , a = +,-, Z,A $ are the on-shell fields, and the respective
representation matrices  $\tau_a $ are obtained by acting with
the orthogonal
matrix $O(\theta_W)$ (\ref{othetaw})  on $\tau_3$
and $G \mathbf 1 $:
\begin{equation}
\tau_a (G) = O^T_{a\a} (\theta_W) \tau_\a + O^T_{a4} G {\mathbf 1}
\end{equation}
Explicitly they read:
 \begin{eqnarray}
\label{tauphys}
\tau_Z(G)  & = & \cw \tau_3 + G \sw {\mathbf 1}  \nonumber \\
\tau_A (G) & = & - \sw \tau_3 +G \cw {\mathbf 1} 
\end{eqnarray}
The abelian parameter $G $ is not fixed by rigid invariance.
Choosing it
\begin{equation}\label{hyperext}
  G = -  \frac{\sin\theta_W}{\cos\theta_W} 
\end{equation}
 one
obtains for vanishing external fields the original gauge fixing with the
 parameters  according to (\ref{riggauge}). The masses of the
would-be Goldstones are  generated by the shift of the external field.
The transformation properties of the trilinear and the mass terms
are now governed by the transformation properties of the external
field $\hat \Phi$.

Modifying the functional operators of rigid transformations
(\ref{wardrig1})
 by the transformations of the
external field according to (\ref{gtextscal})
\begin{eqnarray}
\label{wardgf}
{\cal W}_\a &\to & {\cal W}_\a +  \tilde I_{\a\a'}\int \Bigl(
  i (\hat \Phi + \zeta {\mathrm v}) ^\dagger
\frac { \tau _{\alpha'} } 2  \frac{\overrightarrow 
\delta}{\delta \Phi^\dagger
}
 -  i \frac{\overleftarrow
\delta}{\delta \hat \Phi} \frac { \tau _{\alpha'} } 2  ( 
\hat \Phi +\zeta {\mathrm v}) \Bigr) \qquad \a = +,-,3  \\
{\mathbf w}_4 & \to & {\mathbf w}_4 +
\frac i2 (\hat \Phi +\zeta {\mathrm v} ) ^\dagger   \frac{\overrightarrow 
\delta}{\delta \hat \Phi^\dagger
}  
 -  \frac i2 \frac{\overleftarrow
\delta}{\delta \hat \Phi}  ( \hat \Phi + \zeta {\mathrm v} )   \nonumber 
\end{eqnarray}
we write the invariance properties of $\Gamma_{GSW} + \Gamma_{g.f.}$
in functional form:
\begin{equation}
\label{wardrig}
{\cal W }_\a \bigl(\Ga_{GSW} + \Ga _{g.f.} \bigr)= 0 
\end{equation}
Furthermore it is seen that the gauge transformation of 
the abelian subgroup is broken linearly in propagating fields: 
\begin{equation}
\label{wlocgf}
\bigl(\frac e \cw {\mathbf w}_4 -
 \partial ^\mu \bigl( \sw 
{\delta \over \delta Z  ^\mu} + \cw {\delta \over \delta A ^\mu} \bigr) \Bigr)
\bigl( \Gamma_{GSW} + \Ga_{g.f.} \bigr)
= - \frac 1 \xi \Box (\sw {\cal F} _Z + \cw {\cal F} _A )
\end{equation}
For this reason it is possible
to extend and interpret (\ref{wlocgf}) as a Ward identity for 
Green functions.

This   construction  of the gauge fixing sector 
is essential if one wants
to proceed to higher orders perturbation theory. Especially it
is seen that we need a local Ward identity of the form (\ref{wlocgf})
 for the Green functions in order to fix the weak hypercharge and
 electric charge in
a scheme independent way. 

Finally we want to mention
that  choosing  the parameter $G$ according to 
(\ref{hyperext}) is arbitrary and not
related to any symmetries. It turns out that this parameter as
well as an additional abelian gauge parameter are renormalized
in higher orders of perturbation theory.

\newsubsection{BRS-invariance and Faddeev-Popov ghosts}

The linear $R_\xi$-gauges break also in their covariant form
gauge invariance and especially bring about that the unphysical
fields, the scalar components of the vectors and the would-be
Goldstones, interact with the physical fields violating thereby 
unitarity of the physical S-matrix. For this reason one has to
introduce the Faddeev-Popov ghosts $c_a, a= +,-, Z,A $ with ghost
charge 1 and the respective antighosts $\bar c_a, a = +,-,Z,A $
with ghost charge  -1. They are anticommuting scalars with negative
norm and compensate the unphysical degrees of freedom introduced by
the gauge fixing, if one adds the ghost action in such a way,
that the complete action is invariant under BRS-transformations.

There a several approaches to introduce the Faddeev-Popov fields
\cite{FAD67}
into the perturbative formulation of gauge theories. One
way to proceed is to consider BRS-trans\-for\-mations in a first step
as an alternative way to characterize the Lie algebra of the
gauge group.  This approach is close to
the algebraic analysis which we carry out in the higher
order  construction, and therefore we outline the procedure
in the following: Starting from the gauge transformations of 
the fields as summarized in functional form in (\ref{wardna1}) and 
(\ref{wardab1})
one translates the infinitesimal parameters $\epsilon _\a (x)$ 
 into anticommuting parameters $c_\a (x), \a = +,-,3,4$.
 Considering the gauge transformations on the
on-shell fields (\ref{gaugeward})
 one is lead to carry out the orthogonal transformation $O(\theta_W)$
(\ref{othetaw})
on the ghosts as well
\begin{equation}
\label{onshellghosts}
c_\a = O_{\a a}(\theta_W) c_a \qquad c_a = (c_+,c_-, c_Z, c_A)
\end{equation}
In this procedure
is  quite some arbitrariness,
which has to be exploited in higher order perturbation theory for
a proper definition of massless ghost propagators (see section 5.4).
The BRS-transformations \cite{BRS75}
on the vector bosons $V^\mu_a = (W_+, W_-,Z , A)$,
the scalar doublet $\Phi$ and the fermion doublets and singlets
read in the physical on-shell parameterization:
\begin{eqnarray}
\label{brs}
{\mathrm s} V_{\mu a}&=&\partial _{\mu}c_a + 
\frac e{\sin \theta _W} \tilde I _{aa'} f_{a'bc}V_{\mu
  b}c_c\nonumber{}\\
{\mathrm s} \Phi &=& i \frac e{\sin{\theta_W}}
\frac{{\tau}_a(G_s)}{2} (\Phi+ {\mathrm v}) c_{
 a} \nonumber  \\
{\mathrm s} F^L_{\delta_i} & =& i
\frac e{\sin{\theta_W}}
 \frac{{\tau}_a(G_\delta)}{2} F^L_{\delta_i} c_{ a} \qquad \delta = l,q\\
{\mathrm s}  f_i^R&= & - i e Q_f \frac {\sin{\theta_W}}{\cos{\theta _W}}  
 f_i^R   c_Z-            ie Q_f f_i^R c_A \nonumber 
\end{eqnarray}
The matrices $\tau _a, a = +,-,Z,A $ are given in (\ref{tauphys})
and satisfy the algebra
\begin{equation}
\label{taualg}
  \left[{ {\tau}_a (G)}, { {\tau}_b (G)}\right]= 
  i  {f}_{abc} 
  \tilde{I}_{cc'} { \tau _{c'} (G)}
\end{equation}
with the structure constants
\begin{equation}
\label{fabc}
   {f}_{abc}  = 
O^T_{a\a}(\theta _W ) O^T_{b\b}(\theta _W ) \epsilon _{\a\b\ga}
O_{\ga c}(\theta_W )
 = \left\{
    \begin{array} {ccc}
       {f}_{+-Z}&=&- i \cos\theta_W\\
       {f}_{+-A}& =&i \sin\theta_W 
    \end{array}\right.
\end{equation}
The abelian parameter $G$ appearing in the BRS-transformations
is related to the weak
hypercharge according to the Gell-Mann Nishijima relation, explicitly:
\begin{equation}\label{hyper}
  G_k = - Y^{k}_W \frac{\sin\theta_W}{\cos\theta_W} \qquad
Y_W^{k}  =  \left\{
    \begin{array} {ccc}
       &1 & \hbox{for the scalar ($k=s$)} \\
       &$-1$& \hbox{for the lepton doublets ($k=l$)} \\
       & \hbox{$\frac 13$} & \hbox{for the quark doublets ($k=q$)} 
    \end{array}\right.
\end{equation}

The algebra of the functional operators 
(\ref{localg}), which contains the
complete information about the group structure, is translated into
 the BRS-transformation of the ghosts 
\begin{eqnarray}
\label{brsghost}
{\mathrm s} c_{a}&=&- \frac e{2 \sin \theta_W } \tilde I _{aa'} 
 f_{a'bc}c_bc_c
\end{eqnarray}
The representation equations and also the Jacobi identities are
now encoded in the nilpotency of the BRS-transformations: 
\begin{equation}
{\mathrm s}^2 \varphi_k = 0 \quad \mbox{with} \quad \varphi_k = 
V_{\mu a},\Phi, F_{\delta_i}^L, f_i^R,c_a
\end{equation}
From the construction it is obvious that the gauge invariant part
of the action
(\ref{gagsw}) is
 BRS-invariant:
\begin{equation}
\brs \Gamma_{GSW} = 0
\end{equation}

The gauge fixing (\ref{gaugefix}) breaks gauge invariance;
having introduced the anticommuting fields $c_a$ this breaking
  is absorbed
into the transformation of the antighosts:
\begin{equation}
\int -\frac 1 \xi  {\cal F}_a \tilde I_{aa'} \brs {\cal F}_{a'} - \brs
\bar c_a \tilde I_{aa'}
\brs  {\cal F}_{a'} \stackrel ! = 0
\end{equation}
Therefrom one obtains:
\begin{equation}
{\mathrm s} \bar c_a = - \frac 1\xi {\cal F}_a
\end{equation}
and 
\begin{equation}
\label{brsgf}
\brs \bigl(\Gamma_{g.f.} + \Gamma_{ghost} \bigr) =0 \quad \hbox{with}
\quad
\Gamma_{ghost} = - \int \bar c_a \tilde I _{a b } \brs {\cal F}_b
\end{equation}
The ghost action contains kinetic terms for the Faddeev-Popov fields,
which allows to introduce them as dynamical fields into the theory.

The BRS-transformation of the anti-ghosts is not nilpotent.
To remedy this situation one reformulates the gauge fixing part of
the action by introducing the auxiliary fields $B_a, a= +,-,Z,A$
\begin{eqnarray}
\label{gfixB}
\Gamma_{ {g.f.}}&= 
& \int\frac 12 \xi  B_a \tilde I _{ab} B_{b} + B_a \tilde{I}_{ab} 
{\cal F}_{b}
\end{eqnarray}
It can be transformed into the  usual form of the $R_\xi$
gauges by eliminating the
$B_a$-fields via their equations of motions:
\begin{equation}
\frac{\delta \Gamma}{\delta B _a } 
= \tilde I _{ab} (\xi  B_b +  {\cal F}_b) \, {\buildrel \ast \over =} \, 0
\quad\Longrightarrow \quad
B_a {\buildrel \ast \over =}   - \frac 1 \xi {\cal F}_a
\end{equation}
Therefore the propagators of vectors and scalars are not changed,
but in  addition one has mixed propagators between $B_a$-fields and
vectors and $B_a$-fields and scalars. 
The ghost action is likewise determined from (\ref{brsgf}),
but
the BRS-transformations turn out to be nilpotent also on the 
antighosts:
\begin{equation}
{\mathrm s} \bar c_a = B_a \qquad
{\mathrm s} B_a = 0 
\end{equation}
 For the algebraic characterization it is useful to
have nilpotency of BRS-transformations throughout and we refer 
to this form of the gauge fixing
 in the algebraic proof of renormalizability in higher
orders. Invariance under rigid transformation is maintained, if one
transforms the $B_a$-fields according to the adjoint representation
($\epsilon_\a, \epsilon_4 =$ const.)
\begin{equation}
\label{rigb}
\epsilon_\a \delta _\a B_b  =  
B_c \tilde I_{cc'}\hat\varepsilon _{c'b,\a}(\theta_W) \epsilon_\a  \qquad
\epsilon_4 \delta _4 B_b =  0 
\end{equation}
The tensor $\hat \epsilon _{bc,\a }(\theta_W)$
 is defined in eq.~(\ref{hateabc})

The gauge fixing functions ${\cal F}_a$
depend on the external scalar doublet
$\hat \Phi$ and we have to assign to them also definite transformation
properties  under BRS-transformations.
Transforming $\hat \Phi$  into an external 
anticommuting scalar doublet $\hat {\mathbf q}$ with ghost charge 1
\begin{equation}
{\mathrm s} \hat \Phi =  \hat {\mathbf q} \qquad
{\mathrm s} \hat {\mathbf q} = 0
\end{equation}
does this job and allows to distinguish the propagating and external
scalar fields algebraically.

Explicitly the ghost action is given by
\begin{eqnarray}
\label{gaghost}
\Gamma_{ghost} & = &   \int\biggl(- \bar c _a \Box \tilde I_{ab} c_b
 - \frac e \sw \bar c_a f_{abc} \partial( V_b c_c) \\
   & & \phantom{\int} + i \frac e {2\sw} \bigl(\hat{\mathbf q}
^\dagger \tau_a (G_s) (\Phi + {\mathrm v}) -
(\Phi + {\mathrm v})^\dagger \tau_a (G_s) \hat{\mathbf q} \bigr)
\bar c_a \nonumber \\
& & \phantom{\int}
- \frac e {4\sw} \Bigl( ( \hat \Phi + \zeta {\mathrm v} ) ^\dagger 
\tau_a(G_s) \tau_b(G_s) ( \Phi +  {\mathrm v} )  \nonumber \\
 & & \phantom{\int - \frac e {4\sw}} +
( \Phi +  {\mathrm v} ) ^\dagger \tau_b(G_s) \tau_a(G_s) (
\hat  \Phi +  \zeta {\mathrm v} ) \Bigr)\bar c_a c_b \biggr)\nonumber
\end{eqnarray}
$G_s$ is related to the weak hypercharge of the scalar doublets
according to (\ref{hyper}).
Via the shift of the external and the quantum scalar fields the
charged ghosts as well as the neutral Z-ghost
become massive, whereas the ghost associated with the photon field
remains massless. The bilinear part of the ghost action
\begin{eqnarray}
\label{bilghost}
\Gamma^{(bil)}_{ghost} & = &   \int\Bigl(- \bar c _a \Box \tilde I_{ab} c_b 
-  \zeta M_W^2 (\bar c_+ c_- + \bar c_- c_+) - \zeta M_Z^2 \bar c_Z c_Z
\Bigr) 
\end{eqnarray}
gives rise to free field propagators for the Faddeev Popov fields.

The ghost action is seen to be invariant under rigid transformations
if one assigns the following transformations under $SU(2) \times U(1)$
($\epsilon_\a, \epsilon_Y = $const.)
\begin{equation}
\label{rigghost}
\begin{array}{ccc}
\epsilon_\a \delta_\a \bar c_b & =& 
\bar c_c \tilde I_{cc'}\hat\varepsilon _{c'b,\a}(\theta_W) \epsilon_\a  
\\
\epsilon_4 \delta_4 \bar c_a & = & 0 \end{array}
\qquad 
\begin{array}{ccc}
 \epsilon_\a \delta_\a  c_a & = & 
 c_c \tilde I_{cc'}\hat\varepsilon _{c'b,\a}(\theta_W) \epsilon_\a  
\\
 \epsilon_4 \delta_4  c_a & = & 0 \end{array}
\end{equation}
In particular $\Gamma^{(bil)}_{ghost}$ transforms covariantly in the
same way as $\Gamma^{(bil)}_{GSW}$.

\newsubsection{The tree approximation: the Slavnov-Taylor identity}

In the last sections we have derived the classical action of the
standard model
\begin{equation}
\label{gacl1}
\Gamma_{cl} = \Gamma_{GSW} + \Gamma_{g.f.} + \Ga_{ghost}
\end{equation}
in a way that is invariant under BRS-transformations 
\begin{equation}
\brs \Gacl = 0
\end{equation}
Spontaneously broken rigid $SU(2) \times U(1)$-symmetry has been established 
by introducing an external scalar doublet $\hat \Phi$ into the
gauge fixing part of the action. 

In order to quantize the model in perturbation theory one has to
construct the Green functions of the interacting theory according
to the Gell-Mann  Low formula.
\begin{eqnarray}
\label{GML}
G_{\varphi_{i_1} ...\varphi_{i_n}}(x_1,...,x_n)
& = & \langle T \varphi_{i_1}(x_1) ...\varphi_{i_n}(x_n) \rangle \\
&= & R { \bigl\langle T \varphi^{(o)}_{i_1}(x_1) ...\varphi^{(o)}_{i_n}(x_n)
e^{i\Gamma_{int}(\varphi^{(o)}_k, \hat \Phi,
\hat {\mathbf q})} \bigr\rangle \over \bigl\langle T e^{i\Gamma_{int}
(\varphi^{(o)}_k, \hat \Phi, \hat {\mathbf q})} \bigr\rangle } \bigg| _
{\hat \Phi = 0 \atop
              q = 0} \nonumber
\end{eqnarray}
where $\varphi_k$ denotes the propagating  fields of
the standard model
\begin{equation}
\label{defvarphi}
\varphi_k = \left\{
\begin{array} {cc}  V^\mu_a, B_a, c_a, \bar c_a & \quad a = +,-, Z,A \\
                    \phi_\pm , H , \chi  &\\
                    \nu^L_i , e_i, u_i, d_i & \quad i = 1... N_F
\end{array} \right.
\end{equation}
$\Gamma_{int}$  includes all the interactions
 and the field polynomials
depending on the external fields and is obtained
by splitting off from the classical action
the bilinear part:
\begin{equation}
\Gamma_{cl} = \Gamma^{(bil)} + \Gamma_{int} 
\end{equation}
with
\begin{equation}
\label{gabilcl}
\Gamma^{(bil)} = \Gamma^{(bil)}_{GSW} + \Gamma_{g.f.}
\big|_{\hat \Phi = 0 \atop
        \hat{\mathbf q} = 0} +\Gamma^{(bil)} _{ghost}
\end{equation}
The index $(o)$ stands for free fields.

The formal expansion of the exponential yields
 the Green functions of the interacting theory in expressions
of time ordered vacuum expectation values of free fields.
These expressions are decomposed into a sum of products of free
field propagators and certain vertex factors according to Wick's
theorem. The combinatorics and the vertex factors are summarized
graphically in the Feynman rules. The free field propagators are
determined from $\Gamma^{(bil)}$.
The Feynman rules of the standard model are listed in
the literature and are given e.g.~in \cite{Denner} according to the
conventions we have adopted.

Due to the well-known ultraviolet divergencies the formal expansion 
of the Gell-Mann Low formula is not meaningful
in higher orders of perturbation theory and has to be rendered meaningful
  in the course of renormalization. (This is the sense of R in
eq.~(\ref{GML}).)  In the lowest order, the
tree approximation, the Green functions are well-defined and it has to
be shown, that the physical S-matrix, which is constructed from
these Green functions according to the LSZ reduction formula, is unitary
in the lowest order. This means, that one has  to verify 
that unphysical
particles do not contribute in physical scattering processes, and
that they  are canceled among each other. 
This cancellation mechanism is governed by the Slavnov-Taylor identity, 
which expresses consequences of the classical BRS-symmetry for the
off-shell Green functions.
 In order to derive the Slavnov-Taylor identity in the tree
approximation we introduce the generating functional
of Green functions:
\begin{eqnarray}
\label{genfun}
 & &
Z(j^\mu_a,j^B_a \jmath _a, \bar \jmath _a, J , J^\dagger, \eta _i , \bar \eta _i) 
 \\
 & & \; = \:\Bigl\langle T exp \biggl\{i \int dx \Bigl( \tilde I _{ab}
(j^\mu_a  V_{\mu b} + j^B_a B _b+ \bar\jmath _a c_b +
 \bar  c_a  \jmath _b) + \Phi^\dagger J + J^\dagger \Phi + \bar f_i\eta 
_i 
 + \bar \eta _i f_i \Bigr) \biggr\} \Bigr\rangle \nonumber
\end{eqnarray}
 In (\ref{genfun}) we understand
summation over on-shell field indices $a,b = +,-,Z,A$ and summation
over all fermions $f_i= \nu_i, e_i, u_i, d_i$.
The source functions are commuting ($j^\mu_a,j^B_a, J $) and anticommuting
($\jmath, \bar \jmath,\eta_i$) test functions. Electric
and $\phi\pi$-charge is assigned in such a way that the generating
functional is neutral. 
The Green functions are obtained by differentiation with respect to
the respective source functions.

Although the Green functions are the basic objects of the theory, 
for the purpose of renormalization one better refers to
the building blocks composing them.
These are the connected Green functions and the
one-particle-irreducible (1PI) Green functions.
The generating functional of connected Green functions 
\begin{equation}
Z_c ({j_k})
\equiv
Z_c(j^\mu_a, j^B_a, \jmath _a, \bar \jmath _a, J , J^\dagger, \eta _i , \bar \eta _i) 
\end{equation}
is defined by
\begin{equation}
\label{gfcon}
Z ({j _k}) = e ^{i Z_c ({j_k})}
\end{equation}
and one can show that the differentiation with respect to the sources
yields the connected Green functions in the diagrammatic expansion.
The generating functional of 1PI Green functions is obtained from
$Z_c ({ j_k})$ by Legendre transformation.
For this purpose one introduces the classical fields
\begin{equation}
\label{clfield}
\varphi_k^{cl}(x, j_i) =  {\delta Z_c(j_i)  \over \delta j_k (x)}
\qquad \varphi_k^{cl} (x,0) = 0
\end{equation}
and defines the generating functional of 1PI Green functions 
$\Gamma (\varphi_k^{cl})$ according to
\begin{eqnarray}
\label{gf1pi}
& &
Z_c(j^\mu_a, \jmath _a, \bar \jmath _a, J , J^\dagger, \eta _i , \bar \eta _i) 
 = \Gamma(V^{cl}_{\mu a},B_a^{cl},
 c^{cl}_a , \bar c^{cl}_a, \Phi^{cl} , 
{\Phi^{cl} }^\dagger, f^{cl}_i , \bar f^{cl}_
i) 
\\
& & \; + \int dx \Bigl( \tilde I_{ab}
(j^\mu_a  V^{cl}_{\mu b}+ j^B_a B_b+ \bar\jmath _a c^{cl}_b +
\bar   c^{cl}_a  \jmath _b) + {\Phi^{cl}}^\dagger J + J^\dagger 
\Phi^{cl} + \bar f^{cl}_i\eta 
_i 
 + \bar \eta _i f^{cl}_i \Bigr) \nonumber
\end{eqnarray}
Here the sources $j_k $ are understood as solutions of (\ref{clfield})
\begin{equation}
j_k = j_k(x,\varphi^{cl}_k) \qquad j_k(x, 0) = 0
\end{equation}
The 1PI Green functions are obtained by differentiating the
generating functional with respect to the classical fields $\varphi_k^{cl}$,
and one can show, that they correspond to the 1PI diagrams in
the diagrammatic expansion according to the Feynman rules.

The Slavnov-Taylor identity of the tree approximation can be
derived most simply on the generating functional of
 1PI Green functions, because 
  its lowest  order is seen to be the classical action:
\begin{equation}
\label{pitree}
\Gamma (\varphi^{cl}_k) = \Gamma_{cl} (\varphi ^{cl}_k) \big|_{\hat \Phi= 0 
\atop q = 0} + O(\hbar)
\end{equation}
Therefore we are able to write down the Ward identity of BRS-transformation
 as
\begin{equation}
\label{wibrs}
\sum_{\varphi_k^{cl}} \int dx \, \brs \varphi_k^{cl} (x){\delta \Gamma_{cl} 
(\varphi_i^{cl})
\over \delta \varphi_k^{cl}(x)} = 0
\end{equation}
which is a well defined expression in the tree approximation.
The BRS-transformations are non-linear symmetry transformation in
propagating fields and it is seen that the non-linear symmetry
transformations become insertions into (connected) Green functions,
if one carries out the Legendre transformation.
Roughly speaking one has to replace
\begin{equation}
\brs \varphi_k^{cl} \longrightarrow [\brs \varphi _k ] \cdot
Z_c(j_k)
\end{equation}
where $[\brs \varphi _k ] \cdot
Z_c(j_k) $ is the generating functional of BRS-inserted connected
Green functions. The Green functions with insertions are defined
 according to
\begin{eqnarray}
\label{insgr}
G_{\brs \varphi_k ;\varphi_{i_1} ...\varphi_{i_n}}(x;x_1,...,x_n)
& = & \langle T 
:\brs \varphi_k (x):
\varphi_{i_1}(x_1) ...\varphi_{i_n}(x_n) \rangle \\
&= & R { \bigl\langle T
:\brs \varphi^{(o)}_k (x):
 \varphi^{(o)}_{i_1}(x_1) ...\varphi^{(o)}_{i_n}(x_n)
e^{i\Gamma_{int}(\varphi^{(o)}_k)} \bigr\rangle 
\over \bigl\langle T e^{i\Gamma_{int}
(\varphi^{(o)}_k )} \bigr\rangle } \nonumber
\end{eqnarray}
and are summarized in the functional of BRS-inserted 
(connected) Green functions according to the above definitions.
For setting up the Slavnov-Taylor identity for 
off-shell Green functions one  
does not only have to consider the ordinary Green functions but
also the ones with BRS-insertions. For defining the BRS-inserted
as well as the ordinary Green functions consistently one enlarges
the classical action by the external field part and couples
the non-linear BRS-transformations to external fields:
\begin{equation}
\label{gaclef}
\Gamma_{cl}(\varphi_k) \longrightarrow \Gamma_{cl}(\varphi_k, \Upsilon_k) 
= \Gamma_{cl}(\varphi_k) + \Gamma_{ext.f} (\varphi_k , \Upsilon_k)
\end{equation}
\begin{eqnarray}
\label{gaext}
\Gamma_{ext.f.} & = & \int \Bigl( \rho ^\mu _+ {\brs} W_{\mu,-} +
                           \rho ^\mu _- {\brs} W_{\mu,+ } +
                           \rho ^\mu _3  (\cw {\brs} Z_{\mu} - \sw 
{\brs} A_\mu) \\
               &   & \! \phantom{\int} 
+ \sigma _+ {\brs} c_{-} +
                                        \sigma _- {\brs} c_{+} +
                               \sigma _3 (\cw {\brs}  c_Z - \sw {\brs} c_A) 
\nonumber \\
                              &   & \! \phantom{\int} 
       + Y^\dagger {\brs} \Phi  + ({\brs} \Phi) ^\dagger Y \nonumber \\
    & & \! \phantom{\int}
       + \sum _{i=1}^{N_F} \bigl( \overline {\Psi ^R _{\delta_i}}
  {\brs} F^L _{\delta_i} + 
              \overline {\psi ^L_ {f_i}}{\brs}  f^R_i + \hbox{h.c.} \bigr)
\Bigr) 
\nonumber
\end{eqnarray}
The external fields $\rho^\mu_\alpha$ and $\sigma_\alpha, \alpha = +,-,3$,
are  anticommuting and commuting $SU(2)$-triplets
 with ghost charge $-1$ and $-2$ respectively.
The external field $Y$ is a complex anticommuting scalar doublet  with
ghost charge $-1$, $\psi ^L_{f_i} $
 denotes  external left-handed spinor singlets
with ghost charge $-1$
\begin{equation}
\psi^L_{f_i} \equiv \psi^L _{e_i} , \psi^L _{u_i} , \psi^L _{d_i} 
\end{equation}
whereas $\Psi ^R_{\delta_i} $ denotes  external right-handed spinor doublets
\begin{equation}
\Psi^R _{\delta _i} \equiv \Psi^R _{l_i} ,\Psi^R _{q_i} \qquad
\Psi^R _{l_i}  =
 \left( \begin{array}{c} \psi ^R _{\nu_i} \\
                                \psi^R _{e_i} \end{array} \right)
\qquad
\Psi^R _{q_i}  =
 \left( \begin{array}{c} \psi ^R _{u_i} \\
                                \psi^R _{d_i} \end{array} \right)
\end{equation}
The Green functions with insertions  (\ref{insgr}) are
defined via the external field part:
\begin{eqnarray}
G_{\brs \varphi_k ;\varphi_{i_1} ...\varphi_{i_n}}(x;x_1,...,x_n)
& = & {\delta \over \delta \Upsilon _k(x)}\langle T 
\varphi_{i_1}(x_1) ...\varphi_{i_n}(x_n) \exp \bigl\{i\Gamma_{ext.f}\bigr\}
\rangle \Big| _ {\Upsilon_i = 0}
\end{eqnarray}
 The generating functional of Green functions
\begin{eqnarray}
\label{genfunins}
 & &
Z(j_k, \Upsilon_k)
  = \Bigl\langle T \exp \biggl\{i \int dx \Bigl( j_k \varphi_k +
\Gamma_{ext.f.}(\varphi_k, \Upsilon_k) \Bigr) \biggr\} \Bigr\rangle \nonumber
\end{eqnarray}
includes ordinary and BRS-inserted Green functions, which are obtained
by differentiation with respect to the external fields $\Upsilon_k$.
$\varphi_k j_k$ symbolically denotes the sum over the quantum fields
coupled to their sources as explicitly given in (\ref{genfun}).
Therefrom the connected Green functions are obtained according to
(\ref{gfcon}) and the 1PI Green functions by a Legendre transformation
of the propagating fields as given in (\ref{clfield}) and
(\ref{gf1pi}), where the classical fields depend on the sources and external
fields:
\begin{equation}
\label{gf1piins}
Z_c(j_k,\Upsilon _k)
 = \Gamma(\varphi^{cl}_k, \Upsilon_k)
 + \int dx\:  \varphi^{cl}_k j_k
\end{equation}
Differentiation with respect to external fields on 
$\Gamma(\varphi^{cl}_k, \Upsilon_k) $
reproduces BRS-insertions into 1PI Green functions. Especially one
verifies
\begin{equation}
{\delta \Gamma (\varphi^{cl}_i , \Upsilon _i )\over \delta \Upsilon _k (x)}
= {\delta Z_c (j_i , \Upsilon _i )\over \delta \Upsilon _k( x)}
\end{equation}

With the help of the external field part one is now able to derive
the Slavnov-Taylor identity in a way, that the non-linear BRS-transformations
are properly defined as insertions into Green functions. 
Taking the external field as being invariant under 
classical BRS-transformations
\begin{equation}
\brs \Upsilon_k = 0
\end{equation}
the enlarged classical action (\ref{gaclef}) is BRS-invariant due
to the nilpotency of BRS-trans\-for\-ma\-tions.
\begin{equation}
\brs \Gamma_{cl}(\varphi_k, \Upsilon_k)  = 0
\end{equation}
The lowest order of $\Gamma(\varphi^{cl}_k, \Upsilon_k) $ is the
classical action (cf.~(\ref{pitree})) and the Ward identity
of BRS-transformations (\ref{wibrs}) can be
rewritten into the Slavnov-Taylor (ST) identity of 1PI Green functions
in the tree approximation:
\begin{eqnarray}
\label{ST}
{\cal S}
(\Gacl ) &=& \int \biggl(
\bigl(\sw \partial _\mu c _Z + \cw \partial_\mu c_A\bigr)
             \Bigl(\sw {\delta \Gacl \over \delta Z_\mu } + \cw
       {\delta \Gacl \over \delta A_\mu} \Bigr) \\  
 & & +     {\delta \Gacl \over \delta \rho^\mu_3 }
              \Bigl(\cw {\delta \Gacl \over \delta Z_\mu } - \sw
       {\delta \Gacl \over \delta A_\mu} \Bigr) 
       + {\delta \Gacl \over \delta \sigma _3 }
              \Bigl(\cw {\delta \Gacl \over \delta c_Z } - \sw
       {\delta \Gacl\over \delta c_A} \Bigr) \nonumber \\ 
& &      + {\delta \Gacl \over \delta  \rho^\mu _+ }
               {\delta \Gacl \over \delta W_{\mu,- } }
      + {\delta \Gacl \over \delta \rho^\mu _- }
               {\delta \over \delta W_{\mu,+ } }
+  {\delta \Gacl \over \delta \sigma _+ }
               {\delta \Gacl \over \delta c_{-} }
+  {\delta \Gacl \over \delta \sigma _- }
               {\delta \Gacl \over \delta c_{+} } +
{\delta \Gacl \over \delta Y^\dagger}{ \delta \Gacl \over \delta \Phi } +
{\delta \Gacl \over \delta \Phi^\dagger}{ \delta \Gacl \over \delta Y } 
 \nonumber \\
&& + \sum_{i=1}^{N_F} \Bigl({\delta \Gacl \over \delta \overline{\psi^L_{f_i}}}
{ \Gacl \delta \over \delta f^R_i }
+ {\delta \Gacl \over \delta \overline{\Psi^R_{\delta_i}}}
{ \Gacl \delta \over \delta F^L_{\delta _i }} + 
  \hbox{h.c.} \Bigr)  \nonumber \\
& & + B_a {\delta \Gacl \over \delta \bar c_a }  
   + \hat {\mathbf q}{\delta \Gacl \over \delta \hat \Phi  }  + 
{\delta \Gacl \over \delta \hat \Phi ^\dagger } \hat
{\mathbf q}^\dagger \biggr)  = 0
 \nonumber
\end{eqnarray}
There we have dropped the index classical for the classical fields
appearing in the generating functional of 1PI Green functions
\begin{equation}
\Gamma(\varphi_k, \hat \Phi, q, \Upsilon) = 
\Gacl(\varphi_k, \hat \Phi, q, \Upsilon) + O (\hbar)
\end{equation}
Nonlinear BRS-transformations are now obtained by differentiating
with respect to the external fields. We have included
 the external fields $\hat \Phi$
 into the
definition of the generating functional in order to be able
to derive Ward identities of rigid symmetry for the Green functions.
They produce by differentiation the  mass insertions and their
variations under rigid symmetry of
Goldstone fields. 

The algebraic properties of BRS-transformations are transfered to
nilpotency properties of the Slavnov-Taylor operator:
\begin{eqnarray}
\label{brsnil}
\brs _\Gamma \,{\cal S}
 (\Gamma) &=& 0 \quad \hbox{for any functional}\, \Gamma \\
\brs _\Gamma \, \brs_\Gamma &=& 0 \quad \hbox{if} \quad {\cal S}(\Gamma) =0 
\nonumber
\end{eqnarray}
The operator $\brs _\Gamma $ is the linearized version of the ST identity
and is defined by
\begin{eqnarray}
\label{brsop}
\brs _ \Gamma &=& \int  \biggl(  
\bigl(\sw \partial _\mu c _Z + \cw \partial_\mu c_A\bigr)
             \Bigl(\sw {\delta  \over \delta Z_\mu } + \cw
       {\delta  \over \delta A_\mu} \Bigr)  \\
& & \phantom{\int \biggl(} +  B_a {\delta  \over \delta \bar c_a }  
   + \hat {\mathbf q}{\delta  \over \delta \hat \Phi  }  + 
{\delta  \over \delta \hat \Phi ^\dagger } \hat {\mathbf q}^\dagger
  \nonumber \\
& & \phantom{\int \biggl(} + \sum_{\varphi_k, \Upsilon_k} u_k
\Bigl({\delta \Ga\over \delta \Upsilon_k} {\delta\over \delta \varphi_k}
+ {\delta\Ga\over \delta \varphi_k} {\delta \over \delta \Upsilon_k} \Bigr)
\biggr) \nonumber
\end{eqnarray}
The sum is over all external  and corresponding propagating
fields which gave rise to bilinear appearance of $\Gamma$ 
in the ST identity, $u_k $ denotes
the respective coefficients as $\cw , \sw $ and $1. $

By Legendre transformation one is immediately able 
to give the ST identity for the functional of connected
Green functions $Z_c \equiv Z_c (j_k, \hat \Phi , q , \Upsilon_k)$
in the tree approximation:
\begin{eqnarray}
\label{STc}
{\cal S}
(Z_c ) &=& \int \biggl(
\bigl(\sw \partial _\mu j^\mu_Z + \cw \partial_\mu j^\mu_A\bigr)
             \Bigl(\sw {\delta Z_c \over \delta \jmath_Z } + \cw
       {\delta Z_c \over \delta \jmath_A} \Bigr) \\  
 & & + \Bigl(\cw j^\mu_Z - \sw j^\mu_A \Bigr)     
       {\delta Z_c \over \delta \rho_{\mu 3 } }
       + \Bigl(\cw \jmath _Z - \sw \jmath_A \Bigr)     
{\delta Z_c \over \delta \sigma _3 } 
\nonumber \\ 
& &      + j _{\mu+}{\delta Z_c \over \delta  \rho^\mu _+ }
                  +  j_{\mu-}{\delta Z_c \over \delta \rho^\mu _- }
               +  \jmath _+{\delta Z_c \over \delta \sigma _+ }
               + \jmath_- {\delta Z_c \over \delta \sigma _- }
              + J^\dagger
{\delta Z_c \over \delta Y^\dagger}+ 
{ \delta Z_c \over \delta Y }  J
 \nonumber \\
&& + \sum_{i=1}^{N_F} \Bigl(\overline {\eta^L_i}
{\delta Z_c \over \delta \overline{\psi^L_{f_i}} }
+ \overline {\eta^R_i} {\delta Z_c \over \delta \overline{\psi^R_{\delta_i}}}
+ 
{\delta Z_c \over \delta {\psi^L_{f_i}} } {\eta^L_i}
+ 
 {\delta Z_c \over \delta {\psi^R_{\delta_i}}}  {\eta^R_i}
 \Bigr)  \nonumber \\
& & +
  \bar \jmath _a  {\delta Z_c \over \delta j^B_a}
   + \hat {\mathbf q}{\delta Z_c \over \delta \hat \Phi  }  + 
{\delta Z_c \over \delta \hat \Phi ^\dagger } \hat
{\mathbf q}^\dagger \biggr)  = 0 \nonumber
\end{eqnarray}

The ST identity of the connected Green functions is linear in contrast 
to the one of the 1PI Green functions. It is the starting
point for proving unitarity of the physical S-matrix
\cite{BRS75, BRS76, KUOJ78} . 
Although the
proof of unitarity is beyond the scope of the paper we want to
indicate how the cancellation mechanism  works:
Eliminating the $B_a$-fields and their sources in (\ref{STc})
by
\begin{equation}
{\delta \over \delta j^B_a} \longrightarrow 
-\frac 1 \xi \tilde I _{aa'}{\cal F} _{a'} \Bigl(
            {\delta  \over \delta j^\mu_b} , 
           {\delta  \over \delta J},           
   {\delta  \over \delta J^\dagger},  \hat \Phi\Bigr) 
\end{equation}
with 
\begin{equation}
{\cal F} _b \Bigl(
            {\delta  \over \delta j^\mu_a} , 
           {\delta  \over \delta J},           
   {\delta  \over \delta J^\dagger},  \hat \Phi\Bigr) = \tilde I_{bb'}
\partial_{\mu} {\delta  \over \delta j^\mu_{b'}} - i
 \frac e{\sin \theta_W}
\Bigl((\hat \Phi + \zeta \mbox{v} )^\dagger \frac {\tau^T_{b}} 2
(      {\delta  \over \delta J}            +  \mbox{v} ) -
(      {\delta  \over \delta J^\dagger}  + 
\mbox{v} ^\dagger ) \frac {\tau^T_{b}} 2
( \hat \Phi + \zeta \mbox{v} ) \Bigr) 
\end{equation}
  it is
seen that the ST identity indeed  relates  the
Green functions of ghosts to the ones
with longitudinal vector propagators and would-be  Goldstones at the external
legs. (Due to linear contributions the Green functions of external fields
 include
1-particle reducible ghost propagators.)
 Applying the S-matrix operator the corresponding unphysical
contributions have to be shown to cancel in physical scattering 
processes.

 Renormalization concerns the 1PI
Green functions. Having these well-defined the connected Green
functions exist and are also well-defined  and can be obtained to
all orders by the Legendre transformations (\ref{gf1pi}). For this
reason 
 we are able to restrict all the further considerations to
1PI Green functions. 

In the procedure of renormalization  the ST identity is
 the defining symmetry of the theory,
because it yields unitarity of the physical S-matrix as indicated
above. Due to the abelian subgroup, however,
the ST identity  is not sufficient to fix uniquely the 
Green functions of higher orders.
In addition we have to take into account the Ward identities of
rigid $SU(2) \times U(1)$ invariance and especially the local $U(1)$
Ward identity for being able to fix the electric charges
of the fermions.
 In the tree approximation the Ward identities of rigid symmetry 
are 
immediately derived according to the construction of the gauge fixing 
sector  (cf.~(\ref{wardrig1}),
(\ref{wardrig}), (\ref{rigb}) and (\ref{rigghost})). For consistency
we have to  assign
definite transformation properties under rigid transformation to
the external fields $\Upsilon _k$ in such a way, that the external field part
$\Ga_{ext.f.}$ (\ref{gaclef})
is rigid invariant.
 It is obvious, that the fields 
$\rho_\a $ and $\sigma_\a $ transform under the adjoint
representation, whereas $Y$ and $\Psi^R$ under the fundamental
representation of $SU(2)$. 
We thus arrive at
\begin{equation}
\label{wi}
{\cal W} _\alpha \Gamma _{ cl} (\varphi_k, \Upsilon_k , \hat \Phi,
\hat {\mathbf q} )= 0 
\quad \hbox{and} \quad {\cal W} _4 \Gamma _{cl} (\varphi_k,
 \Upsilon_k , \hat \Phi,
\hat {\mathbf q} )= 0  
\end{equation}
where $\Ga_{cl}$ is understood to be the lowest order of the generating
functional of 1PI Green functions.
\begin{equation}
\label{gacl}
\Gamma_{cl} = \Gamma_{GSW} + \Gamma_{g.f.} + \Ga_{ghost} + \Ga_{ext.f}
\end{equation}
 The Ward operators of rigid $SU(2)$-transformations include all the
propagating and external fields we have introduced:
\begin{eqnarray}
\label{wardna} 
{\cal W}_\alpha =  \tilde I_{\alpha\alpha'} 
\int & \hspace{-3mm} \Biggl( & \hspace{-3mm} V^\mu _b 
     O^T_{b\beta}(\theta_W) \hat\varepsilon _{\b \gamma \alpha} 
     O_{\gamma c}(\theta_W) \tilde I _{cc'}
\frac{\delta}{\delta V^\mu _{c'}} 
 + \{  c_a, B_a, \bar c_a \}  \\
& &
+\rho^\mu _\beta \varepsilon _{\beta\gamma \alpha'}  \tilde I _{\gamma \gamma'}
\frac{\delta}{\delta \rho^\mu _{\gamma'}} + \{\sigma_\a \}  \nonumber \\
&  & + 
  i (\Phi + {\mathrm v}) ^\dagger
\frac { \tau _{\alpha'} } 2  \frac{\overrightarrow 
\delta}{\delta \Phi^\dagger}
 -  i \frac{\overleftarrow
\delta}{\delta \Phi} \frac { \tau _{\alpha'} } 2  (\Phi +{\mathrm v})  
+  \{ Y , \hat \Phi + \zeta {\mathrm v}, \hat{\mathbf q} \} \nonumber \\
& &  + \sum_{i = 1}^{N_F} \sum_{\delta = l,q} \Bigl(
 i \overline{F ^L_{\delta_i}} \frac {\tau _{\alpha'} } 2  
\frac{\overrightarrow 
\delta}{\delta \overline{ F^L_{\delta_i}} }
 -  i \frac{\overleftarrow
\delta}{\delta F^L_{\delta_i}} \frac { \tau _{\alpha'} } 2  F^L_{\delta_i}  
+  \{ \Psi^R _{\delta_i} \} \Bigr)\Biggr)   \nonumber
\end{eqnarray}
The abelian Ward operator  comprises the doublets and right-handed
fermions together with the external fields coupled to their
BRS-variations.
\begin{eqnarray}
\label{wardab}
{\cal W}_4 & = & \int \Biggl(
\frac i2 (\Phi +{\mathrm v} ) ^\dagger   \frac{\overrightarrow 
\delta}{\delta \Phi^\dagger}
 -  \frac i2 \frac{\overleftarrow
\delta}{\delta \Phi}  ( \Phi + {\mathrm v} )  
+  \{ Y , \hat \Phi + \zeta {\mathrm v} , q \}  \\
& & \phantom{\int \biggl(} 
+ \sum_{i =1}^{N_F} \biggl( \sum_{\delta = l,q} 
Y_W^\delta \Bigl( \frac i2 
 \overline{F _{\delta_i}^L}   \frac{\overrightarrow 
\delta}{\delta \overline{ F_{\delta_i}^L} }
 -  \frac i2 \frac{\overleftarrow 
\delta}{\delta F_{\delta_i}^L}    F_{\delta_i}^L  
+  \{ \Psi^R _{\delta_i} \} \Bigr)    \nonumber \\
& & \phantom{\int \biggl( + \sum_{i =1}^{N_F} }
 - \sum_f  Q_f \Bigl( i \overline{f_i ^R}   \frac{\overrightarrow 
\delta}{\delta \overline{ f_i^R} }
 -  i \frac{\overleftarrow
\delta}{\delta f_i^R}   f_i^R  
+  \{ \psi^R _{f_i} \} \Bigr) \biggr) \Biggr)  \nonumber 
\end{eqnarray}
The Pauli matrices $\tau_\a$ are defined in (\ref{pauli}),
the antisymmetric tensor $\varepsilon_{\a\beta\ga}$ in (\ref{eabc}).
The Ward operators of rigid symmetry satisfy the $SU(2) \times U(1)$
algebra:
\begin{eqnarray}
\label{wardalg}
\bigl[ {\cal W }_\alpha , {\cal W } _\beta \bigr] & =& \varepsilon 
_{\alpha\beta\gamma} \tilde I_{\gamma \gamma'} {\cal W} _{\gamma'} \\
\bigl[ {\cal W }_\alpha , {\cal W } _4 \bigr] &=& 0 \nonumber
\end{eqnarray}

In connection with the abelian Ward identity of rigid symmetry 
there exists also
a local version (cf.~(\ref{wlocgf})), which reads in $B_a$-gauges:
\begin{equation}
\label{wardloc}
{\mathbf w}_4 \Gamma_{cl}- \frac 1e 
\cw \Bigl(\sw \partial {\delta \Gamma_{cl}\over \delta Z} + \cw \partial 
{\delta \Gamma _{cl}\over \delta
A}\Bigr) = 
\frac 1e \cw (\sw \Box B_Z + \cw \Box B_A )
\end{equation}
The local operator ${\mathbf w}_4$
 is defined by dropping the integration from the 
rigid operator:
\begin{equation}
{\cal W}_4 = \int {\mathbf w}_4
\end{equation}
The ST identity (\ref{ST}), the Ward identities of rigid symmetry 
(\ref{wi})  and
the local abelian Ward identity 
(\ref{wardloc}) are the algebraic symmetries of
the standard model in the tree approximation. It has to be
shown, that these symmetries can be continued to higher orders
and that they together with appropriate normalization conditions uniquely
define the Green functions of the standard model to all orders.

\newpage

\newsection{The construction of higher orders: The algebraic method}

In the procedure of renormalization one has to make meaningful
the undefined expressions, which are obtained in the formal
expansion of the Gell-Mann Low formula according to Wick's theorem
(cf.~(\ref{GML}) and (\ref{insgr})).
As we have already mentioned renormalization concerns the
1PI Green functions summarized in the generating functional
(\ref{gf1piins})
\begin{equation}
\Gamma (\phi_k, \Upsilon, \hat \Phi , \hat{\mathbf q}) =
\Gamma (V_a,\Phi, f_i, c_a , \bar  c_a,\rho_\a, Y, \psi_i ,   \sigma_\a,
        \hat \Phi, \hat{\mathbf q} )
\end{equation}
It depends on the external fields and the ``classical'' fields defined
by the Legendre transformation
(\ref{clfield}). For simplification we have dropped the
index `classical'. The 1PI Green functions are divergent according
to their degree of divergencies:
\begin{equation}
d_\Gamma = 4 - \sum_{\mathrm ext. legs} d_E - \sum_{\mathrm vertices} (d_i -4)
\end{equation}
Here $d_E$ is the ultraviolet (UV) dimension of the fields appearing
at the external (amputated)
legs: They include propagating as well as external
fields. $d_i$ denotes the UV-dimension of the vertices. The UV-dimensions
of the fields are listed in the appendix.
There are different schemes, which remove the divergencies consistently.
For practical calculations it is convenient to use dimensional
regularization in connection with a prescription for subtracting the
D-dimensional poles in the limit to 4 dimensions
\cite{BEMA77}. For abstract renormalization
one better refers to the momentum subtraction scheme in the version
of BPHZL \cite{ZIM69, LOW76, CLLO76}. 

For higher orders the Gell-Mann Low formula has to be modified taking
in the interaction part not only the vertices of the tree approximation
but also the counterterms of higher orders. In the QED-like on-shell
schemes the counterterms are power series in the electromagnetic coupling $e$.
All terms are collected in a  $\Gamma_{eff}$:
\begin{equation}
\label{gaeff}
\Gamma_{eff} = \Ga_{cl} + O(\hbar) = \Gamma^{(bil)} + \Ga^{(int)}_{eff}
\end{equation}
The bilinear parts are defined in (\ref{gabilcl}).
At first $\Ga_{eff}$ contains all field polynomials in external and
quantum fields which are compatible with the power counting analysis
of renormalizable quantum field theory, i.e.~they have UV-dimension less
than of equal to 4. The explicit form of $\Ga_{eff}$ depends on the
renormalization scheme one has used to remove the divergencies. Therefore
rather than relying on properties of an explicit $\Ga_{eff}$ and a 
subtraction scheme, one deals in the construction of 1PI Green functions
with finite renormalized Green functions and their properties with 
respect to the symmetries of the standard model. (For an introduction to
algebraic renormalization see \cite{PISO95}

In the last sections we have given the tree approximation and the
symmetries of the tree approximation, the Slavnov-Taylor identity
(\ref{ST}),
the Ward identities of rigid symmetry 
(\ref{wi}) and the local $U(1)$-Ward identity (\ref{wardloc}).
Having readily defined the lowest order, the Green functions of the
1-loop approximation are calculated with a specific renormalization 
scheme leading to a finite result $\Gamma_{ren}$.
 Different schemes differently dispose  of the local contributions
of the next order, whereas the non-local contributions are uniquely
defined. Therefore after subtraction the  Green functions
are well-defined up to local contributions. In order to determine these
local contributions  one has to adjust those which break
the symmetry, in a way that the symmetries of the lowest order
are restored in the 1-loop order. The remaining (symmetric)
ones have to be fixed by normalization conditions. Then one is able to proceed
to higher orders by induction repeating the above steps from
order $n$ to order $n+1$. 

The symmetries of the lowest order
can be also violated  by anomalies. Anomalies 
arise from non-local contributions and cannot be removed by adjusting
local contributions. They have then explicitly to be proven to
be absent to all orders of perturbation theory.

 In the standard model  restoration of symmetries and the
setting of proper normalization conditions 
  are deeply connected
with each other: 
It has to be shown that one is able to impose  normalization conditions
on the 2-point functions in such a way that the 2-point Green functions
have one particle properties in the LSZ limit
(apart from the problem of unstable particles). Thereby special attention has
 to be paid to the massless particles: In order not to introduce
off-shell infrared divergent diagrams to the next order 
 the 2-point functions of massless particles as well as
 also the mixed 2-point functions of massive
and massless particles have to be required to vanish at $p^2 = 0$
to all orders of perturbation theory:
\begin{eqnarray}
\label{infra}
\Gamma_{ZA}(p^2=0) = \Gamma_{AA}(p^2 = 0) &=& 0 \\
\Gamma_{\bar c _A c _Z}(p^2=0) =
\Gamma_{\bar c _Z c _A}(p^2=0) = \Gamma_{\bar c_ Ac _A}(p^2 = 0) &=& 0 
\nonumber
\end{eqnarray}
These normalization conditions have to be proven to be in accordance
with the symmetries of the standard model and will be shown to
lead to higher order corrections of the weak mixing angle in
the ST identity and the Ward identities. 

The 1PI Green function of the standard model summarized in
the generating functional 
are defined in order $n$ by 
\begin{equation}
\label{defren}
\Gamma^{(\le n)} = \Gamma^{(\le n)}_{ren} + \Gamma^{(n)} _{inv} 
+ \Gamma ^{(n)}_{break} 
\end{equation}
and have to be shown to have
well defined normalization properties and to satisfy the
symmetries
\begin{equation}
\label{sym}
\Bigl({\cal S} (\Gamma)\Bigr)^{(\le n)} 
= 0 \qquad
\Bigl({\cal W}_\a \Gamma\Bigr) ^{(\le n)} = 0
\end{equation}
and a local $U(1)$ Ward identity.
The ST operator and the Ward operators
 are thereby established via their algebraic characterization
 (\ref{brsnil}) and (\ref{wardalg}).
In (\ref{defren}) $\Gamma^{(n)}_{inv}$ and
$\Gamma^{(n)}_{break}$ denote purely local field polynomials.
They depend on propagating and external fields introduced in
the classical approximation and constitute a complete basis
of field polynomials with UV-dimension
less than or equal 4. In a specific scheme
the local contributions are governed  by the counterterms appearing
in a $\Gamma_{eff}$.
Discrete symmetries are not affected by
renormalization, we are therefore able to restrict the analysis
to field polynomials which are neutral with respect to electric
and Faddeev-Popov charge and are CP-even, due to the fact, that
we did not introduce family mixing in the classical approximation.
The quantum numbers of the fields under the discrete symmetries are
listed in the appendix.

As indicated by the notation (\ref{defren})
 local contributions are algebraically divided into two classes:
the invariant 
 and non-invariant  
field polynomials. The invariant field polynomials
appearing in $\Ga _{inv}$
 constitute together
 with $\Gacl $  the general classical solution
$\Ga_{cl}^{gen}$, i.e.~the 
general field polynomials, which are solutions of
the Slavnov-Taylor identity and are rigid invariant 
\begin{eqnarray}
\label{invc}
 \Ga_{cl}^{gen} = \Gacl + \sum_{n=1}^\infty \Ga^{(n)}_{inv}
\qquad
\begin{array}{lcc}
{\cal S} (\Gamma_{cl}^{gen}) & = & 0  \\
{W_\a } (\Gamma_{cl}^{gen}) & = & 0 \end{array} 
 \end{eqnarray} 
The free parameters
of $\Ga_{cl}^{gen}$ are determined by the normalization conditions
and the local $U(1)$ Ward identity order by order in perturbation
theory.

The non-invariant field polynomials
 $\Ga_{break}$ are used to remove the breakings
of the symmetries, which have been introduced
  by an implicit 
scheme dependent
adjustment of finite counterterms in the subtraction procedure.
The abstract construction of $\Ga_{break}$ is carried out with the 
algebraic method which is based on the  action principle in its
quantized version valid for off-shell Green functions
\cite{LAM73, CLLO76}. It is most 
easily formulated on the general functional of the respective Green
functions and relates variations with respect to sources
or classical fields, respectively, and external
fields  to insertions with a well-defined UV and IR-degree. 
Especially the action principle
 states that the symmetries of the tree approximation
are broken at most by a integrated  field polynomial in 1-loop order and
proceeds to higher orders by induction: 
\begin{equation}
\begin{array}{lcr}
\bigl({\cal S} (\Gamma )\bigr)^{(\le n-1)} & = &0  \\
\bigl({\cal W}_\a \Gamma \bigr)^{(\le n-1)} & = &0\end{array}
\quad \Longrightarrow \quad \begin{array}{lcl}
\bigl({\cal S} (\Gamma )\bigr)^{(\le n)} & = &\Delta_{brs}^{(n)} \\
\bigl({\cal W}_\a \Gamma \bigr)^{(\le n)} & = &\Delta^{(n)}_\a 
\end{array}
\end{equation}
The breakings have well-defined properties with
respect to the discrete symmetries. For example $\Delta_{brs}$ has
$\phi\pi$-charge 1, is neutral with respect to electric charge
and even under CP, if the classical action is CP-invariant.
Furthermore they have a well-defined ultraviolet and infrared
degree. Up to this point the analysis has been completely
scheme independent just being founded on properties of
renormalized perturbation theory, but in classifying the breakings 
according to their UV- and IR-degree
 we assume that the renormalized Green functions $\Ga_{ren}$
have been constructed within the BPHZL scheme. 
In the BPHZL scheme the normalization conditions (\ref{infra}),
which otherwise have to be established by hand, 
are immediately
implemented by the subtraction procedure,
guaranteeing, that nowhere
infrared divergent contributions are introduced by the subtraction scheme.
Infrared divergent
contributions are detected by a pure power counting analysis.
Especially counterterms of IR dimension less than 4 are
forbidden in the BPHZL-scheme because  they would destroy
the normalization conditions (\ref{infra}).
Therefore 
we adopt the UV and IR degrees of fields as given in the appendix.
They  are uniquely determined  
by the behaviour of the free-field propagators for $p^2 \to
\infty $ and $p^2 \to 0$, respectively. Then by 
an analysis of the ST identity and the Ward operators
it is derived that the breakings of the
symmetries have the following UV and IR degrees:
\begin{eqnarray}
\dim^{UV} \Delta_{brs} \le 4 & \qquad &\dim^{IR} \Delta_{brs} \ge 3 \nonumber \\
\dim^{UV} \Delta_\a \le 4 & \qquad & \dim^{IR} \Delta_\a \ge 2 \nonumber
\end{eqnarray}
and it is seen, that all symmetries have to be carefully constructed
concerning the infrared.

Apart from the IR and UV degree we do not refer to further properties of
the BPHZL scheme as the $\Ga_{eff}$, but classify 
the breakings  by the algebra of
symmetry transformations, the nilpotency of BRS-transformations
and the algebra of ${\cal W}_\a $. (For details see section~7.)
Especially we have to show that all the breakings of the ST identity
are variations of  the linear ST operator (\ref{brsop}) and that
they can be absorbed into local contributions of $\Gamma_{break}$
without spoiling the normalization conditions especially (\ref{infra}): 
\begin{equation}
\label{break}
\brs_{\Gamma_{cl}} \Gamma_{break} = - \Delta_{brs} 
\end{equation}
Via equs.~(\ref{invc}) and (\ref{break}) the local contributions
are  uniquely fixed.

The  construction as outlined above is not only interesting from
an abstract point of view for having properly defined the standard
model but it passes through all the steps, which have been carried
out in explicit calculations, too. Especially the construction
of the symmetries in 1-loop order including the rigid Ward identity
is essential if one wants to proceed to higher orders of 
perturbation theory.  Dimensional regularization makes the analysis
of $\Ga_{break}$  easier, because it is an invariant scheme
for parity conserving
gauge theories, but also there it is well-known that
the normalization conditions and especially the infrared conditions
have to be established by an explicit adjustment of naively non-invariant
counterterms, which lead to corrections  of the weak mixing angle in the
ST identity and the Ward identities. 

According to the procedure we have outlined in this section
we will proceed for constructing the 1PI Green functions of
the standard model as follows:

\begin{enumerate}
\item We construct the most general ST operator and Ward operators
       of rigid symmetry,
     which are in accordance with the algebraic characterization.
       (Section 4)

\item We impose normalization conditions according to the on-shell schemes
 which allow to define proper 2-point functions in the LSZ-limit 
 apart from the problem of unstable particles and derive the most
general classical solution, which is in accordance with the
normalization conditions and the symmetries. (Section 5)

\item We classify the breakings according to the symmetries and
 show that they can be absorbed into local contributions to
 the 1PI Green functions. (Section 7)

\end{enumerate}

\newsection{The algebraic characterization of the symmetry transformations}
\newsubsection{The general ansatz and discrete symmetries}

In section 2.4 we have derived the symmetries of the standard model
for the 1PI-Green functions in the tree approximation: the Slavnov-Taylor
identity (\ref{ST}), the Ward identities of rigid symmetries 
(\ref{wardna}) and the local
$U(1)$-Ward identity (\ref{wardab}). The functional operators depend in lowest
order explicitly on the weak mixing angle $\theta_W$, which is in
the on-shell schemes defined by the ratio of the W- and Z-mass
\cite{SIMA80}
\begin{equation}
\cw \equiv {M_W \over M_Z}
\end{equation}
It is obvious and can be seen also from explicit 1-loop calculations
that the lowest order gets higher order corrections. These higher
order corrections depend on the normalization conditions, one has
chosen for fixing the 2-point Green functions.
Since the standard model includes massless particles, especially
the photon and the corresponding $\phi\pi$-ghosts, it is even not possible
to define the Green functions of higher orders by the symmetries as
given in lowest order. For this reason we construct in a first
step towards quantization the symmetry operators in a most general 
form and characterize them by  their  algebraic properties. 
We restrict the analysis to the generating functional of 1PI Green
functions as defined in (\ref{gf1piins}), which depends on the classical
fields as well  as on the external fields:
\begin{equation}
\Ga \equiv \Gamma (V_a,B_a,\Phi, f_i, c_a ,\bar  c_a,\rho_\a, Y, \psi_i ,
                    \sigma_\a,
        \hat \Phi, \hat{\mathbf q} )
\end{equation}
The vectors, $\phi\pi$-ghost and the $B$-fields have on-shell field
indices $a=+,-, Z,A$
\begin{equation}
\begin{array}{lcl}
V^\mu_a &= & ( W^\mu_+,W^\mu_-, Z^\mu, A^\mu) \\ 
B_a &= & ( B_+,B_-, B_Z, B_A) \end{array}
\quad 
\begin{array}{lcl}
c_a &= & ( c_+,c_-, c_Z, c_A) \\ 
\bar c_a &= & ( \bar c_+,\bar c_-, \bar c_Z, \bar c_A) \end{array}
\end{equation}
Since the theory is spontaneously broken, it is more adequate to
introduce the scalars 
 as a 4-vector 
with indices $a= +,-, H, \chi$
\begin{equation}
\begin{array}{lcl}
\phi_a &= & ( \phi_+,\phi_-, H, \chi) \\ 
Y_a &= & ( Y_+,Y_-, Y_H, Y_\chi) \end{array}
\quad 
\begin{array}{lcl}
\hat \phi_a &= & ( \hat \phi_+,\hat \phi_-, \hat H, \hat\chi) \\ 
q_a &= & ( q_+,q_-, q_H, q_\chi) \end{array}
\end{equation}
The external fields $\rho_\a$ and $\sigma_\a$  are three component
fields
\begin{equation}
\rho_a =  ( \rho_+,\rho_-, \rho_3 )  \quad
\sigma_a =  ( \sigma_+,\sigma_-, \sigma_3 ) 
\end{equation}
The charged vectors and scalars are complex fields with
\begin{equation}
\varphi_+ ^* =  \varphi_- \, ,
\end{equation} 
whereas the neutral vectors and scalars are real fields
\begin{equation}
\varphi_a^* = \varphi_a \qquad a= Z,A, H, \chi, 3 \, .
\end{equation}
We group the fermions into a vector
according to
\begin{equation}
\begin{array}{lcl}
f^L_{i} &=& (\nu_i^L, e_i^L, u_i^L,d_i^L) \\
f^R_{i}  &=& ( e_i^R, u_i^R,d_i^R)
\end{array}\qquad
\begin{array}{lcl}
\psi^R_{f_i} &=& (\psi_{\nu_i}^R, \psi_{e_i}^R, \psi_{u_i}^R,\psi_{d_i}^R) \\
\psi^L_{f_i}  &=& (\psi_ {e_i}^L, \psi_{u_i}^L,\psi_{d_i}^L)
\end{array}\qquad
\end{equation}
  $i= 1,..., N_F$ denotes the family index.
The quantum number of fields are summarized in  the appendix.

The algebraic properties of the functional operators acting
on $\Gamma $ are the nilpotency of the Slavnov-Taylor operator
(\ref{brsnil})
\begin{eqnarray}
\label{brsnil2}
\brs _\Gamma \,{\cal S}
 (\Gamma) &=& 0 \quad \hbox{for any functional}\, \Gamma \\
\brs _\Gamma \, \brs_\Gamma &=& 0 \quad \hbox{if}\quad {\cal S}(\Gamma) =0 
\nonumber
\end{eqnarray}
and the algebra of rigid operators
(\ref{wardalg})
\begin{eqnarray}
\label{wardalg2}
\bigl[ {\cal W }_\alpha , {\cal W } _\beta \bigr] & =& \hat \varepsilon 
_{\alpha\beta\gamma} \tilde I_{\gamma \gamma'} {\cal W} _{\gamma'} 
\end{eqnarray}
with $\alpha,\beta, \gamma = +,-,3,4$ and $\hat \varepsilon_{\a\b\ga}$
denotes the structure constants of $SU(2) \times U(1) $, which are
taken as
completely antisymmetric in all 3 indices
\begin{equation}
\label{hatve}
\hat \varepsilon_{\a\b\ga} = \left\{ \begin{array}{ccl}
\hat \varepsilon_{+-3} & = & -i \\
\hat \varepsilon_{+-4} & = & 0 \nonumber
 \end{array}\right.
\end{equation}
In addition to the algebra  the Ward operators are specified 
by their transformation
with respect to complex conjugation:
\begin{equation}
\label{cc}
\begin{array}{ccc}
{\cal W}_+^\dagger & = & {\cal W}_- \\
{\cal W}_-^\dagger & = & {\cal W}_+
\end{array}
\quad
\begin{array}{ccc}
{\cal W}_3^\dagger & = & {\cal W}_3 \\
{\cal W}_4^\dagger & = & {\cal W}_4
\end{array}
\end{equation}
which is assigned in agreement with the tree approximation.

Since the functional $\Gamma $ will be constructed as a simultaneous
solution of rigid transformations and the ST identity the 
respective functional
operators have to
satisfy the consistency relation
\begin{equation}\label{cons}
{\cal W}_\a {\cal S}(\Ga) - \brs _\Ga {\cal W}_\a \Gamma = 0
\quad \hbox{for any functional}\, \Gamma 
\end{equation}

Discrete and
global unbroken symmetries, we want to impose on the functional $\Gamma$,
can be imposed to all orders and are not affected by renormalization.
These symmetries are electric and $\phi\pi$-charge neutrality:
\begin{eqnarray}
\label{chargecons}
{\cal W}_{em} = - i \int dx &\hspace{-3mm}\biggl( & \hspace{-3mm}
W_+{\delta \over \delta W_+} -W_-{\delta \over \delta W_-} +
B_+{\delta \over \delta B_+} -B_-{\delta \over \delta B_-} +
c_+{\delta \over \delta c_+} -c_-{\delta \over \delta c_-} 
\nonumber\\
&\hspace{-3mm}+& \hspace{-3mm}
\bar c_+
{\delta \over \delta \bar c_+} -\bar c_-{\delta \over \delta \bar c_-} +
\rho_+{\delta \over \delta \rho_+} -\rho_-{\delta \over \delta \rho_-} + 
\sigma_+{\delta \over \delta \sigma_+} -\sigma_-{\delta \over \delta 
\sigma_-} \nonumber \\
&\hspace{-3mm}+& \hspace{-3mm}
\phi_+{\delta \over \delta \phi_+} -\phi_-{\delta \over \delta \phi_-} +
Y_+{\delta \over \delta Y_+} -Y_-{\delta \over \delta Y_-} \nonumber \\
&\hspace{-3mm}+& \hspace{-3mm}
\hat \phi_+{\delta \over \delta \hat \phi_+} -\hat \phi_-{\delta \over 
\delta \hat \phi_-} +
q_+{\delta \over \delta q_+} -q_-{\delta \over \delta q_-} \nonumber \\
&\hspace{-3mm}-& \hspace{-3mm}
 \sum _{i=1}^{N_F} 
\Bigl( Q_e
(\bar e_i {\delta \over \delta \bar e_i} - {\delta \over \delta e_i} e_i  +
\bar \psi_{e_i} {\delta \over \delta \bar \psi_{e_i}} -
 {\delta \over \delta \psi_{e_i} \psi _{e_i}} ) \nonumber\\
&\hspace{-3mm}& 
+ Q_u 
(\bar u_i {\delta \over \delta \bar u_i} - {\delta \over \delta u_i} u_i 
+ \bar \psi_{u_i} {\delta \over \delta \bar \psi_{u_i}}
 - {\delta \over \delta \psi_{u_i}} \psi_{u_i} ) \nonumber \\
&\hspace{-3mm}& 
+ Q_d 
(\bar d_i {\delta 
\over \delta \bar d_i} - {\delta \over \delta d_i} d_i +
\bar \psi_{d_i} {\delta 
\over \delta \bar \psi_{d_i}} - {\delta \over \delta \psi_{d_i}} \psi_{d_i}
 ) \Bigr) \biggr)  \nonumber \\
{\cal W}_{\phi\pi} =  \int dx &\hspace{-3mm}\biggl( & \hspace{-3mm}
c_a{\delta \over \delta c_a} -
\bar c_a{\delta \over \delta \bar c_a} - 
\rho_\a{\delta \over \delta \rho_\a}  - 2\sigma_\a{\delta \over \delta 
\sigma_\a}- Y_a{\delta \over \delta Y_a} + q_a{\delta \over \delta q_a}\\
&\hspace{-3mm}-& \hspace{-3mm}
 \sum_{i=1}^{N_F}\Bigl(\bar \psi_{m_i} {\delta 
\over \delta \bar \psi_{m_i}} + {\delta \over \delta \psi_{m_i}} \psi_{m_i}
\Bigr)\biggr)  \nonumber
\end{eqnarray}
Since fermions and quarks cannot be mixed and since we forbid
explicitly family mixing we have some further symmetries
which correspond to lepton and quark family number conservation:
\begin{eqnarray} 
\label{qlfcons}
{\cal W}_{l_i} = i \int dx &\hspace{-3mm}\biggl( & \hspace{-3mm}
\bar e_i {\delta \over \delta \bar e_i} - {\delta \over \delta e_i} e_i  +
\bar \psi_{e_i} {\delta \over \delta \bar \psi_{e_i}} -
 {\delta \over \delta \psi_{e_i}} \psi _{e_i} \\ 
&\hspace{-3mm}+& \hspace{-3mm}
\overline {\nu^L_i} {\delta \over \delta \overline {\nu^L_i}} 
- {\delta \over \delta \nu^L_i} \nu^L_i
  +\overline {\psi^R_{\nu_i}} {\delta \over \delta \overline
 {\psi^R_{\nu_i}}} -
 {\delta \over \delta \psi^R_{\nu_i}} \psi^R _{\nu_i} \biggr)\nonumber\\
{\cal W}_{q_i} = i \int dx &\hspace{-3mm}\biggl( & \hspace{-3mm}
\bar d_i {\delta \over \delta \bar d_i} - {\delta \over \delta d_i} d_i  +
\bar \psi_{d_i} {\delta \over \delta \bar \psi_{d_i}} -
 {\delta \over \delta \psi_{d_i}} \psi _{d_i} \\ 
&\hspace{-3mm}+& \hspace{-3mm}
\bar u_i {\delta \over \delta \bar u_i} 
- {\delta \over \delta u_i} u_i
  + \bar \psi_{u_i} {\delta \over \delta \bar \psi_{u_i}} -
 {\delta \over \delta \psi_{u_i}} \psi _{u_i} \biggr)\nonumber
\end{eqnarray}
The functional $\Ga$ is to all orders invariant under these global
symmetries by construction:
\begin{equation}
\begin{array}{lcl}
{\cal W}_{em} \Ga& = & 0  \\
{\cal W}_{\phi\pi} \Ga & = &  0
\end{array}
\qquad
\begin{array}{lcl}
{\cal W}_{l_i} \Ga& = & 0  \\
{\cal W}_{q_i} \Ga & = &  0
\end{array}
\end{equation}
In addition we have also colour SU(3)-invariance, which we do not
consider explicitly. 
All the functional operators when applied on $\Ga $ are
seen to be restricted  with respect
to these global symmetries, especially
\begin{equation}
\label{emcons}
\begin{array}{ccc}
\bigl[{\cal W}_{em} , {\cal W}_+ \bigr] & = & - i {\cal W}_+ \\
\bigl[{\cal W}_{em} , {\cal W}_- \bigr]  & = & + i {\cal W}_-
\end{array}
\qquad
\begin{array}{ccc}
\bigl[{\cal W}_{em} , {\cal W}_3 \bigr] & = & 0\\
\bigl[{\cal W}_{em} , {\cal W}_4 \bigr]  & = & 0
\end{array}
\end{equation}
and
\begin{equation}
\label{lqcons}
\bigl[{\cal W}_{l_i} , {\cal W}_\a \bigr]  =  0 \qquad
\bigl[{\cal W}_{q_i} , {\cal W}_\a \bigr]   =  0
\end{equation}

The functional $\Ga$ we consider in this paper is also invariant
with respect to CP-trans\-for\-mations. Therefrom it is derived
that the rigid operators as well as the Slavnov-Taylor operator
have definite transformation properties with respect to CP.
Explicitly it is possible to characterize them by their transformation
as given in the tree approximation:
\begin{equation}
\label{cpcons}
\begin{array}{lcl}
{\cal W}_{+} & \stackrel{CP}{\longrightarrow}
 & -{\cal W}_-  \\
{\cal W}_{-}  & \stackrel{CP}{\longrightarrow} &  -{\cal W}_+  
\end{array}
\qquad
\begin{array}{lcl}
{\cal W}_{3} & \stackrel{CP}{\longrightarrow} & -{\cal W}_3    \\
{\cal W}_{4}  & \stackrel{CP}{\longrightarrow} &  -{\cal W}_4  
\end{array}
\end{equation}
whereas the ST operator is CP-even.

We therefore make for the ST operator the ansatz:
\begin{eqnarray}
\label{STgen}
{\cal S}
(\Ga ) &=& \int \biggl( Z_4 
\bigl(\sgt \partial _\mu c _Z + \cgt \partial_\mu c_A\bigr)
             \Bigl(\svf {\delta \Ga \over \delta Z_\mu } + \cvf
       {\delta \Ga \over \delta A_\mu} \Bigr) \\  
 & & +     {\delta \Ga \over \delta \rho^\mu_3 } z^\rho
              \Bigl(\cvt {\delta \Ga \over \delta Z_\mu } - \svt
       {\delta \Ga \over \delta A_\mu} \Bigr) 
       + {\delta \Ga \over \delta \sigma _3 } z^\sigma
              \Bigl(\cgt {\delta \Ga \over \delta c_Z } - \sgt
       {\delta \Ga\over \delta c_A} \Bigr) \nonumber \\ 
& &      + {\delta \Ga \over \delta  \rho^\mu _+ }
               {\delta \Ga \over \delta W_{\mu,- } }
      + {\delta \Ga \over \delta \rho^\mu _- }
               {\delta \Gamma\over \delta W_{\mu,+ } }
+  {\delta \Ga \over \delta \sigma _+ }
               {\delta \Ga \over \delta c_{-} }
+  {\delta \Ga \over \delta \sigma _- }
               {\delta \Ga \over \delta c_{+} } +
{\delta \Ga \over \delta Y_a}\tilde I_{aa'}
{ \delta \Ga \over \delta \phi_{a'} } 
 \nonumber \\
&& + \sum_{i=1}^{N_F} \Bigl({\delta \Ga \over \delta \overline{\psi^L_{f_i}}}
{ \Ga \delta \over \delta f^R_{i} }
+ {\delta \Ga \over \delta \overline{\psi^R_{i}}}
{ \Ga \delta \over \delta f^L_{i} } + 
  \hbox{h.c.} \Bigr)  \nonumber \\
& & + B_a \hat g _{ab} {\delta \Ga \over \delta \bar c_b }  
   + q_a{\delta \Ga \over \delta \hat \phi_a  }  \biggr)  = 0
 \nonumber
\end{eqnarray}
It is neutral with respect to electric charge, commutes with the
operators ${\cal W}_{l_i}$ and ${\cal W}_{q_i}$ and arises $\phi\pi$
charge by one unit. Defining its
linear version $\brs_\Ga$ as given in (\ref{brsop}) one
immediately checks the nilpotency properties (\ref{brsnil2}).
The three independent angles $\Theta^V_3, \Theta^V_4$
and $\Theta^g_3$ describe, how vectors and ghosts
are rotated with respect to the abelian subgroup.
In the writing of (\ref{STgen}) we have  already anticipated that we 
are able to absorb constants in external fields  at will. 
The coefficients $z^\rho$ as well as
$z^\sigma $ could be reabsorbed into the external fields
$\rho_3$ and $\sigma _3$, but we
take them as arbitrary for a proper adjustment later on. 
For similar reasons we also keep 
$Z_4$ in the ansatz.
The matrix $\hat g _{ab}$ is an arbitrary neutral
matrix, it can be introduced into the ST
identity
without spoiling nilpotency and rigid symmetry (see (\ref{rigantig})). 
Its explicit form will be considered, when we give the general
classical solution of the gauge-fixing ghost sector.

For the Ward  operators of rigid symmetry we take the most general
ansatz, which is linear in fields:
\begin{eqnarray}
\label{wardriggen}
{\cal W}_\a = \tilde I_{\a\a'}\int dx &\hspace{-3mm}\biggl( & \hspace{-3mm}
V^\mu_b  \hat a^V_{bc,\a'} \tilde I_{cc'} {\delta \over \delta V^\mu_{c'} }
+B_b  \hat a^B_{bc,\a'} \tilde I_{cc'} {\delta \over \delta B_{c'} }
+c_b \hat a^g_{bc,\a'} \tilde I_{cc'} {\delta \over \delta c_{c'} }
+\bar c_b  \hat a^{\bar g}_{bc,\a'} \tilde I_{cc'} {\delta \over \delta \bar
c_{c'} } \nonumber \\
&\hspace{-3mm}+& \hspace{-3mm}
\phi_b  b^{\phi}_{bc,\a'} \tilde I_{cc'} {\delta \over \delta \phi_{c'} }
+Y_b  b^{Y}_{bc,\a'} \tilde I_{cc'} {\delta \over \delta Y_{c'} }
+\hat\phi_b  b^{\hat \phi}_{bc,\a'} \tilde I_{cc'} {\delta \over \delta 
\hat\phi_{c'} }
+q_b  b^{q}_{bc,\a'} \tilde I_{cc'} {\delta \over \delta q_{c'} } \nonumber \\
&\hspace{-3mm}+&\hspace{-3mm}
v_{c\a'}\tilde I_{cc'} {\delta \over \delta \phi_{c'}}
+ \hat v_{c\a'}\tilde I_{cc'} {\delta \over \delta \hat \phi_{c'}} \nonumber\\
&\hspace{-3mm}+&\hspace{-3mm}
\rho_\b   a^{\rho}_{\b\ga,\a'} \tilde I_{\ga \ga'} 
{\delta \over \delta \rho_{\ga'} }
+\sigma_\b   a^{\sigma}_{\b\ga,\a'} \tilde I_{\ga \ga'} 
{\delta \over \delta \sigma_{\ga'} } \nonumber \\
&\hspace{-3mm}+&\hspace{-3mm}
\sum_{i=1}^{N_F} \Bigl(
\overline {f^L_{i}}   h^{f_i}_{ff',\a'} 
{\delta \over \delta \overline {{f'}^L_{i} }} +
{\delta \over \delta  {f'}^L_{i} } {h^{f_i\dagger}_{f'f,\b}} 
f^L_{i} \tilde I_{\b\a'}\nonumber \\
&\hspace{-3mm} & \hspace{-3mm} \phantom{\sum} 
+\overline {\psi^R_{f_i}}   h^{\psi_i}_{ff',\a'}  
{\delta \over \delta \overline {\psi^L_{f'_i}} }  +
{\delta \over \delta  \psi_{f'_i}^R } h^{{\psi_i}\dagger}_{f'f,\b}
\psi_{f_i}^R \tilde I_{\b\a'}\nonumber \\
&\hspace{-3mm} & \hspace{-3mm} \phantom{\sum} +
\overline {f^R_{i}}   \tilde h^{f_i}_{ff',\a'} 
{\delta \over \delta \overline {f'^R_{i} }} +
{\delta \over \delta  f'^R_{i} } \tilde h^{f_i\dagger}_{f'f,\b}
f^R_{i} \tilde I_{\b\a'} \nonumber \\
&\hspace{-3mm} & \hspace{-3mm} \phantom{\sum} +
\overline {\psi^L_{f_i}}  \tilde h^{\psi_i}_{ff',\a'} 
{\delta \over \delta \overline {\psi^L_{f'_i}} } +
{\delta \over \delta  \psi_{f'_i}^L } \tilde h^{\psi_i\dagger}_{f'f,\b}
\psi_{f_i}^L \tilde I_{\b\a'}
\Bigr)\biggr)
\end{eqnarray}
The coefficients are restricted by the prescription for complex conjugation
(\ref{cc}) and by electric charge conservation (\ref{emcons}).

In the notation these properties are taken into account by having
neutral index structure throughout and changing 
 $+$ and $-$ by complex conjugation. 
 Well-defined transformation properties under CP
(\ref{cpcons}) 
yields furthermore that $\hat a _{bc,\a}$ and $  a_{\b\ga,\a}$ as
well as $\tilde h_{ff',\a}$ and $h_{ff',\a}$ are imaginary.
Similar restrictions are derived for the coefficients $b_{bc,\a}$.
Family mixing as well as lepton quark mixing are forbidden according
to (\ref{lqcons}). 

In the following we will solve the algebra 
(\ref{wardalg2}) as well as the consistency
equation (\ref{cons}) in all generality.
 Since we construct the Green functions  in perturbation theory,
 it would be also sufficient to
start from the tree approximation and consider  its possible
perturbations. Such a treatment, however, would disguise the simple 
algebraic structure of the final solution. 

\newsubsection{The  vector-ghost sector}

Evaluating the algebra of rigid operators (\ref{wardalg2}) 
for the vectors, $\phi\pi$-ghosts, the B-fields and the
external fields $\sigma_\a$ and $\rho_\a$ yields the following
representation equations for the coefficients:
\begin{eqnarray}
\label{adrep}
\hat a^\varphi _\a\, \tilde I\,\hat a^\varphi _\b - 
\hat a^\varphi _\b\, \tilde I\,\hat a^\varphi _\a &=&
- \hat \varepsilon _{\a \b \ga}
\tilde I_{\ga \ga'}\hat a^\varphi_{\ga'} \qquad 
\varphi \equiv V^\mu_a, c_a, \bar c_a ,
B_a \nonumber \\
 a^\Upsilon _\a \tilde I a^\Upsilon _\b - 
 a^\Upsilon _\b \tilde I a^\Upsilon _\a &=&
-  \varepsilon _{\a \b \ga}
\tilde I_{\ga \ga'} a^\Upsilon_{\ga'}  \qquad \Upsilon \equiv
 \rho_\a, \sigma_\a
\end{eqnarray}
Here we have introduced a matrix
notation:  ${\left(\hat a _\a\right)} _{bc} 
= \hat a_{bc,\a}$ denotes  $4\times 4$ matrices and
  $ {( a_\a)}_{\b\ga} = a_{\b\ga,\a}$ $3\times 3 $ matrices.
Due to CP-invariance the non-trivial solutions of (\ref{adrep}) are
uniquely related to the adjoint representation:
\begin{equation}
\label{adrepsol}
\hat a^\varphi _{\a} \sim \hat \varepsilon _\a \qquad  
\hat a^\Upsilon _{\a} \sim  \varepsilon _\a 
\end{equation}
with ${(\hat \varepsilon_{\a})}_{\b \ga} = \hat \varepsilon_{\b \ga\a}$  
defined by the structure constants of $SU(2) \times U(1)$
(\ref{hatve})
and ${( \varepsilon_{\a})}_{\b \ga} =  \varepsilon_{\b \ga\a}$  by the
structure constants of $SU(2)$.
 From this special solution one obtains the general solution
by the following equivalence transformations:
\begin{eqnarray}
{(\hat a _\a)}_{bc} &= &
\hat y_{b\b}\, \hat \varepsilon_{\b \ga \a}
\,{\hat y^{-1}}_{\ga c}  \nonumber \\
{( a _\a)}_{\b\ga}  & = &
 y_{\b\b'} \, \varepsilon_{\b' \ga' \a} \,
(y^{-1})_{\ga' \ga} 
\end{eqnarray}
The matrices $y$ and $\hat y$ have to be chosen in accordance with
the discrete symmetries. When we consider the general
classical solution of the standard model
it is seen that the equivalence transformations  are  related to 
field redefinitions. Therefore we parameterize
them in the following way:
\begin{equation}
\label{fred}
( \hat y^T)_{\a a} \equiv \hat z_{\a a} = \left(\begin{array}{cccc}
            \hat  z_W& 0& 0 \\
               0       &   \hat z_W & 0& 0\\
               0 & 0 &  \hat z_Z \cos \theta_Z & - \hat
                            z_A \sin \theta_A    \\
              0 & 0 &   \hat z_Z \sin \theta_Z &   \hat  z_A \cos \theta_A 
                   \end{array} \right)  
\qquad
y _{\b\ga} \equiv z_{\b \ga} = \left(\begin{array}{ccc}
               z_W &0& 0 \\
               0       &  z_W & 0\\
               0 & 0 &  z_3 
        \end{array} \right)
\end{equation}
We have suppressed the field indices, but one has to keep in mind, that
the algebra allows independent field redefinitions for each field
we consider.

In the tree approximation the representation matrices are
given by
\begin{equation}
\hat a^{(0)}_\a = O^T(\theta_W)\hat 
\varepsilon _{\a} O (\theta_W ) \quad \hbox{and} 
\quad a^{(0)}_\a = \varepsilon_\a
\end{equation}
for all fields in question, i.e. one has in perturbation theory
for the propagating fields
\begin{equation}
\hat z^\varphi_{\a b} = O (\theta _W )_{\a a}(
{\mathbf 1}+ \delta \hat z^\varphi)_{ab}  \qquad
\hbox{with} \quad (\delta \hat z^\varphi)_{ab} = O(\hbar)
\end{equation}
The matrix $O(\theta_W)$ is the orthogonal matrix, which transforms
 the $SU(2) \times U(1)$ gauge fields into on-shell fields
(\ref{othetaw}).
The matrix $\hat z$ is however not completely specified by
the representation matrices, indeed it is seen that equivalence 
transformations with diagonal
matrices $\hat z_{inv}$ leave the adjoint representation invariant:
\begin{equation}
\hat \varepsilon _{\a} = \hat z_{inv} \hat \varepsilon_{\a} 
\hat z^{-1}_{inv}  \qquad 
  \varepsilon _{\a} =   z_{inv}   \varepsilon_{\a} 
 z^{-1}_{inv}
\end{equation}
if 
\begin{equation}
\hat z_{inv} \equiv \left(\begin{array}{cccc}
              \hat z_2& 0& 0 \\
               0       &   \hat z_2 & 0& 0\\
               0 & 0 &   \hat z_2 & 0    \\
              0 & 0 &   0 &   \hat  z_1 
                   \end{array} \right)  
\qquad
z_{inv} = \left(\begin{array}{ccc}
               z_2 &0& 0 \\
               0       &  z_2 & 0\\
               0 & 0 &  z_2 
        \end{array} \right)
\end{equation}

There are several possibilities to parameterize the remaining parameters.
A symmetric  parameterization, which  is well adapted   to treat
higher order corrections of the vectors, is given by
\begin{equation}
\label{par}
 r_A = {z_Z \over z_W} {\cos\theta_Z \over \cos \Theta} 
\qquad
r_Z = { z_A \over z_W} {\sin\theta_A \over \sin \Theta}  
\end{equation}
$$
  \cos \Theta  =  {1\over\sqrt{ 1 + \tan \theta_Z \tan \theta_A}} 
$$
In this parameterization the general solution of 
(\ref{adrep}) reads explicitly 
\begin{eqnarray}
\label{adjpar}
\hat a_+ & = &\left( \begin{array}{cccc}
            0 & 0 & 0 & 0 \\
           0&  0 &  -i r_Z ^{-1} \cos \Theta & i r_A^{-1}  \sin \Theta \\
            0 & i r_Z \cos \Theta & 0 & 0 \\
          0    &-i r_A \sin \Theta  &  0 & 0 
\end{array} \right) \quad  \quad \hat a_3 = \hat \varepsilon _3 \\
\hat a_- & = & \left( \begin{array}{cccc}
                 0&  0  & i r_Z^{-1}  \cos \Theta & - i r_A^{-1} \sin \Theta \\
                 0 & 0 & 0 & 0\\
                 - i r_Z \cos \Theta & 0 & 0 & 0\\
                 i r_A  \sin \Theta  &  0 & 0 & 0 
\end{array} \right)
 \quad \quad \hat a_4 = 0
\end{eqnarray}
One has to determine three independent parameters for each field
in the rigid Ward identity. They are fixed
by the normalization conditions imposed on the 2-point Green
functions.
Vice versa it is seen that for the vectors we could also
choose $r_A = 1$ and $ r_Z = 1 $ replacing two normalization conditions
by the Ward identities of rigid symmetry.
Such a choice
corresponds to the minimal on-shell scheme \cite{BOHO86}.

Finally  the consistency equation between
the Ward operators of rigid symmetry and the ST operator
(\ref{cons}) relates the angles appearing in the ST operator
to the parameters of rigid Ward operators.
In the parameterization (\ref{par}) one gets 
\begin{eqnarray}
\label{stwardvec}
 & \tan \Theta^V_{3} = {r^V_A \over r^V_Z} \tan \Theta ^V \qquad
\tan \Theta^V_{4} = {r^V_Z \over r^V_A} \tan \Theta ^V & \nonumber \\
&z^\rho \, r_3^\rho = \sqrt {\frac 1 {{(r_Z^V)}^2} \cos^2\Theta^V
+ \frac 1 {{(r_A^V)}^2} \sin^2\Theta^V} &
\end{eqnarray}
and similar equations for the parameters of
the ghosts and $\sigma$-fields
\begin{eqnarray}
\label{stwardghost}
\tan \Theta^g_3 = {r^g_A \over r^g_Z} \tan \Theta ^g \qquad
z^\sigma \, r_3^\sigma = \sqrt {\frac 1 {{(r_Z^g)}^2} \cos^2\Theta^g
+ \frac 1 {{(r_A^g)}^2} \sin^2\Theta^g} &
\end{eqnarray}
It has to be proven, that these relations can be consistently maintained
to all orders of perturbation theory. In the tree approximation
they are obviously fulfilled.
Furthermore it is seen, that
the normalization constants $z^\sigma$ and $z^\rho$ can be fixed
by the Ward identity of rigid symmetry. Requiring that the external
fields transform to all orders just as in the tree approximation
\begin{equation}
\label{extffix}
a^\rho _{\b\ga \a} = \varepsilon_{ \b \ga \a} \qquad 
a^\sigma _{\b\ga\a} = \varepsilon_{\b \ga \a}
\end{equation}
  the parameters $z^\sigma$ and $z^\rho$ are uniquely determined.

In order to determine the transformation matrices of the
$B$-fields $\hat a^B_\a$, it has to be observed, that the gauge fixing is
linear in propagating fields. 
 Differentiating the functional
of 1PI Green functions with respect to $B_a$ therefore yields a local
expression to all orders of perturbation theory, which allows to
fix the normalization of the $B$-fields  on the longitudinal
parts of the vectors:
\begin{equation}
\label{gfixho}
{\delta \Gamma \over \delta B} = \xi_{ab} B_b + \tilde I_{ab} \partial^\mu
V_{\mu b} + r_{bc,a} \hat \phi_b \phi_c  +
         w_{ca} \phi_c + \hat w_ {ca} \hat \phi_c
\end{equation}
Applying the Ward operators of rigid symmetry on this local
expression it is seen, that
the transformation of the $B_a$-fields is  completely governed
by the vectors:
\begin{equation}
(a^B_\a) = - {(a^V _\a)}^T
\end{equation}
which reads for the parameters introduced above (\ref{par})
\begin{equation}
r^B_A = {1\over r^V_A} \qquad r^B_Z = {1\over r^V_Z} \qquad  \tan \Theta^B
= \tan \Theta^V
\end{equation}
One is able to establish rigid symmetry quite trivially on the
B-dependent part of the generating functional. 
 Accordance with rigid symmetry directly restricts the independent
parameters appearing in (\ref{gfixho}). The explicit form is
given in section 5.4.

Finally the consistency condition (\ref{cons}) relates the matrix $\hat g
_{ab}$ to 
the rigid transformations of anti-ghosts:
\begin{equation}
\label{rigantig}
\hat a ^{\bar  g}_{bc,\a} = - {\hat g ^T}_{bb'} a^V_{b'c',\a} 
{\hat g ^{-1T}}_{c'c}
= - (\hat g z^V)^T_{b\beta} \ve_{\b\ga\a} {(\hat g z^V)}^{-1 T}
_{\ga c}
\end{equation}
From rigid invariance it is therefore allowed to
introduce an arbitrary matrix into the BRS-trans\-for\-mation of
ghosts. From (\ref{rigantig}) it is obvious that such a  general
ansatz is related to  different field redefinitions of $B$-fields 
and anti-ghosts and, finally, vectors and anti-ghosts.

\newsubsection{The scalar sector}

The  algebra for the coefficients of the scalar fields has
the same form as the one for the vectors
\begin{eqnarray}
\label{funrep}
 b^s _\a\, \tilde I\,b^s _\b - 
b^s _\b\, \tilde I\,b^s _\a &=&
- \hat \varepsilon _{\a \b \ga}
\tilde I_{\ga \ga'}b^s_{\ga'} \qquad 
s \equiv \phi_a , Y_a, \hat \phi_a , q_a 
\end{eqnarray}
with $(b_\a)_{bc} = b_{bc,\a}$.
The solution, however, is distinguished from the one of the vector
representation equations due to a different transformation behaviour
of scalars with respect to CP: The general solution of
the scalar representation equations (\ref{funrep}) 
 is the fundamental representation  with
its equivalence class:
\begin{equation}
\label{hattdef}
b^s _{\a} \sim \hat t _\a 
\end{equation}
with 
\begin{eqnarray}
\label{that}
\hat t _+ = \frac 12 \left(\begin{array}{cccc} 0 & 0 & 0 & 0 \\
                                          0 & 0 & i & -  1 \\
                                          0 & - i& 0  & 0   \\
                                          0 &  1 & 0 &  0
                                                   \end{array}\right)
& \qquad &
\hat t _3 = \frac 12 \left(\begin{array}{cccc} 0 & -  i&0 & 0\\
                                           i& 0 & 0 & 0 \\
                                          0 & 0 & 0 &   1  \\
                                          0 & 0 & -  1 & 0
                                                         \end{array}\right) 
\nonumber \\
\hat t _- = \frac 12\left(\begin{array}{cccc} 0 & 0 & - i & - 1 \\
                                          0 & 0 & 0  & 0 \\
                                           i & 0 & 0 & 0  \\
                                           1 & 0 & 0 & 0
                                                     \end{array}\right)
& \quad & G^s
\hat t _4 = \frac {G^s} 2\left(\begin{array}{cccc}0 & -  i&0  & 0 \\
                                           i& 0 & 0 & 0 \\
                                          0 & 0 & 0 &  - 1 \\
                                          0 & 0 &   1 & 0
                                                         \end{array}\right)
\end{eqnarray}
It involves in the abelian component an undetermined parameter
for each field. The 4-dimensional representation we have chosen
here is equivalent to the complex 2-dimensional representation, which
is usually assigned to the scalars in the tree approximation
and which we have introduced in section~2.
The general solution is obtained from the special solution
(\ref{that}) by an equivalence
transformation. Because the Ward operators have to be CP-odd, 
the transformation matrices have to be real and diagonal. This means, that
mixing between Higgs- and the neutral would-be Goldstone is forbidden
in a CP-invariant theory:
\begin{eqnarray}
\label{bsol}
{( b _\a)}_{bc} &= &
 z_{bb'}\, {(\hat t_{\a})}_{bc}
\, z^{-1}_{c'c}  
\end{eqnarray}
with 
\begin{equation}
z_{a b} = \left(\begin{array}{cccc}
              z_+& 0& 0 & 0 \\
               0       &   z_+ & 0& 0\\
               0 & 0 &   z_H & 0      \\
              0 & 0 &  0& z_\chi     
                   \end{array} \right)  
\end{equation}
We have again suppressed the scalar field indices.
The dependence of the representation matrices on these parameters
is quite simple, it is seen that the representation matrices
only involve the ratios
\begin{equation}
\label{rscal}
r_+ = {z_+ \over z_H } \quad \quad  r_\chi = {z_\chi \over z_H} \, .
\end{equation}
As in the vector sector, rigid symmetry allows  independent field
redefinitions 
for each scalar field. Likewise one can fix the field redefinitions of the
charged and CP-odd components by the Ward identity of the tree
approximation:
\begin{equation}
r_+ = 1 + \delta r_+ \quad r_\chi = 1 + \delta r_\chi \qquad
\mbox{with}\quad \delta r_a = O(\hbar)
\end{equation}

Finally the consistency condition (\ref{cons}) relates the transformation
of the external fields $Y_a$ to the transformation of the
propagating fields $\phi_a$, and the transformation of $q_a$
to the transformation of $\hat \phi_a$:
\begin{equation}
b^Y_{bc,\a} = b^\phi_{bc,\a}  \qquad b^q_{bc,\a} = b^{\hat \phi}_{bc,\a}
\end{equation}
which reads for the free parameters involved
\begin{eqnarray}
r_a^Y = r_a^\phi& \qquad & G^Y = G^\phi \nonumber \\
r_a^q = r_a^{\hat \phi}& \qquad & G^q = G^{\hat \phi} \nonumber 
\end{eqnarray}
Because we are free to dispose over the external fields at will, as long
as we do not find any restrictions in the procedure of quantization,
we restrict the transformation of the fields $\hat \phi_a$ to be
the same as the one of the quantum fields. 
\begin{equation}
b^\phi_{bc,\a} = b^{\hat \phi}_{bc,\a}
\end{equation}

The representation equations of the shifts ($\tilde v_{b\a}\equiv
 v_{b\a}, \hat v_{b\a}$)
\begin{equation}
\label{shiftrep}
{\tilde v} _{b\a} \tilde I_{bb'} \, b^{ \phi} _{b'c\b} - 
{\tilde v} _{b\b} \tilde I_{bb'} \, b^{ \phi} _{b'c\a}  
=
- \hat \varepsilon _{\a \b \ga}
\tilde I_{\ga \ga'}{ \tilde v_{c \ga'} }
\end{equation}
are solved by
\begin{equation}
\label{shiftsol}
\begin{array}{cclccclcccl}
 v_{+-} = - v_{-+} & =& \frac i2 v r^{-1}_+ &
\quad & v_{\chi 3} &  =  & \frac 12 r^{-1}_\chi v & \quad &
v_{\chi 4 } & =& - \frac 12 G^\phi r^{-1}_\chi v  \\
\hat v_{+-} = - \hat v_{-+} & = & \frac i2 \zeta
v r^{-1}_+ &\quad & \hat v_{\chi 3} & = & \frac 12 r^{-1}_\chi \zeta v &\quad &
\hat v_{\chi 4 } & = & - \frac 12 G^{\phi} r^{-1}_\chi \zeta v  
\end{array}
\end{equation}
All the other components vanish according to charge neutrality and
CP-invariance. The free parameters are the shift of the quantum field
$v$ and the shift of the external field $\zeta v$:
\begin{equation}
v = 2 {M_Z\over e} \cw \sw + O(\hbar)
\end{equation}

\newsubsection{The fermion sector}

The algebra for representation matrices of fermions has the
following form:
\begin{eqnarray}
\label{fermrep}
h^{f_i}_\a \, h^{f_i}_\b - h^{f_i}_\b \, h^{f_i}_\a 
= - \hat \ve_{\a \b \ga} \tilde I_{\ga \ga'} \, h^{f_i}_{\ga'}
\nonumber \\
\tilde h^{f_i}_\a \,\tilde h^{f_i}_\b 
- \tilde h^{f_i}_\b \,\tilde h^{f_i}_\a = - 
\hat \ve_{\a \b \ga} \tilde I_{\ga \ga'} \, \tilde h^{f_i}_{\ga'} 
\end{eqnarray}
From the consistency equation with the Slavnov-Taylor operator
it is seen, that the transformation of external fields is
governed by the transformation of propagating fields:
\begin{equation}
h^{f_i}_\a = h^{\psi_i}_\a \qquad \tilde h^{f_i}_\a = \tilde h^{\psi_i}_\a 
\end{equation}
The matrices ${(h^{f_i}_{\a})}_{ff'}, f,f' = \nu , e, u, d,$ are $4\times 4$
 matrices and  ${(\tilde h^{f_i}_{\a})_{ff'}}, f,f' =   e, u, d,$ 
are $3\times 3$ matrices.
CP-invariance implies, that they are imaginary.

Lepton and quark number conservation
enables one to treat quarks and leptons separately and, actually, one only
has to  consider 2-dimensional representation matrices.
The non-trivial solution of the algebra is  
represented by the Pauli-matrices completed by
the unit matrix and its equivalence representations.
We know from the tree approximation, that left-handed fermions transform
according to doublets, whereas right-handed fields transform trivially
under $SU(2)$. This transformation behaviour cannot be spoiled in
perturbation theory. Therefore we assign in accordance with the
tree approximation
\begin{equation}
\label{doubsol}
h^{\delta_i}_\a \sim i\frac {\tau_\a}2 \qquad
 \tilde h^{\delta_i}_\a \sim 0 \quad\hbox{with}\quad \a = +,-,3
\end{equation}
$\delta_i= l_i, q_i$ is the index for quarks and leptons.
The abelian component is not well-defined by the algebra, but 
contains some free parameters.
For the nontrivial solution one finds always
one undetermined parameter for 
left-handed leptons and quarks of each family 
\begin{equation}
\label{doubsolab}
h^{f_i}_4 =  \left(\begin{array}{cc}i G^{l_i}{\mathbf 1} & 0 \\
                                  0  & i G^{q_i}{\mathbf 1} \end{array}
                  \right)
\end{equation}
The singlet solution involves undetermined parameters for each
right-handed
fermion.
\begin{equation}
\label{singsol}
\tilde h^{f_i}_4 = \left(\begin{array}{ccc} i G^{e_i} & 0 & 0   \\
0 & i G^{u_i} & 0   \\ 0 & 0 &  i G^{d_i}    
\end{array} \right)
\end{equation}
Due to charge conservation and CP-invariance the equivalence transformations
are carried out by diagonal real matrices, which are related in the
course of quantization to the field redefinitions of right- and left-handed
fields:
\begin{equation}
h^{l_i}_\a = i z^{l_i} \frac {\tau_\a}2 (z^{l_i})^{-1} \qquad
h^{q_i}_\a = i z^{q_i} \frac {\tau_\a}2 (z^{q_i})^{-1}
\end{equation}
with 
\begin{equation}
z^{l_i} = \left(\begin{array}{cc} z^{\nu_i} & 0 \\
                              0      & z^{e_i} \end{array}\right)
\qquad
z^{q_i} = \left(\begin{array}{cc} z^{u_i} & 0 \\
                              0      & z^{d_i} \end{array}\right)
\end{equation}
The singlet representation is independent from field redefinitions.
 The charged components
of the doublet representation  depend
 on the ratio of  field redefinitions carried
out for left-handed up-and down type quarks 
and left-handed neutrinos and charged leptons, respectively.
\begin{equation}
\label{rferm}
r_{l_i} = {z^{e_i} \over z^{\nu_i}} \qquad
r_{q_i} = {z^{d_i} \over z^{u_i}}
\end{equation}

Having analysed the general structure of $SU(2)\times U(1)$ operators
it is obvious, that rigid symmetry  does not restrict the
number of independent field redefinitions. Therefore it is allowed
to impose independent normalization conditions for the propagating
physical fields as well as for the would-be Goldstones and 
$\phi\pi$-ghosts without spoiling rigid symmetry.

\newsubsection{The algebraic characterization of an abelian local 
Ward operator}

The algebraic analysis of the last sections has shown that
the $SU(2)$-components of the rigid Ward operators are uniquely fixed
up to equivalence transformations, which are related to field
redefinitions of the different fields in question. The abelian
component ${\cal W}_4$, however, involves several free parameters, which in
higher orders appear as instabilities of the abelian subgroup and
have to be determined. If we assume now, that the
instabilities of the Ward operator are indeed the only breakings
which appear in higher orders, then one is able to fix
some of the free parameters 
to all orders of perturbation theory. 
But there are left the
parameters which correspond to lepton and quark family conservation,
and it is obvious that they remain independent
parameters of the abelian rigid Ward operator.

When one constructs the electroweak standard model from gauge
invariance these parameters are determined on the gauge
transformation by the Gell-Mann Nishijima relation, which
ensures that the photon couples to the electromagnetic current.
In the course of renormalization the gauge symmetries are broken
and the role of gauge symmetry is taken over by BRS-symmetry
and the Slavnov-Taylor identity. Via the nilpotency properties
it contains also the algebraic structure of the group in the
external field part. When solving the Slavnov-Taylor identity
it is seen that one is lead to representation equations for
the BRS-transformations, which have the same form as the ones
we have solved for establishing rigid operators.
It turns out, that the abelian
component is also not uniquely
defined in the  solution of the Slavnov-Taylor identity.
 In fact one finds the free
parameters which correspond to lepton and quark family conservation
to be undetermined as well. Leaving them as free parameters
the photon will not couple properly to the electromagnetic current
but also on the currents associated with lepton and quark
family conservation. For this reason we have to use a local
Ward identity in addition to the
Slavnov-Taylor identity for defining the gauge transformations of the
abelian component in an appropriate way. 
 The local Ward identity of electromagnetic symmetry has
non-abelian components and does not exist in renormalizable gauges.
Therefore we have to use the abelian Ward operator for fixing
the undetermined parameters 
continuing the Gell-Mann Nishijima relation on a functional level
to all orders of perturbation thery.

As we have already mentioned the Ward identities, which correspond
to charge conservation and conservation of lepton and quark family
number are not affected by renormalization. Therefore 
the identity
\begin{equation}
\label{conseq}
\Bigl({\cal W}_{em} + \sum_{i=1}^{N_F}
\left( g_{l_i} {\cal W}_{l_i}+ g_{q_i} {\cal W}_{q_i} \right)
\Bigr) \Gamma = 0
\end{equation}
is valid to all orders of
perturbation theory with arbitrary parameters $g_{l_i}$ and $g_{q_i}$.
Adding the general Ward operators ${\cal W}_3$ and ${\cal W}_4$
 in a way that 
for vectors and scalars the electromagnetic Ward operator arises
and the shifts vanish
\begin{equation}
{\cal W} = {\cal W}_3 + \frac 1{G^s} {\cal W}_4
\end{equation}
one gets by using (\ref{conseq}) the following identity, when
acting on the functional $\Ga$:
\begin{eqnarray}
{\cal W} \Gamma  =  \int & \biggl( & 
   \frac 1 {G^\phi}\sum_{i=1}^{N_F}
\Bigr( \,
 i ( G ^{u_i} - G^{q_i} -G^\phi ) \overline {u_i^R} {\delta \over \delta 
\overline{u_i^R} } +
 i ( G ^{d_i} - G^{q_i} ) \overline {d_i^R} {\delta \over \delta 
\overline{d_i^R} }  \nonumber \\
& & \phantom{\frac 1 {G^\phi}\sum_{i=1}^{N_F}} +
i ( G ^{e_i} - G^{l_i} -G^\phi ) \overline {e_i^R} {\delta \over \delta 
\overline{e_i^R} }  + \hbox{h.c.} \Bigl) \biggr) \Gamma
\end{eqnarray}
If one assumes, that these are the only breakings of 
the rigid Ward operators, which arise in higher orders, then
  the coefficients appearing therein  have to vanish to all orders
of perturbation theory,
because they can be independently tested on non-vanishing
vertices of the classical action, namely on the scalar interaction,
the fermion masses and the gauge fixing:
\begin{equation}
\label{chargerel}
  G^{u_i} = G^{q_i} + G^\phi \qquad
G^{d_i} = G^{q_i} \qquad 
G^{e_i} = G^{l_i}
\end{equation}
These relations just state
that the charges of leptons and quarks of each family differ
by one unit, which is determined by the charge of the $W_+$.
 The final abelian Ward operator acting on
$\Gamma$ takes the form:
\begin{equation}
\label{relab}
{\cal W}_4 \Gamma = \Bigl(
 {\cal W}_{em} -{\cal W}_3 +  \sum_{i=1}^{N_F} \bigl(
g_{l_i} {\cal W}_{l_i}+ g_{q_i} {\cal W}_{q_i} \bigr) \Bigr)\Gamma
\end{equation}
with undetermined parameter $g_{l_i}$ and $g_{q_i}$. 
Here we have also chosen the overall normalization appropriately,
i.e.~$G^\phi =1 $. The problem of deriving a local Ward identity
in connection with the abelian subgroup is therefore not well-posed,
but has to be restated by requiring to have a local
Ward identity in connection with the electromagnetic current.
Defining the local Ward operator connected with electromagnetic
current conservation by
\begin{equation}
\label{wqdef}
{\mathbf w}_4^Q \equiv {\mathbf w}_{em} - {\mathbf w}_3
\quad\hbox{with}\quad {\cal W}_3 - {\cal W}_{em} = \int 
( {\mathbf w}_3 - {\mathbf w}_{em} )
\end{equation}
it is seen that it is algebraically unique up to
a total derivative acting
on  the differentiation with respect to the abelian combination
of vector fields. The operator
\begin{equation}
\label{wqloc}
\hat {\mathbf w}_4^{Q} = g_1 {\mathbf w}_4^Q 
- \frac 1{r^V_Z} \partial{\delta \over \delta Z} \sin \Theta ^V 
- \frac 1{r^V_A} \partial{\delta \over \delta A} \cos \Theta ^V 
\end{equation}
commutes with the Slavnov-Taylor operator and the Ward operators
of rigid symmetry:
\begin{equation}
\label{wqcons}
\left[
\hat {\mathbf w}^Q_4 , {\cal W}_\a \right] = 0
\qquad \brs_\Gamma  \hat {\mathbf w}_4 \Gamma - \hat {\mathbf w}^Q_4 
{\cal S}(\Gamma) =0 \quad\hbox{for any }\Gamma
\end{equation}  
The final version of the abelian Ward identity we have to prove to
all orders of perturbation theory takes the form
\begin{equation}
\label{wardidho}
\left( g_1 {\mathbf w}_4^Q 
- \frac 1{r^V_Z} \partial{\delta \over \delta Z} \sin \Theta ^V 
- \frac 1 {r^V_A} \partial{\delta \over \delta A} \cos \Theta ^V \right)
\Gamma =\frac 1{r^V_Z} \Box B_Z \sin \Theta ^V +
\frac 1 {r^V_A} \Box B_A \cos \Theta ^V 
\end{equation}
with $ r_Z^V ,r_A^V $ and $\Theta^V$ determined on the charged  rigid $SU(2)$
Ward identities (\ref{par}). It involves an overall normalization parameter,
which  depends on the parametrization one has chosen and in higher
orders on the normalization condition of the coupling. In the QED-like
on-shell schemes it is given by
\begin{equation}
g_1 = \frac e \cw + O(\hbar) 
\end{equation}

The Ward identity (\ref{wardidho}) has to be established in higher
orders of perturbation theory, in order to fix the undetermined
parameters appearing in the action as a consequence of the
instability of the abelian subgroup. These instabilities
 are connected with the fact, that it is not possible to
algebraically   distinguish between gauging the electromagnetic
current and the currents associated with lepton and quark familiy
conservation. 
If we were not able to establish the abelian local Ward identity to all orders
we had to impose a normalization condition for one charged-fermion
photon vertex of each family, but we would loose thereby the control 
if the gauge symmetry is indeed the electromagnetic symmetry  and 
not the current associated with lepton and quark family 
conservation, which one gauges in higher orders.

These observations have important consequences for the construction
of the gauge fixing and ghost sector: In order to identify the abelian
Ward identity according to 
(\ref{wqdef}) and (\ref{wqcons})
rigid  $SU(2)\times U(1)$ Ward identities have to
be established. The gauge fixing
sector
 has therefore to be constructed with the help
of the external scalar fields as introduced in section 2.2.
In this procedure the
number of independent gauge parameters is restricted. In order to
avoid infrared divergent counterterms for the $\phi\pi$-ghosts one
is forced to introduce in higher orders an
independent ghost angle, which appears in the Slavnov-Taylor identity
and the Ward identities of rigid symmetry via different field redefinitions
of vectors and antighosts  as derived in (\ref{rigantig}). 

\newsection{The general local solution of the Slavnov-Taylor identity
and rigid symmetries}
\newsubsection{The normalization conditions}

As we have outlined in section 2, the construction of higher orders
proceeds  by proving, that it is possible
to adjust local contributions in such a way, that the functional
of 1PI Green functions is  invariant under the ST identity
and the Ward identities of rigid symmetry. Local contributions are
algebraically separated into two classes: invariants of the symmetry
and non-invariant contributions (cf.~(\ref{defren})).
 The coefficients of the invariants
have to be fixed by appropriate normalization conditions and 
vice versa it has to be shown, that the normalization conditions
one wants to impose for a proper particle interpretation only
dispose of invariant coefficients. 

We impose for all physical fields on-shell conditions as given
in the literature (see e.g.~\cite{AOHI82}).
 For vectors and scalars they read on the
2-point functions
\begin{equation}
\label{normmass}
\begin{array}{lccclcc}
{\mathrm Re} \Gamma^T_{+-} (p^2)\big|_{p^2  = M_W^2} & = & 0 &\qquad &
{\mathrm Re} \Gamma^T_{ZZ} (p^2)\big|_{p^2  = M_Z^2} & =& 0 \\
\phantom{\mathrm Re} \Gamma^T_{AA} (p^2)\big|_{p^2  = 0} & = & 0 &\qquad &
{\mathrm Re} \Gamma_{HH} (p^2)\big|_{p^2  = m_H^2} &  = &  0
\end{array}
\end{equation}
The photon and the Z-boson are not distinguished by quantum numbers
and mix from 1-loop order onwards. Therefore they have to be
separated on-shell:
\begin{equation}
\label{normmix}
\Ga^T_{ZA} (p^2)\big|_{p^2  = 0} = 0 \qquad
{\mathrm Re} \Ga^T_{ZA} (p^2)\big|_{p^2  = M_Z^2} = 0
\end{equation}
The Higgs and the neutral would-be Goldstone are distinguished by their
transformation properties under CP. For this reason the respective
conditions for scalars 
are valid by construction in a CP-invariant theory.
The transversal part of the vector
2-point functions is defined according to
\begin{equation}
\Gamma_{V^\mu_a V^\nu _b} \equiv - \bigl( \eta ^{\mu \nu} - 
{p^\mu p^\mu \over p^2 } \bigr) \Gamma^T_{ab}
- {p^\mu p^\mu \over p^2 }  \Gamma^L_{ab}
\end{equation}
For the unstable particles the counterterms are fixed by the requirement
that the real part of the 2-point functions is vanishing. This
prescription has to be reanalysed 
\cite{STU97}, if one constructs the S-matrix 
and especially wants to prove gauge parameter independence of physical
quantities. For the construction of finite Green functions it is certainly
a well-defined normalization condition, which continues the tree 
approximation of the on-shell scheme  to higher orders in a proper form.

Because the residua are finally  canceled when constructing the S-matrix,
there is quite some arbitrariness involved in the 
respective normalization
conditions.
In the complete on-shell scheme the residua of all 
physical particles are
fixed at the pole position. In order to avoid on-shell infrared divergencies
we modify the complete on-shell scheme by introducing a normalization point
$\kappa_a^2$ for each vector   and impose for 
the transversal part of the vectors:
\begin{equation}
\label{normres}
{\mathrm Re}
\partial _{p^2}\Gamma^T_{+-} (p^2)\big|_{p^2  = \kappa_W^2} = 1 \quad
{\mathrm Re}
\partial _{p^2}\Gamma^T_{ZZ} (p^2)\big|_{p^2  = \kappa_Z^2} = 1 \quad
{\mathrm Re}
\partial _{p^2}\Gamma^T_{AA} (p^2)\big|_{p^2  = \kappa_A^2} = 1
\end{equation}
In this form they allow to switch between different normalization 
conditions by adjusting $\kappa^2_a$.
As we have already mentioned two of these
normalization conditions can be replaced by the Ward identities
of rigid symmetry, which corresponds to the
minimal on-shell scheme. 

Finally one has to specify normalization conditions for the
residua of the scalars, which is carried out similarly as
above. As it will be seen from the general solution of the ST identity
also the residua of unphysical Goldstone bosons are not fixed
by the ST identity, but could be fixed on the Ward identities of
rigid symmetry. Because they are not considered in external legs
in physical scattering processes, the divergencies appearing
in dimensional regularization are often subtracted according to
 the MS scheme. In order to remain quite general in the construction
we impose normalization conditions on arbitrary normalization points
$\mu _a^2$:
\begin{equation}
\label{normresscal}
{\mathrm Re}
\partial _{p^2}\Gamma_{+-} (p^2)\big|_{p^2  = \mu_W^2} = 1 \quad
{\mathrm Re}
\partial _{p^2}\Gamma_{HH} (p^2)\big|_{p^2  = \mu_H^2} = 1 \quad
{\mathrm Re}
\partial _{p^2}\Gamma_{\chi\chi} (p^2)\big|_{p^2  = \mu_\chi^2} = 1
\end{equation}

The normalization conditions for fermions are listed in the literature
quite generally 
also for the case, that there is CP-violation via
the CKM matrix \cite{AOHI82}. They simplify considerably, if one assumes
lepton and quark family conservation. 
Decomposing the fermion 2-point functions according to
\begin{eqnarray}
\Ga_{\bar f_if_i} & =& \ps \Ga^L_{f_i}(p^2)\hfrac 12 (1-\ga_5) +
              \ps \Ga^R_{f_i}(p^2)\hfrac 12 (1+\ga_5) 
              - m_{f_i} {\mathbf 1} \Ga^m_{f_i}(p^2) \nonumber \\
 & \equiv &\ps - m_{f_i} + \ps \Sigma^L_{f_i}(p^2)\hfrac 12 (1-\ga_5) +
              \ps \Sigma^R_{f_i}(p^2)\hfrac 12 (1+\ga_5) 
              - m_{f_i} {\mathbf 1} \Sigma^m_{f_i}(p^2)
\end{eqnarray}
the on-shell conditions read
\begin{equation}
{\mathrm Re}\bigl(
\Ga^L_{f_i} (p^2) - \Ga^m_{f_i} (p^2)\bigr)\big|_{p^2 = m_{f_i}^2} = 0 \qquad
{\mathrm Re}\bigl(
\Ga^R_{f_i} (p^2) - \Ga^m_{f_i } (p^2)\bigr)\big|_{p^2 = m_{f_i}^2} = 0
\end{equation}
They impose pole conditions on the Dirac spinors and forbid parity
violation for the on-shell propagators.
On-shell residua are endangered by infrared divergencies as it is the case 
in QED with a massless photon. 
We therefore introduce off-shell conditions: 
\begin{equation}
{\mathrm Re}\partial_{p^2}\bigl( p^2
\Ga^L_ {f_i} (p^2) + p^2 \Ga^R_{f_i} (p^2)\bigr) \big|_{p^2 = \kappa_i^2} = 1
\end{equation}

Just by construction of vertex functions it is ensured that
\begin{equation}
\Gamma_H = 0
\end{equation}
which forbids to introduce linear Higgs field terms  into
the local contributions of higher
order corrections.

For proving unitarity of the physical S-matrix we have also to
impose normalization conditions on the unphysical fields. The poles of
propagators of the longitudinal parts of the vectors, of the unphysical
would-be Goldstones and the Faddeev-Popov ghosts are seen to be
related by the ST identity. The normalization conditions on
the poles of unphysical particles are most easily established
on the Faddeev-Popov fields and read:
\begin{equation}
\label{ghostmass}
{\mathrm Re}
\Ga_{\bar c_+ c_-} (p^2) \big|_{p^2 = \zeta_W M_W^2} = 0 \quad
{\mathrm Re}
\Ga_{\bar c_Z c_Z} (p^2) \big|_{p^2 = \zeta_Z M_Z^2} = 0 \quad
\Ga_{\bar c_A c_A} (p^2) \big|_{p^2 = 0} = 0
\end{equation}
Furthermore one has to require on-shell separation for neutral ghosts
\begin{equation}
\label{ghostdemix}
\begin{array} {lccclcc}
{\mathrm Re}
\Ga_{\bar c_Z c_A} (p^2) \big|_{p^2 = \zeta_Z M_Z^2} & = & 0 & \qquad &
\Ga_{\bar c_Z c_A} (p^2) \big|_{p^2 = 0} & =&  0 \\
{\mathrm Re}
\Ga_{\bar c_A c_Z} (p^2) \big|_{p^2 = \zeta_Z M_Z^2} & = & 0  & \qquad &
\Ga_{\bar c_A c_Z} (p^2) \big|_{p^2 = 0}&  = & 0
\end{array}
\end{equation}
Finally we impose also normalization conditions on the residua
of the ghost propagators:
\begin{equation}
\label{ghostres}
{\mathrm Re}
\partial _{p^2}\Gamma_{\bar c_+ c_-} (p^2)\big|_{p^2  = \kappa_W^2} = 1 \quad
{\mathrm Re}
\partial _{p^2}\Gamma_{\bar c_ Z c_Z}
 (p^2)\big|_{p^2  = \kappa_Z^2} = 1 \quad
{\mathrm Re}
\partial _{p^2}\Gamma_{\bar c_A c_A} (p^2)\big|_{p^2  = \kappa_A^2} = 1
\end{equation}

Solving the ST identity for the most general local action which is
compatible with UV dimension 4, the parameters, which are fixed
order to order by normalization conditions, have to be free 
parameters in terms of which all the other couplings  are determined.
 The local contributions which
are fixed by the above normalization conditions are given by
\begin{eqnarray}
\label{gagenbil}
\Ga^{gen}_{bil} & = &
\int \biggl(-\hfrac 14 
\bigr(\partial^\mu V^\nu_a - \partial^\nu V^\mu_a \bigl)  Z^V_{ab}
\bigr(\partial_\mu V_{\nu b} - \partial_\nu V_{\mu b} \bigl)  
+ \hfrac 12 V^\mu_a  {\cal M}^V_{ab} V_{\mu b} \nonumber \\
& & \phantom{\int } \:
+ \hfrac 12 \partial^{\mu} \phi_a Z_{ab}^S \partial_\mu\phi_b 
- \hfrac 12 M_H^2 H^2(x) \nonumber \\
& & \phantom{\int } \:
+ i Z_{f_i}^R\bar f_{i}^R \ds f_{i}^R +
i Z_{f_i}^L\bar f_{i}^L \ds f_{i}^L
- M_{f_i} (\bar f_{i}^R  f_{i}^L +\bar f_{i}^R  f_{i}^L) \nonumber \\
& & \phantom{\int } \:
- \bar c_a Z_{ab}^g \Box c_b 
- \bar c_a {\cal M}^g_{ab} c_b
\biggr)
\end{eqnarray}
The matrices and parameters are chosen in accordance with charge
neutrality and CP-invariance, especially $ Z^S_{ab}$ is a diagonal
matrix. In perturbation theory the parameters are order by order
determined by the above normalization conditions:
\begin{equation}
\begin{array}{lclclcl}
Z^V_{ab} & = &\tilde I_{ab} + \delta Z^V_{ab} 
& \qquad & Z^R_{f_i} &=& 1+ \delta Z^R_{f_i}  \\
Z^S_{ab} & = & \tilde I_{ab} + \delta Z^S_{ab}
&\qquad  & Z^L_{f_i} & = & 1+ \delta Z^L_{f_i} 
\end{array}
\end{equation}
and respective expressions for the Higgs mass and fermion masses
\begin{equation}
M_H^2 = m_H^2 + \delta m_H^2 \qquad
M_{f_i} = m_{f_i} + \delta m_{f_i}      
\end{equation}
 The vector mass matrix is non-diagonal and can be decomposed into
an orthogonal matrix and a diagonal matrix:
\begin{equation}
{\cal M}^V_{ab} = O^T (\theta)
\left( \begin{array}
{cccc}
0 &M_{+-} &  0 & 0 \\
M_{+-} & 0 & 0 & 0 \\
0 & 0 & M_{ZZ} & 0 \\
 0  & 0 & 0 & M_{AA}  \end{array}
\right) O (\theta )\quad \hbox{with} \quad 
\begin{array}{lcl}
M_{+-} & = & M_W^2 + \delta M_W^2 \\
M_{ZZ}  & =&  M_Z^2 + \delta M_Z^2 \\
M_{AA} & = & 0  \\                                
\theta & =&  0 + \delta \theta
\end{array}
\end{equation}
$M_W^2, M_Z^2, m_H^2$ and $m_{f_i}$ are the physical masses of the
particles.
The explicit form of the local counterterms is of course dependent
on the way one has constructed the finite renormalized 1PI Green functions.
The  objects we are able to talk about in a scheme independent
way are the finite Green functions.
 Constructing them in accordance with
the symmetries, they are finally   governed by the normalization conditions and
are independent of the scheme, one has used for subtracting the divergencies.
Especially
   the conditions for separating massless
and massive particle at $p^2 = 0$ (\ref{infra})
\begin{eqnarray}
\label{infra2}
\Gamma_{ZA}(p^2=0) = \Gamma_{AA}(p^2 = 0) &=& 0 \\
\Gamma_{\bar c _A c _Z}(p^2=0) =
\Gamma_{\bar c _Z c _A}(p^2=0) = \Gamma_{\bar c_ Ac _A}(p^2 = 0) &=& 0 
\nonumber
\end{eqnarray}
 have 
 to be established on the finite 2-point functions
in order to be able to carry out infrared finite
higher order calculations.
In the BPHZL scheme, which
treats massless particles quite systematically,
 these normalization conditions  are implemented in the
scheme. One has therefore $\delta \theta ^{BPHZL} = 0 $. In dimensional
regularization these normalization conditions have to be carefully implemented
by adjusting e.g.~$\delta \theta ^{dim}$.

\newsubsection{The symmetry transformations and the general action}

For finding the invariant counterterms,  which  are added order
by order  in perturbation theory to the nonlocal contributions, we have
 to solve the ST identity and the Ward identities of rigid
symmetry  for the most general local action
$\Ga^{gen}_{cl}$, which is compatible with renormalizability by
power counting (cf.~(\ref{defren}) and (\ref{invc})). 
\begin{equation}
\label{gagensym}
{\cal S} (\Gamma_{cl}^{gen}) =0 \qquad {W_\a } (\Gamma_{cl}^{gen}) = 0 
\end{equation} 
For solving these equations
one could proceed  in a perturbative
expansion, but as  for the Ward operators of rigid symmetry
such a treatment
disguises the simple algebraic structure of the final solution,
hence we proceed differently.

For the ST operator and the Ward operators we take the general
operators as they are determined by consistency and by the 
$SU(2) \times U(1)$-algebra in section 3 from the general ansatz 
 (\ref{STgen}) and (\ref{wardriggen}).
The ST operator is written in the following form: 
\begin{eqnarray}
\label{STgen2}
{\cal S}
(\Ga ) &\hspace{-3mm}= &\hspace{-3mm} \int \biggl(  
\bigl(r^ g_{4Z} \partial _\mu c _Z + r^ g_{4A} \partial_\mu c_A\bigr)
             \Bigl(\frac 1{r_Z} \sin
\Theta {\delta \Ga  \over \delta Z_\mu } + \frac 1{r _A} \cos \Theta 
       {\delta \Ga   \over \delta A _\mu} \Bigr) \\  
 &\hspace{-3mm}
 & +     {\delta  \Ga \over \delta \rho^{\mu}_3 } 
              \Bigl(\frac 1 {r _Z} \cos \Theta 
       {\delta \Ga 
  \over \delta Z _\mu } - \frac 1 {r _A} \sin \Theta 
       {\delta \Ga   \over \delta A _\mu} \Bigr) 
       + {\delta \Ga   \over \delta \sigma _3 } {1 \over \det r^g}
              \Bigl(r^ g_{4A} {\delta \Ga   \over \delta c_Z } - r^ g_{4Z}
       {\delta \Ga  \over \delta c_A} \Bigr) \nonumber \\ 
&\hspace{-3mm} &      + {\delta \Ga   \over \delta  \rho^{\mu} _+ }
               {\delta \Ga   \over \delta W _{\mu,- } }
      + {\delta \Ga   \over \delta \rho^{\mu} _- }
               {\delta \Ga  \over \delta W _{\mu,+ } }
+  {\delta \Ga   \over \delta \sigma _+ }
               {\delta \Ga   \over \delta c_{-} }
+  {\delta \Ga   \over \delta \sigma _- }
               {\delta \Ga   \over \delta c_{+} } +
{\delta \Ga   \over \delta Y _a}\tilde I_{aa'}
{ \delta \Ga   \over \delta \phi _{a'} } 
 \nonumber \\
&\hspace{-3mm}& + \sum_{i=1}^{N_F} 
\Bigl({\delta \Ga   \over \delta \overline{\psi^{L}
_{f_i}}}
{   \delta \Ga \over \delta f^{R}_{i} }
+ {\delta \Ga   \over \delta \overline{\psi^{R}_{f_i}}}
{   \delta \Ga \over \delta f^{L}_{i} } + 
  \hbox{h.c.} \Bigr)  \nonumber \\
& \hspace{-3mm}& 
+ B_a (r^{V})^{-1}_{a\a}  \delta  \hat 
g_{\a b} {\delta \Ga   \over \delta \bar c_b }  
   + q_a{\delta \Ga  
 \over \delta \hat \phi_a  }  \biggr) 
 \nonumber
\end{eqnarray}
The Ward operators involve the representation matrices of the
fundamental and adjoint representation with their equivalence
classes (cf.~(\ref{adrepsol}),(\ref{hattdef}) and (\ref{doubsol})).
Because the abelian Ward operator  is related to the nonabelian
neutral Ward operator ${\cal W}_3$ and to the operators of global
unbroken symmetries ${\cal W}_{em}$ and ${\cal W}_{l _i},{\cal W}_{q _i}$ 
according to equ.~(\ref{relab}),
we only have to consider the non-abelian Ward operators 
for establishing rigid symmetry ($\a = +,-,3$).
\begin{eqnarray}
\label{wardopequivclass}
{\cal W}_\a = \tilde I_{\a\a'}\int dx &\hspace{-3mm}\biggl( & \hspace{-3mm}
V^{\mu}_b ( r^{V})^T_{b\b}
 \hat \ve_{\b\ga \a'} ( r^{V})^{-1T}_{\ga c}
\tilde I_{c c'} {\delta \over \delta V^{\mu}_{c'} }
+B_b (r^{V})^{-1}_{b\b} 
\hat \ve_{bc,\a'}( r^{V})_{\ga c}  
\tilde I_{cc'} {\delta \over \delta B_{c'} }  \\
&\hspace{-3mm}+& \hspace{-3mm}
c_b (r^{g})^{T}_{b\b}  \ve_{\b\ga\a'}
(r^{g})^{-1T}_{\ga c}  \tilde I_{cc'} {\delta \over \delta c_{c'} }
+\bar c_b (\delta g)^{-1}_{b\b}  \hat \ve_{\b \ga \a'}
 \delta g_{\ga c}  \tilde I_{cc'}
 {\delta \over \delta \bar 
c_{c'} } \nonumber \\
&\hspace{-3mm}+& \hspace{-3mm}
({r^S_b} \phi_b + \delta_{Hb} v) 
\hat t_{bc,\a'} {r^S_c}^{-1} 
 \tilde I_{cc'} {\delta \over \delta \phi_{c'} }
+Y_b  {r^S_b}^{-1} 
\hat t_{bc,\a'} {r^S_c}  \tilde I_{cc'} {\delta \over \delta Y_{c'} }
\nonumber \\
&\hspace{-3mm}+& \hspace{-3mm}
 ( {r^S_b} \hat\phi_b + \hat\zeta v \delta_{Hb} ) 
\hat t_{bc,\a'} {r^S_c}^{-1}  \tilde I_{cc'} {\delta \over \delta 
\hat\phi_{c'} }
+q_b  {r^S_b}
\hat t_{bc,\a'} {r^S_c}^{-1} 
 \tilde I_{cc'} {\delta \over \delta q_{c'} } \nonumber \\
&\hspace{-3mm}+&\hspace{-3mm}
\rho_\b   \ve_{\b\ga,\a'} \tilde I_{\ga \ga'} 
{\delta \over \delta \rho_{\ga'} }
+\sigma_\b   \ve_{\b\ga,\a'} \tilde I_{\ga \ga'} 
{\delta \over \delta \sigma_{\ga'} } \nonumber \\
&\hspace{-3mm}+&\hspace{-3mm}
\sum_{i=1}^{N_F} \sum_{\delta = l,q} \Bigl(
\overline {F^L_{\delta_i}} 
 r^{\delta_i}   \frac {i\tau_{\a'}}2  (r^{\delta_i})^{-1}
{\delta \over \delta \overline {F^L_{\delta_i} }} -
{\delta \over \delta  F^L_{\delta_i} }  r^{\delta_i} \frac {i\tau_{\a'}}2
(r^{\delta_i})^{-1} F^L_{\delta_i} \nonumber \\
&\hspace{-3mm} & \hspace{-3mm} \phantom{\sum \sum} 
+\overline {\Psi^R_{\delta_i}} (r^{\delta_i})\inv \frac { i \tau_{\a' }} 2
r^{\delta_i}
{\delta \over \delta \overline {\Psi^L_{\delta_i}} }  -
{\delta \over \delta  \Psi_{\delta_i}^R }
 (r^{\delta_i})\inv \frac { i \tau_{\a' }} 2 r^{\delta_i}
\Psi_{\delta_i}^R 
\Bigr)\biggr) \nonumber 
\end{eqnarray}
There
 we have parameterized the equivalence classes of
 rigid transformations by  $r^V ,r^S,r^ g$ and $ r^{l_i},
 r^{q_i}$  taking into account, that we are able
to determine field redefiniton matrices up to invariant matrices.
According to (\ref{par})
and (\ref{adjpar}) we define the matrix $r^V$  by
\begin{equation}
\label{defrv}
r^V_{\a a} = \left(\begin{array}{cccc}
                 1& 0 & 0 & 0\\
                 0& 1 & 0 & 0 \\
                 0& 0 &
            r_Z \cos \Theta & -r_A\sin \Theta \\
                 0 & 0 & r_Z \sin \Theta & r_A \cos \Theta
                 \end{array} \right)
\end{equation}
The equivalent transformations for scalars and fermions are chosen
as in (\ref{rscal}) and (\ref{rferm}):
\begin{equation}
\label{defrs}
r^S  = \left( \begin{array}{cccc} r^S_+ & 0 & 0 & 0 \\
                           0 & r^ S_+ & 0 &  0 \\           
                           0 & 0 & 1 &  0 \\           
                           0 & 0 & 0 & r^S_\chi   \end{array} \right) 
\qquad        
r^{\delta_i}  = \left( \begin{array}{cc} 1 & 0  \\
                           0 & r_{\delta_i}             
                              \end{array} \right)         
\end{equation}
Furthermore  the vector transformations in the
ST operator (\ref{STgen2}) are parametrized in agreement with the relations
gained from the consistency between the general Ward operators
and the ST operator (cf.~(\ref{stwardvec}) and (\ref{extffix})).
The transformation matrix of ghosts is defined by
\begin{equation}
\label{rghost}
r^g_{\a a} = \left(\begin{array}{cccc}
                 1& 0 & 0 & 0\\
                 0& 1 & 0 & 0 \\
                 0& 0 &  r^ g_{3Z}  & r^ g_{3A} \\
                 0 & 0 &   r^ g_{4Z}  & r^ g_{4A} \\
                 \end{array} \right)
\end{equation}
Here we have disposed over  the invariant abelian
parameter by taking the linear
BRS-transformation of vector fields into the transformation matrix.
The parameters in the nonlinear transformations of the neutral ghosts
 are then chosen in
accordance with nilpotency and the consistency relation.
Finally we have splitted the matrix $\hat g_{ab}$ which governs
the BRS-trans\-for\-mations of anti\-ghosts into the     rigid
transformation matrices of $B$-fields and antighosts:
\begin{equation}
\hat g_{ab} = (r^V)^{-1}_{a \beta} \delta \hat g_{\beta b}
\end{equation}
The rigid transformations of antighosts are determined from the
consistency condition as related to $\delta \hat g_{\a b}$.

With these operators we have to act on 
$\Ga_{cl}^{gen}$, which consists of all local field polynomials compatible
with UV-dimension 4. (See the appendix for quantum numbers.) For finding
the invariant counterterms we do not use a specific scheme for
treating massless particles. If  the parameters of the bilinear
action (\ref{gagenbil}) are indeed  free parameters, it is ensured,
  that we are
able  to establish all normalization conditions, especially the
ones for separating massless and massive particles at $p^2 = 0$.
Further restrictions on $\Ga_{cl}^{gen}$ are  neutrality
with respect to electric and Faddeev-Popov charge
(\ref{chargecons}) and 
lepton and quark family conservation (\ref{qlfcons})
\begin{equation}
\label{consgagen}
\begin{array}{lcl}
{\cal W}_{em} \Ga^{gen}_{cl}& = & 0  \\
{\cal W}_{\phi\pi} \Ga^{gen}_{cl} & = &  0
\end{array}
\qquad
\begin{array}{lcl}
{\cal W}_{l_i} \Ga^{gen}_{cl}& = & 0  \\
{\cal W}_{q_i} \Ga^{gen}_{cl} & = &  0
\end{array}
\end{equation}
According to lepton and quark family conservation  one is able to
restrict the analysis to CP-invariant field polynomials.
From formal unitarity it is required, that local contributions
are hermitean:
\begin{equation}
\label{comconj}
\Ga_{cl}^{gen} = {\Ga_{cl}^{gen}}  ^\dagger
\end{equation}
The general ansatz for $\Ga^{gen}_{cl}$ is quite lengthy. For the
purpose of the present paper we explicitly give only the  
most general  external field part. When solving (\ref{gagensym})
it is seen that the solution is
 traced back  to  representation equations for the general
couplings of the external field part.
Finally 
 it has to be shown that these couplings
 as well as the ST identity
and the Ward operators of rigid symmetry are uniquely determined
as functions of those parameters of which one disposes
by the normalization conditions (\ref{gagenbil}). 
Vice versa the bilinear part
of the action cannot possibly be restricted by the ST identity
and rigid symmetry, because otherwise unitarity and particle
interpretation of the field theory is endangered.

We start the presentation of the general classical solution
in the vector, scalar and fermion
part of the action. Having established there the ST identity,
rigid invariance follows as a consequence with well determined 
coefficients of rigid
transformations for vectors, scalars
and fermions. The gauge fixing sector can then be established
in accordance with rigid invariance and allows to compute the
ghost sector and at the same time the transformation parameters
of ghosts and antighosts.   
The procedure for adjusting symmetric
local contributions as outlined here in the abstract approach has to
be done exactly the same way in practice when calculating  order
by order in perturbation theory.

\newsubsection{The vector-scalar and fermion part of the action}

In this section we present  the general solution of the
ST identity in the vector, scalar and fermion part of the
action, which can be solved in combination with the external
field part self-consistently.
  The bilinear part of the most general local action
$\Ga_{cl}^{gen} $ is given
in (\ref{gagenbil}).
Because the parameters appearing therein  are fixed by normalization
conditions, they should not be determined in the course of the calculations.
For this reason we redefine the vector and scalar  fields 
in such a way, that the bilinear part takes a simple form:
\begin{equation}
\label{barefields}
\begin{array}{lcl}
V^o_{\mu a} & = & z^V_{ab} V^o_{\mu a} \\
\phi_a\bare  & = & z^S _a \phi_a
\end{array} \qquad
\begin{array}{lcl}
f^{oR}_{i} = \tilde z_{f_i} f_{i}^R \\
f_{i}^{oL} =  z_{f_i} f_{i}^L
\end{array} 
\end{equation}
In a first step we have to show that these parameters 
 are uniquely determined from the
parameters appearing in the bilinear part of the action
$\Ga^{gen}_{bil}$, we fix by the normalization conditions.
 Due to CP-invariance all  field redefinitions can be
chosen real. Fermion und scalar field redefinitions are  determined
on the kinetic parts up to a sign, which is irrelevant in perturbation
theory and can finally be adjusted in the tree approximation:
\begin{equation}
\label{fieldredscf}
Z^S_{ab} = 
\left(\begin{array}{cccc}
       0 & z_+^2 & 0 & 0 \\
       z_+^2 & 0 & 0 & 0 \\
        0& 0& z_H^2 &  0 \\
       0 & 0 & 0 & z^2_\chi \end{array} \right)
\qquad
\begin{array} {ccc}
Z^R _{f_i} & = &\tilde z_{f_i}^2 \\
Z^L _{f_i} & = & z_{f_i}^2
\end{array}
\end{equation}
Because photon and Z-boson are not distinguished by any quantum numbers,
the 
 vector redefinition matrix  nondiagonal in the neutral sector.
We parameterize this matrix as in (\ref{fred}):
\begin{equation}
 z^V_{ a b} = \left(\begin{array}{cccc}
              z_W& 0& 0 \\
               0       &    z_W & 0& 0\\
               0 & 0 &   z_Z \cos \theta_Z & - z_A \sin \theta_A    \\
              0 & 0 &    z_Z \sin \theta_Z &     z_A \cos \theta_A 
\end{array}\right)
\end{equation}
 On the kinetic parts  $z^V_{ab}$ is determined up to 
an orthogonal matrix:
\begin{equation}
Z^V_{ab} = {z^V _{aa'}}^T \tilde I_{a'b'} z^V _{b'b}
         = \bigl(  O(\theta){z^V}\bigr)^T_{aa'}  \tilde I_{a'b'}
 \bigl(O (\theta) z^V\bigr)_{b'b}
\end{equation}
 This remaining orthogonal matrix can be fixed on the vector
mass matrix.
Requiring ${\cal M}^o_{ab}$ to be diagonal
\begin{equation}
\label{baremassmatrix}
{\cal M}^o_{ab} = (z^V)_{aa'}^{-1T} {\cal M}_{a'b'} (z^V)_{b'b}\inv \quad
\hbox{with} \quad
{\cal M}^o_{ab} =
\left(\begin{array}{cccc}
       0 & {M_W^o}^2 & 0 & 0 \\
       {M_W^o}^2 & 0 & 0 & 0 \\
        0& 0& {M_Z^o}^2 &  0 \\
       0 & 0 & 0 & {M_A^o}^2 \end{array} \right)
\end{equation}
finally determines $z^V_{ab}$ uniquely up to signs.
Transforming likewise the masses of the fermions and the Higgs
into bare masses:
\begin{equation}
m^o_{f_i} = \tilde z_{f_i}\inv z_{f_i}\inv M_{f_i}
\qquad {m^o_H}^2 = M_H^2 z_H^{-2}
\end{equation}  
the bilinear part of the action is transformed into the standard form
expressed in terms of bare quantities, which depend by definition
on the arbitrary field redefinitions $z^V_{ab}, z^S_a, z_{f_i} $
and $\tilde z_{f_i}$. 
\begin{eqnarray}
\label{gagenbilo}
\Ga^{gen}_{bil} & = &
\int \biggl(-\hfrac 14 
\bigr(\partial^\mu V^\nu_a - \partial^\nu V^\mu_a \bigl)  Z^V_{ab}
\bigr(\partial^\mu V_{\nu b} - \partial^\nu V_{\mu b} \bigl)  
+ \hfrac 12 V^\mu_a  {\cal M}^V_{ab} V_{\mu b} \nonumber \\
& & \phantom{\int } \:
+ \hfrac 12 \partial^{\mu} \phi_a Z_{ab}^S \partial_\mu\phi_b 
- \hfrac 12 M_H^2 H^2(x) \nonumber \\
& & \phantom{\int } \:
+ i Z_{f_i}^R\bar f_{i}^R \ds f_{i}^R +
i Z_{f_i}^L\bar f_{i}^L \ds f_{i}^L
- M_{f_i} (\bar f_{i}^R  f_{i}^L +\bar f_{i}^R  f_{i}^L) \biggr) 
\nonumber\\
& = &
\int \biggl(-\hfrac 14 
\bigr(\partial^\mu V^{o\nu}_a - \partial^\nu V^{o\mu}_a \bigl)  \tilde I_{ab}
\bigr(\partial^\mu V^o_{\nu b} - \partial^\nu V^o_{\mu b} \bigl)  
+ \hfrac 12 V^{o\mu}_a  {\cal M}^o_{ab} V^o_{\mu b} \nonumber \\
& & \phantom{\int } \:
+ \hfrac 12 \partial^{\mu} \phi^o_a \tilde I_{ab} \partial_\mu\phi^o_b 
- \hfrac 12 m_H^{o2} H^{o2}(x) \nonumber \\
& & \phantom{\int } \:
+ i \bar f^{oR}_{i} \ds f_{i}^{oR} +
i \bar f_{i}^{oL} \ds f_{i}^{oL}
- m^o_{f_i} (\bar f_{i}^{oR}  f_{i}^{oL} +\bar f_{i}^{oR}  f_{i}^{oL}) \biggr)
\end{eqnarray}
 These redefinitions are carried out throughout in $\Ga_{cl}^{gen}$
by redefining also all the arbitrary couplings appearing therein
as we did it for the masses. 
At the same time we have to transform the original fields into
bare fields in the
ST identity and the Ward identities of
rigid symmetry. The arbitrary field redefinitions appearing 
thereby are absorbed into a redefinition of parameters and
external fields:
\begin{equation}
\label{bareextfields}
\begin{array}{lcl}
\rho^o_{\mu\a} &= & z_W\inv \rho_{\mu\a} \\
Y^o_a & = & {z_a^S} \inv Y_a \end{array} \qquad
\begin{array}{lcl}
{\psi ^{oR}_{f_i}} & = & z_{f_i}\inv \psi^R_{f_i} \\
{\psi ^{oL}_{f_i}}  & = & \tilde z_{f_i}\inv  \psi^L_{f_i}
\end{array}
\end{equation}
The bare parameters are defined via the representation matrices of 
rigid invariance:
\begin{eqnarray}
\label{barepara}
(r^{oV})\inv
 \hat \ve_{\a} (r^{oV})
\is         {z^V}( r ^{V}) \inv \hat \ve_{\a}  r^{V}(z^V) \inv \nonumber \\
(r^{oS})\inv \hat t_\a (r^{oS})
\is  {z^S}( r ^{S}) \inv \hat t_{\a}  r^{S}(z^S) \inv \nonumber \\
(r^{o \delta_i})\inv   \tau_\a (r^{o\delta_i})
\is  {z^\delta_i}( r ^{\delta_i}) \inv  
\tau_{\a}  r^{\delta_i}(z^\delta_i) \inv
\end{eqnarray}

We are now ready
 to apply the ST operator on $\Ga^{gen}_{cl} $. The computation
is quite lengthy, therefore we only quote the final result and
the crucial equations. Most important for the solution is
the external field part of the general action:
\begin{eqnarray}
\label{extfgen}
\Ga^{gen}_{ext.f.} =  \int &\hspace{-3mm} \biggl(&\hspace{-3mm}
 -\frac 12 \sigma _\a f_{\a,bc} c_b c_c \nonumber \\
&\hspace{-3mm} &\hspace{-3mm}
                     + \rho^o_{\mu\a} (a'^g_{\a b} \partial^\mu c_b +
                                    \hat a' _{a, bc} V^{o\mu} _b c_c \bigr)
                      + Y^o_a ( t'_{ab,c}
 \phi^o_b c_c + v'_{ab} c_b ) \nonumber\\
&\hspace{-3mm} &\hspace{-3mm}
                    + \sum_{i=1}^{N_F}
 \bigl( \bar \psi^{oR}_{f'_i} f^{oL}_i  h'^i_{f'f,a} c_a  -
 \bar f^{oL}_i \psi^{oR}_{f'_i}   h'^i_{ff',a} c_a  \nonumber \\
&\hspace{-3mm} &\hspace{-3mm} 
                \phantom{+ \sum_{i=1}^{N_F}}
                  + \bar \psi^{oL}_{f'_i} f^{oR}_i \tilde h'^i_{f'f,a} c_a  -
 \bar f^{oR}_i \psi^{oL}_{f'_i}   h'^i_{ff',a} c_a  \bigr) \biggr)
\end{eqnarray}
For simplicity we have suppressed
 the interaction polynomials of external scalars
$\hat \phi_a$. These polynomials are considered in the context of
the gauge fixing and ghost sector.

The arbitrary coupling matrices are restricted by the global
symmetries (\ref{consgagen}), complex conjugation (\ref{comconj})
and CP-invariance.  
We have already carried out the transformation into bare fields
for vectors, scalars and fermions and the respective
external fields and have transformed
at the same time the original couplings into primed couplings 
(see (\ref{barecoup})).

Via the ST identity the couplings of the vector, scalar and fermion
part of the general action
are determined as functions of
the coupling matrices appearing in the external field part 
and of the parameters of the ST identity.  
Explicitly they depend on the following combinations:
\begin{eqnarray}
\label{barecoup}
\hat a^o_{a, bc} & =  & (r^{oV})\inv_{a\a} \hat a'_{\a bc'} \tilde z^{g -1}_{cc'}
                 = z_W  (r^{oV})\inv_{a\a} \hat a_{\a bc'} 
                     z^{V-1}_{b'b}  \tilde z^{g -1}_{cc'} \nonumber \\ 
t\bare _{ab,c} &  =  & t'_{ab,c'} \tilde z^{g -1}_{cc'} 
             = z^S_{aa'} 
 t_{a'b',c'} z^{S-1} _{b'b} \tilde z^{g -1}_{cc'}  \nonumber \\
v\bare_{ac} & = & v'_{ac'} \tilde z^{g -1}_{c'c} 
           = z^S _{aa'}v_{a'c'} \tilde z^{g -1}_{c'c} \nonumber \\
h^{oi} _{ff'c} & = & h'^i_{ff'c'} \tilde z^{g -1}_{c'c} 
               =  z_f z_{f'}\inv h^i_{ff'c'} \tilde z^{g -1}_{c'c}  \nonumber \\
\tilde h^{o i} _{ff'c} & = & \tilde h'^i_{ff'c'} \tilde z^{g -1}_{c'c}
             = \tilde z_f \tilde z_{f'}\inv \tilde
                                  h^i_{ff'c'} \tilde z^{g -1}_{c'c} 
\end{eqnarray}
 $\tilde z^g_{ab}$ denotes a ghost transformation matrix which arises from
the linear part of vector transformations and the matrix $r^{oV}$:
Vector transformations consist of the linear part
appearing in the ST identity $r^g_{4b}$ and
the linear part of the nonlinear
vector transformations, $a'^g_{\a b} = z_W a^g_{\a b}, \a = +,-,3$:
\begin{equation} 
\label{ghostred}
\tilde z^g_{ab} = 
\sum_{{\a, \a'=} \atop {+,-,3}}
(r^{oV})\inv_{a\a} \tilde I_{\a'\a}a'_{\a b} + (r^{oV})\inv_{a4} r^g_{4b} 
\end{equation}

Evaluating the ST identity one finds, that
the couplings defined in (\ref{barecoup}) have to satisfy the following
equations:
\begin{itemize}
\item 
On the part containing the 4-dimensional vector polynomials
$\hat a \bare _{abc} $ is determined to be completely antisymmetric
and is seen to be the solution of the Jacobi identity:
\begin{equation}
\label{Jacobi}
\hat a \bare _{abc} = - \hat a \bare _{bac} =
 \hat a \bare _{bca} 
\end{equation}
\[  \hat a \bare _{abc} \tilde  I_{aa'}  \hat a \bare _{b'a'c'} 
    + \hat a \bare _{acc'} \tilde  I_{aa'}  \hat a \bare _{b'a'b} 
    + \hat a \bare _{ac'b} \tilde  I_{aa'}  \hat a \bare _{b'a'c} = 0
\]

\item 
On the scalar-vector part, which contains the bare mass matrix as
 defined in (\ref{baremassmatrix}), the matrix $v^o_{ab}$ is determined in
terms of the bare masses:
\begin{equation}
\label{vmassrel}
{\cal M}^o_{ab} = \tilde I _{a'b'}v^o _{a'a} v^o_{b'b} 
\Longrightarrow \left\{ \begin{array}{lcl} |v^o_{+-}|^2 = {M_W\bare}^2 \\
                                           (v^o_{\chi Z})^2 = {M_Z\bare}^2 \\
                                           (v^o_{\chi A})^2 = 0 \end{array} 
                                       \right. 
\end{equation}
Therefrom it follows that the mass of the photon has  to vanish and is
not an independent parameter of the theory:
\begin{equation}
{M_A\bare}^2 = 0
\end{equation}
The matrices $t\bare_{ab,c}$ have to be antisymmetric in the scalar
indices
\begin{equation}
\label{tas}
t\bare_{ab,c} = - t\bare_{ba,c}
\end{equation}
and satisfy the following representation equations:
\begin{equation}
\label{funrep2}
t\bare_{ba,b'} \tilde I_{aa'} t\bare_{a'c,c'}
-t\bare_{ba,c'} \tilde I_{aa'} 
t\bare_{a'c,b'} = - \hat a\bare _{b'c' a'} \tilde I_{a'a}
                                        t\bare_{bc,a}
\end{equation}
and
\begin{equation}
\label{shiftrep2}
t\bare_{ba,b'} \tilde I_{aa'} v\bare_{a'c'}
-t\bare_{ba,c'} \tilde I_{aa'} v\bare_{a'b'} = - \hat a\bare
 _{b'c' a'} \tilde I_{a'a}
                                        v\bare_{ba}
\end{equation}

\item 
In the fermion part of the action we find on the bare mass terms
\begin{equation}
\label{sdrel}
\tilde h^{oi}_{ffA} = h^{oi}_{ffA}
\end{equation}
for all massive fermions $f = e,u,d$.
Because the kinetic terms  of bare fields
are normalized, one has furthermore
\begin{equation}
\label{has}
h^{oi}_{ff'+} = - h^{oi}_{f'f-} \qquad \begin{array}{lcll} f& = & \nu,\, & u \\
                                                          f'& = & e,\, & d  
                                       \end{array} 
\end{equation}
 In addition one gets the
following representation equations for each family:
\begin{eqnarray}
\label{fermrep2}
h^{oi}_{ff',b} h^{oi}_{f'f'',c}
-h^{oi}_{ff',c}  h^{oi}_{f'f'',b} 
& = & - \hat a^o _{bc a'}  \tilde I_{aa'}  h^{oi}_{ff'',a} \\
\tilde h^{oi}_{ff',b} \tilde h^{oi}_{f'f'',c}
 - \tilde h^{oi}_{ff',c} \tilde  h^{oi}_{f'f'',b} 
& = & - \hat a^o _{bc a'} \tilde I_{a'a}  \tilde h^{oi}_{ff'',a} 
\nonumber
\end{eqnarray}
\end{itemize}

 It is  straightforward to solve these equations:
Due to (\ref{Jacobi})  $\hat a^o_{abc}$ are qualified 
as   structure constants
of $SU(2) \times U(1) $ and are related to the structure constants
$\hat \ve_{\a \b\ga}$ by:
\begin{equation}
\label{strucc}
\hat a^o_{abc} = g^o_2 \hat \ve_{\a\b\ga} O_{\a a} (\theta_W^o) O_{\b b} 
(\theta_W^o)
                                    O_{\ga c} (\theta_W^o)
\end{equation}
 $\theta_W^o$ and $g^o_2$ are at this stage two 
arbitrary parameters, which parameterize 
the two remaining parameters of the  coupling matrix
$\hat a^o_{abc}$. Therefore
the representation equations (\ref{funrep2}) and (\ref{shiftrep2})
are equivalent to 
 equs.~(\ref{funrep}) and (\ref{shiftrep}) 
and their solutions can be read off from the solutions (\ref{shiftsol})
and
(\ref{bsol}):
\begin{equation}
t\bare_{bc,a} = g_2\bare\hat t_{bc,\a} O_{\a a} (\theta_W\bare)
\qquad v\bare _{ba} = v^o_{b\a} O_{\a a} (\theta_W\bare)
\end{equation} 
Here we have already used that
antisymmetry (\ref{tas}) singles out from the equivalence class
the antisymmetric solution. Most important are the solutions of the
shift equations, which relate  the remaining undetermined
parameters to the masses of Z-boson and W-boson.
Inserting the relations (\ref{vmassrel}) into
(\ref{shiftsol}) determines $ G^\phi, \theta^o_W
$ and $v^o$:
\begin{equation}
\label{barerel}
\cos \theta_W^o = {M_W^o \over M_Z^o} \quad
   G^\phi = - {\sin \theta_W^o \over \cos \theta_W\bare} \quad
v^o = {2 \over g\bare _2} M^o_Z cos\theta_W\bare 
\end{equation}
(The signs of $\cos \theta_W^o$ and $v^o$ are
 chosen in accordance with the tree approximation.)
Only the nonabelian coupling $g^o_2$ remains undetermined.

In the same way
the representation equations of the fermion matrices (\ref{fermrep2}) are
equivalent to equs.~(\ref{fermrep}) we have solved in equs.~(\ref{doubsol}),
(\ref{doubsolab}) and (\ref{singsol}).
The relation (\ref{has})
singles out  the
 antisymmetric solution:
\begin{eqnarray}
{h^{o}_+}^{\delta_i} &= &ig\bare _2  \frac {\tau_+}2  \qquad
{h^{o}_-}^{\delta_i}\hspace{3mm} = \hspace{3mm} ig \bare_2 \frac {\tau _-}2 \\
{h^{o}_Z}^{\delta_i}  & = & ig_2 \bare (\cos \theta_W\bare \frac {\tau_3}2
                + \sin \theta_W\bare G^{\delta_i}\frac {\mathbf 1}2 )
\nonumber \\
{h^{o}_A}^{\delta_i} & = & ig_2 \bare (- \sin \theta_W\bare \frac {\tau_3}2
                + \cos \theta_W\bare G^{\delta_i}\frac {\mathbf 1}2 )
\nonumber
\end{eqnarray}
is the solution of the doublet representation and
\begin{eqnarray}
\tilde h^{oi}_Z & =  & ig_2 \bare \sin \theta_W\bare 
\left(\begin{array}{ccc} i G^{e_i} & 0 & 0   \\
0 & i G^{u_i} & 0   \\ 0 & 0 &  i G^{d_i}    
\end{array} \right) \nonumber \\
\tilde h^{oi}_A &  = & i g_2 \bare \cos \theta_W\bare 
\left(\begin{array}{ccc} i G^{e_i} & 0 & 0   \\
0 & i G^{u_i} & 0   \\ 0 & 0 &  i G^{d_i}    
\end{array} \right)
\end{eqnarray}
is the solution of the singlet representation.
There we have introduced a matrix notation as in section~3.4 ($\delta_i
= l_i , q_i $). Although it is not relevant for perturbation theory,
we want to mention, that on the mass terms  doublet and singlet
representations are distinguished.
 The ST identity is only solved, if we assign to one
chirality the doublet representation and to the second the singlet 
representation. Inserting furthermore the relation (\ref{sdrel}),
which relates the algebraically undetermined parameters of the
abelian subgroup, yields immediately the relations (\ref{chargerel}).
With (\ref{barerel}) they read 
\begin{equation}
\frac 12 (- \tan \Theta_W^o + G^{q_i}) = G^{u_i} \qquad
\frac 12 ( \tan \Theta_W^o + G^{q_i}) = G^{d_i}
\qquad \frac 12 ( \tan \Theta_W^o + G^{l_i}) = G^{e_i}
\end{equation}
As expected we remain with one undetermined parameter for each
lepton and quark family. Parameterizing $G^{\delta_i}$
by the electric charge, $Q_e = -1, Q_d = -\frac 1 3$,
 and an remainder $g_{\delta_i},$ 
\begin{equation}
\label{lqfcoupl}
G^{q_i} = - \tan \theta_W^o(2 Q_d +1 + 2 g_{q_i}) \qquad
G^{l_i} = - \tan \theta_W^o(2 Q_e +1 + 2 g_{l_i}) 
\end{equation}
it is seen
that the free parameters correspond to coupling the Noether
currents of
lepton and quark family conservation to the photon.

Finally the angle $\Theta^{o}$, which appears
as a free parameter in the ST identity is determined 
as function of the bare vector mass ratio
by inverting the relation 
between $\hat a^o_{abc}$ and $\hat a'_{\a bc}, \a = +,-,3$:
\begin{eqnarray}
\hat a^o_{Z+-} &=& -i g_2\bare \cos \theta_W^o = 
\frac 1{r^{o}_Z} cos \Theta^{o} \hat a'_{3+-} \frac 1 {\tilde z^g_{++}}\\
\hat a_{A+-} & = & i g_2\bare \sin \theta_W^o =
- \frac 1{r^{o}_A} \sin \Theta^{o} \hat a'_{3+-} \frac 1 {\tilde z^g_{++}}
\end{eqnarray}
i.e.
\begin{equation}
\tan \Theta^{o}  = \frac {r^{o}_A}{r^{o}_Z} \tan \theta^o_W 
\end{equation}
On the ST identity it is not possible to fix   $r^{oV}_A$ and $r^{oV}_Z$,
but they are determined by using rigid symmetry.

Having solved the above equations  the action of vectors, scalars
and fermions and the external field part
is determined from the ST identity 
\begin{equation}
{\cal S} (\Ga^{gen}_{GSW} + \Ga^{gen}_{ext.f} ) = 0
\end{equation}
without having
specified the ghost redefinition matrices and without using
rigid or local gauge symmetry. Ingredients are only
nilpotency of the ST identity, the
bare form of the bilinear action, which states 
especially that there  are
massive vector bosons, and the global symmetries as charge neutrality
and lepton and quark family number conservation.
Explicitly we find as solution of the ST identity in the
vector, fermion and scalar sector
the following general action expressed in terms of bare fields:
\begin{equation}
\label{gagengsw}
\Ga^{gen}_{GSW} (V^o_a, \phi^o_a, f_i^o) = 
\Ga_{YM}(V^o_a) + \Ga_{scalar}(\phi_a^o, V_a^o) + 
\Ga_{Yuk}(\phi_a^o, f_i^o) + \Ga_{matter} (V_a^o, f_i^o)
\end{equation}
with 
\begin{eqnarray}
\label{gagenym}
\Gamma_{YM} & = &
  -\frac 1 4 \int  G_\a^{o\mu\nu}\tilde{I}_{\a \a'}G^o_{\mu\nu \a'} \\
\Gamma _{scalar}    &=&\int \Bigl( (D^\mu\Phi^o 
 )^\dagger D_{\mu}\Phi^o -g_2 ^o\frac 1 4\frac{m_H^{o2}} {M_W^{o2}}
({\Phi^o}^\dagger\Phi^o +  4 {M_W^o \over g_2^o}  H^o )^2 \Bigr) \nonumber\\
\Gamma_{matter}& = &  \sum_{i=1}^{N_F} \int \Bigl( \overline
 {F^{oL}_{l_i}} i \Ds F^{oL}_{l_i}
       + \overline {F^{oL}_{q_i}} i \Ds F^{oL}_{q_i} + \overline
 {f^{oR}_i} i \Ds f^{oR}_{i} \Bigr) \nonumber \\
\Gamma_{Yuk}  & = &- \sum_i^{N_F} \int \Bigl( m_{f_i}^o \bar f^o_i f^o_i +
\frac {g_2^o}{\sqrt 2 M_W^o}\bigl( 
   m^o_{e_i}  
\overline
 {F^{oL}_{l_i}} \Phi^o  e^{oR}_i \nonumber \\
        & & \phantom{\sum \int \frac{sqrt 2} v}
         +  m^o_{u_i}  \overline
 {F^{oL}_{q_i}} \Phi^o u^{oR}_i 
                 + m^o_{d_i}  \overline {F^{oL}_{q_i
}} \tilde \Phi^o  d^{oR}_i + \hbox{h.c.} \bigr) \Bigr) \nonumber
\end{eqnarray}
For notational convenience we have rewritten the 4-vector of scalars
$\phi^o_a$
into the complex scalar doublet $\Phi^o$ and $\tilde \Phi^o$.
The structure of the individual ST-invariant terms is the same as
in the tree approximation.
Therefore it is seen that the information on the 
$SU(2) \times U(1)$ algebra is completely transfered to
the ST identity. Because the bare form of the action has been fixed,
the covariant derivatives are immediately computed  as functions
of $M_W^o$ and $M_Z^o$. The weak mixing angle in its bare form
 $ \theta _W^o $ is defined by the 
bare vector mass ratio 
\begin{equation}
\label{weinbare}
\cos{\theta_W^o } = {M_W^o \over M_Z^o}
\end{equation}
and is not an independent parameter of the theory 
(cf.~(\ref{barerel})). From the above construction it is obvious,
that the broken theory is considered and characterized in its own 
right and one never refers to the underlying symmetric theory.
\begin{eqnarray}
\label{gencov}
 G^{o\mu\nu}_\a & = & O_{\a a} (\theta_W^o) \bigl(
\partial ^\mu V_a^{o\nu} - \partial^{\nu} V_a^{o\mu} \bigr)
  +  g^o_2 \tilde{I}_{\a\a'} 
  {\hat \varepsilon}_{\a\b \gamma} 
 O_{\b b} (\theta_W^o)   O_{\ga c} (\theta_W^o) 
V^{o\mu}_b V^{o\nu}_c\\
  D_{\mu}\Phi^o & = & \partial_{\mu}\Phi^o - i   g^o_2 \Bigl(
 \frac{{\tau}_\a}{2} O_{\a a} (\theta_W^o)
  - \tan \theta_W^o O_{4a}(\theta_W^o) \Bigr)
 V^o_{\mu a}  ( \Phi^o + \frac {\sqrt 2} {g_2^o}{0\choose  M_W^o} ) 
\nonumber \\
D^{\mu} F^{oL}_{\delta_i} &=& \Bigl( \partial^\mu  - i 
g^o_2 \bigl(
 \frac{{\tau}_\a}{2}
O_{\a a} (\theta_W^o)
 V^{o\mu }_a +   \frac {G^{\delta_i}} 2 O_{4 a} (\theta_W^o)
  V^{o\mu}_ a \bigr) \Bigr)
F^{oL}_{\delta_i}\qquad \delta= l,q \nonumber \\
  D^{\mu} f_i^{oR} &=& \bigl( \partial^\mu   + 
 i g_2^o \frac 12 (\tan \theta_W^o + G^{\delta_i} ) \bigr)  f_i^R O_{4a}
(\theta_W^o) V^{o\mu}_a  \qquad f_i= e_i, d_i \nonumber \\
  D^{\mu} f_i^{oR} &=& \bigl( \partial^\mu   +
 i g_2^o\frac 12 (-\tan \theta_W^o + G^{q_i} ) \bigr) f_i^{oR} O_{4a}
(\theta_W^o) V^{o\mu}_a  \qquad f_i= u_i \nonumber 
\end{eqnarray}
The external field part depends on the ghost redefinition matrices
$a'^g_{\a b}$ and $r^g_{4 b}$.  In accordance with
rigid symmetry (\ref{rghost}) we introduce the following notation 
\begin{equation}
\label{zgdef}
 a'^g_{\a b} = \hat z^g_W\tilde I_{\a\b} r^g_{\b b} \\
\end{equation}
These parameters will be finally fixed in the ghost sector
on the bilinear parts of the ghosts:
\begin{eqnarray}
\label{gagenextf}
\Ga^{gen}_{ext.f.}  =   \int &\hspace{-3mm} \biggl(&\hspace{-3mm}
 -\hat z^g_{W}\frac {g_2^o} {  2} \sigma _\a \hat \ve_{\a \b \ga } 
{r}^g_{\b b} c_b {r}^g_{\ga c} c_c \nonumber \\
&\hspace{-3mm} &\hspace{-3mm}
                     + \rho^o_{\mu\a} \hat z_W^ g( \tilde I_{\a\b}
\partial^\mu {r}^g_{\b b}
  c_b + g_2 ^o \hat \ve _{\a \b \ga} 
O_{\b b} (\theta _W^o) 
V^{o\mu} _b {r}^g_{\ga c} c_c \bigr) \nonumber \\
&\hspace{-3mm} &\hspace{-3mm}
            + g_2^o \Bigl( Y^{o\dagger }  
\bigl( i 
 \frac{{\tau}_\a}{2} \hat z^g_{W}{r}^g_{\a a}
  - i \frac {\mathbf 1}2 \tan \theta_W^o {r}^g_{4a}) \bigr) c_a
   ( \Phi^o + \frac {\sqrt 2}{g_2^o}{0\choose  M_W^o} ) + \hbox{h.c.} \Bigr)
  \\
&\hspace{-3mm} &\hspace{-3mm}
                    + \sum_{i=1}^{N_F}\Bigl( \sum_{\delta = l,q}
  \bar \Psi^{oR}_{\delta_i} i 
g^o_2 \bigl(
 \frac{{\tau}_\a}{2} \hat z^g_{W}
 {r}^g_{\a a} 
  +    {G^{\delta_i}}\frac {\mathbf 1} 2 {r}^g_{4 a} \bigr) c_a
   F^{oL}_{\delta_i} \nonumber \\
&\hspace{-3mm} &\hspace{-3mm} 
                \phantom{+ \sum_{i=1}^{N_F}}  + 
                   \bar \psi^{oL}_{e_i} 
i g_2^o \frac 12 (\tan \theta_W^o + G^{l_i} )   e_i^{oR} 
 {r}^g_{4a} c_a  \nonumber \\
&\hspace{-3mm} &\hspace{-3mm} 
                \phantom{+ \sum_{i=1}^{N_F}}  + 
                   \bar \psi^{oL}_{d_i} 
i g_2^o \frac 12 (\tan \theta_W^o + G^{q_i} )   d_i^{oR} 
 {r}^g_{4a} c_a  \nonumber \\
&\hspace{-3mm} &\hspace{-3mm} 
                \phantom{+ \sum_{i=1}^{N_F}}
                  + \bar \psi^{oL}_{u_i} 
 i g_2^o\frac 12 (-\tan \theta_W^o + G^{q_i} )  u_i^{oR} 
 {r}^g_{4a} c_{a} + \hbox{h.c.}\Bigr)
 \biggr) 
  \nonumber 
\end{eqnarray}
The general form of the ST-invariant action is obtained by transforming
the bare field into the original fields
 according to
(\ref{barefields}). The  parameters 
 of the bilinear action remain arbitrary as we did not to
have to dispose on them when solving the ST identity.  Only
 the bare mass of the photon is determined as zero
from the ST identity and is not an independent parameter of the theory.

Besides the ghost redefinition matrix $r^g_{\a b}$ (\ref{ghostred})
there remain  undetermined 
  the nonabelian coupling $g_2^o$  and the 
fermion couplings $G^{l_i} $ and $G^{q_i}$. 
In order to embed the structure of quantum electrodynamics into
the standard model they have to be fixed on the 
local abelian Ward identity
as given in (\ref{wqdef}) and (\ref{wqloc}),
remaining with one free parameter $g^o_2$, which can be finally
adjusted to the fine structure constant in the Thompson limit.
For this reason the Ward identities of rigid invariance have to
be established.

The solution of the ST identity $\Ga^{gen}_{GSW}$ 
in the vector, scalar and fermion sector
as given above  is immediately seen to
be  invariant under
rigid symmetry. Applying the Ward operators 
${\cal W }_\a$ (\ref{wardopequivclass}) on $\Ga_{GSW}^{gen}$
determines uniquely the 
matrices of equivalence transformations:
\begin{equation}
r^{oV} = O(\theta^o_W) \qquad r^{oS} = {\mathbf 1} \qquad
r^{o\delta_i} = {\mathbf  1}
\end{equation}
For the parameter $v^o$, which appears as a free parameter in
the Ward operators, one gets
\begin{equation}
v^o = {2 M_W^o \over g_2^o} 
\end{equation}
Inverting the relations (\ref{barepara})
finally yields $r^S, r^V, r^{\delta_i}$
as functions of masses and field redefinitions:
\begin{eqnarray}
\label{rdetv}
r_Z & = & { z_Z \over z_W} \cos(\theta_W^o + \theta _Z)
  \sqrt {1 + \tan (\theta_W^o + \theta _Z)
\tan (\theta_W^o + \theta _A) } \nonumber \\
r_A & = &  {z_A \over z_W} \sin(\theta_W^o + \theta _A) 
 \sqrt {1 + \cot (\theta_W^o + \theta_Z)
\cot (\theta_W^o +\theta _A )} \nonumber \\ 
\tan \Theta & = & \sqrt{ \tan(\theta _W^o + \theta_Z) \tan (\theta_W ^o
+\theta_A) }
\end{eqnarray}
and
\begin{equation}
\label{rdets}
r_a^S = {1 \over z_H} z_a^S  \qquad r_{l_i} = {z^{e_i}\over z_{\nu_i}} 
\qquad r_{q_i} = {z^{d_i}\over z_{u_i}} 
\end{equation}
The shift parameter $v$ as defined in (\ref{shiftsol}) is 
determined on the general classical action to
\begin{equation}
v= {2 M_W^o \over g_2 ^o z_H }
\end{equation}
The Ward identities of rigid symmetry hold then without further
restrictions on $\Ga^{gen}_{GSW}$ and $\Ga ^{gen}_{ext.f}$.

One is now ready to apply the local Ward operator $
{\mathbf w}^Q$ (\ref{wqloc}).
 Requiring the Ward identity (\ref{wardidho})
to be valid on the $\Ga^{gen}_{GSW} + \Ga^{gen}_{ext.f}$ 
\begin{equation}
\label{wqlocgen}
\biggl( g_1 {\mathbf w}_4^Q 
- \frac 1{r^V_Z} \partial{\delta \over \delta Z} \sin \Theta ^V 
- \frac 1{r^V_A} \partial{\delta \over \delta A} \cos \Theta ^V 
\biggr)(\Ga^{gen}_{GSW} + \Ga^{gen}_{ext.f}) = 
0
\end{equation}
determines $G^{l_i}$ and $G^{q_i}$ as functions of the electric
charge of the charged leptons ($Q_e = -1$) and down-type quarks
($Q_d = -\frac 1 3$):
\begin{equation}
\label{lqfcouplch}
G^{q_i} = - \tan \theta_W^o(2 Q_d +1 ) \qquad
G^{l_i} = - \tan \theta_W^o(2 Q_e +1 ) 
\end{equation}
The overall normalization constant $g_1$ is determined on the
general classical action as function of the nonabelian coupling $g^o_2$,
the wave function renormalization and the bare masses:
\begin{equation}
\label{ovnorm}
g_1 = z_W g^o_2 \tan \theta _W^o \sqrt {\tan (\theta_W^o + \theta_A)
\over \tan (\theta_W^o + \theta_Z)}
\end{equation}
After having applied the local Ward identity there remains only one
coupling
 $g^o_2$, which is  not fixed on the mass terms and by symmetries.
In a QED-like parametrization
this coupling 
 is determined by the fine structure constant,
which measures the interaction strength of the photon to
 the electromagnetic current in the Thompson limit:
\begin{equation}
\label{coupl}
\bar u (p)  \Gamma _{ee{A_\mu}}(p,p,0) u(p) \big| _{p^2 =  m_e^2} 
= i e \bar u(p) \gamma_\mu u(p)
\end{equation}
In the tree approximation this relation yields
\begin{equation}
g^o_2 = {e \over \sin \theta_W} + O(\hbar) 
\end{equation}
Respectively one can parameterize the bare coupling $g_2^o$ by
the electromagnetic bare coupling
 $e^o$ and the bare mass ratio of $W $- and $Z$-boson
\begin{equation}
g_2^o =
{e^o \over \sin \theta^o_W}  \qquad \hbox{with} 
\qquad \hbox{with} 
\quad e^o = e + \delta e
\end{equation}

Transforming the bare fields of the general action back to original
fields and expanding the free parameters
in perturbation theory determines the symmetric local contributions
$\Ga^{(n)}_{inv}$ (\ref{defren}), which are in agreement with the ST identity,
rigid symmetry and the local Ward identity, as specified by algebra 
and nilpotency.
Since some combinations
of parameters (\ref{rdetv}) explicitly enter as parameters
the symmetry operators,  also the explicit form of the ST identity
and the Ward identities is modified in higher orders of perturbation
theory.

In concrete calculation it is wideley used that dimensional regularization
is an invariant scheme if parity is conserved. Under such
circumstances
only the symmetric counterterms appearing in
the general ST-invariant action  would appear as counterterms in the
 $\Ga_{eff}$, which governs the calculation of Green functions
in the Gell-Mann Low formula. 
The Feynman rules of the vector, scalar and fermion part
 derived from such a symmetric $\Ga_{eff}$ are
listed for example in \cite{Denner} and have  exploited in
constructing ST-invariant 1-loop Green functions in the physical
sector. Furthermore, if one had an invariant scheme, 
the parameters of the ST identity
and the Ward identity would be related to the respective counterterms
 in renormalized perturbation theory as derived in (\ref{rdetv}),
(\ref{rdets}) and
(\ref{ovnorm}). 
Of course a necessary prerequisite of exploiting such relations 
is the complete construction of the ST identity, the rigid symmetries
and the local abelian gauge Ward identity order by order in perturbation
theory for
all Green functions involved, especially also for the ones
of the ghost sector. In the next sections
we outline the construction of
  the gauge fixing and ghost sector
in the classical approximation and in higher orders, taking care in
 preserving the Ward identities of
rigid and local gauge invariance.

\newsubsection{The gauge fixing and ghost sector}

Classically the ghost sector has been completely determined
as BRS-variation of the gauge fixing function (cf.~(\ref{brsgf})). The
respective relation is  immediately derived for the generating functional
of 1PI Green functions by differentiating  the ST identity (\ref{STgen2})
with respect to the $B$-fields:
\begin{equation}
\label{ghosteq}
\brs _{\Ga} \left( {\delta \Ga \over \delta B_a} \right)= -
(r^V )^{-1}_{a\a} \delta \hat g _{\a b} {\delta \Ga \over \delta \bar c_b}
\end{equation}
In the linear gauges, moreover,  $\delta \Ga \over \delta B_a $ is  local
(\ref{gfixho}),
because there are no vertices which could constitute loop
diagrams with external $B$-legs.
Then  eq.~(\ref{ghosteq}) yields the linear
ghost equations, which have to be established to all orders of 
perturbation theory.

In the tree approximation we have constructed the gauge fixing
sector to be invariant under Ward identities of rigid symmetry
by introducing the external scalar fields $\hat \phi_a$
(see (\ref{gfrig}) -- (\ref{wardrig})). As the abelian fermion couplings
are not well determined from the ST identity,  Ward identities 
of rigid symmetries have to
be maintained for the generating functional of Green functions. 
For this reason we have to choose the gauge fixing sector as
being invariant under rigid transformations  as specified in the 
vector and scalar part of the action. To the external scalars
we assign the transformation behaviour of the propagating scalars 
and derive for the most general rigid invariant gauge-fixing
sector the following expression:
\begin{eqnarray}
\label{gab}
\Ga^B & = & \int \Bigl( \frac 1 2 \xi
B_a (r^V)^{-1}_{a\a} \tilde I _{\a \b} (r^V)^{-1 T} _{\beta b} B_b
+ 
\frac 1 2\hat \xi B_a (r^V)^{-1}_{a4}  (r^V)^{-1 T} _{4 b} B_b 
+ B_a\tilde I_{ab} \partial^\mu V_{\mu b} \\
& & \phantom{\int}
+ g_\xi \Bigl( \sum_{\a = \atop +,-,3}
 B_a  (r^V)^{-1}_{a\a} (r_b ^S \phi_b + \delta_{Hb} v)
\hat t_{ bc,\a}
(r_c^S \hat \phi _c + \delta _{Hc} \hat \zeta v )  \nonumber \\
& & \phantom{\int + g_\xi}
+ \hat G B_a  (r^V)^{-1}_{a4} (r_b ^S \phi_b + \delta_{Hb} v)
\hat t_{ bc,4} (r_c^S \hat \phi _c + \delta _{Hc} \hat \zeta v )  \bigr) \Bigr)
\nonumber
\end{eqnarray}
The matrix $r^V_{\a a}$ 
is the three parameter matrix, which parameterizes
the rigid transformations of vectors (\ref{defrv})
and $r^S_b$ are the three parameters
of the scalar rigid transformations (\ref{defrs}), $v$ denotes the
shift parameter of the scalar field as it appears as  parameter
in the Ward operators. 
 These parameters are determined by the 
normalization conditions on the vector and scalar 2-point functions
and one cannot dispose of them in the gauge fixing sector anymore.
$\Ga^B$ as given in (\ref{gab})  
holds for the  B-dependent
part of the generating functional of 1PI Green 
functions in linear gauges to all orders. The free parameters of the
gauge fixing sector are 
\begin{equation}
\label{gfixpar}
\xi ,\, \hat \xi,\, \hat \zeta , \, \hat G, \, g_\xi 
\end{equation}
Finally a further free parameter is the overall normalization of
the external scalar field $\hat \phi_a$, which can be used to fix 
the parameter $g_\xi$ at will. In QED-like parameterizations
it is convenient to adjust 
\begin{equation}
\label{gxi}
g_\xi = {e \over \sin \theta_W }
\end{equation}
 (\ref{gab}) yields  the linear ghost equations,
which are valid in this form to all orders
(see \cite{PISO95}
for details) ($\a = +,-,3$): 
\begin{eqnarray}
\label{geq}
\partial ^\mu {\delta \Gamma \over \delta \rho^\mu_\a}
+ g_\xi  {\delta \Ga \over \delta Y_b'} r^S_{b'} \tilde I_{b'b}
\hat t_{bc,\a} ( r^S_c \hat \phi_c + 
\delta_{Hc} \hat \zeta v )   &        &\nonumber \\
+ (r^ S_b\phi_b + \delta_{Hb} v)     \hat t_{bc, \a} r^S_c \hat q_c
 & = & - {\delta \hat g _{\a b}}{ \delta \over \delta \bar c_b}
\nonumber \\
 \Box (r^g_{4Z} c_Z + r^g_{4A} c_A )
+ g_\xi \hat G {\delta \Ga \over
\delta Y_b'}r^S_{b'} \tilde I_{b'b }\hat t_{bc,4}( r^S_c \hat \phi_c + 
\delta_{Hc} \hat \zeta v )  & & \nonumber \\
+ (r^S_b \phi_b +
\delta_{Hb} v )  \hat t_{bc, 4} r^S_c
\hat q_c & = & -\delta \hat g _{4 b}{ \delta \over \delta \bar c_b}
\end{eqnarray}
Using rigid symmetery the ghost  equations
are immediately integrated  yielding that the
generating functional of 1PI Green functions depends on specific 
combinations between the external fields $\rho^\mu_\a $ and $Y_a$ and
antighosts, whereas the remaining contributions are local. 
 Splitting off also the local $B$-dependent part 
 the  generating functional of 1PI Green functions can be decomposed 
in the following way:
\begin{eqnarray}
\label{geqint}
&\hspace{-3mm}  & \hspace{-3mm}
 \Ga (V^\mu_a, B_a, c_a,\phi_a, \hat \phi_a, \sigma_\a ,
         \rho^\mu_\a , Y_a , \bar c_a , f_i, \psi_{f_i} ) \nonumber \\
&\hspace{-3mm}  & \hspace{-3mm}
=  
\Ga^{nl} (V^\mu_a , c_a , \phi_a , \hat \phi_a, \sigma_\a ,
        \rho'^\mu_\a , Y'_a , f_i , \psi_{f_i} ) +
\Ga^B (B_a , V^\mu_a, \phi_a , \hat \phi_a) \nonumber \\
&\hspace{-3mm}  & \hspace{-3mm}
-    \bar c _a (\delta \hat g) _{a4}\inv  r^g_{4b}\Box c_b  
      -  g_\xi\bar c_a
\Bigl(\sum_{\a=\atop +,-,3}
  (\delta \hat g)_{a \a}\inv\hat t_{bc,\a }
  +   ( \delta \hat
g)\inv_{a  4  } \hat G \hat t_{bc,4} \Bigr) 
\bigl( \phi_b + \delta _{Hb} v\bigr)                  \hat q_c
\end{eqnarray}
with 
\begin{eqnarray}
\rho'^\mu_\a &= &
\rho^\mu_\a + \partial^\mu \bar c_a (\delta \hat g)\inv_{a \a} \nonumber\\
Y'_b   & = & 
Y_b -  
\bar c_a g_\xi \tilde I_{bb'}\Bigl(\sum_{\a=\atop +,-,3}
(  \delta \hat g)_{a \a}\inv \hat t_{b'c,\a } 
  +  \hat G ( \delta \hat
g)_{a  4  }  \inv\hat t_{b'c,4} \Bigr) (\hat \phi_c + \delta_{Hc} \hat\zeta v)
\end{eqnarray}
For simplification we have absorbed  the irrelevant scalar redefinitions,
i.e.~$r^S_b = 1 $.
All non-local contributions are contained in the part $\Ga^{nl}$.
The proof that all breakings of the ST identity
can be absorbed by
adjusting local contributions, 
can be finally restricted to this functional (see section 7).
The solution of the ghost equations is quite trivial in a massless
symmetric gauge. This is not the case for the standard model where 
the ghosts are massive.
 There one not only has
to solve the ghost equation, but also one has to show, that 
 the on-shell normalization conditions for
the ghost 2-point functions  and in particular the infrared conditions
\begin{equation}
\label{infraghost}
\Gamma_{\bar c _A c _Z}(p^2=0) =
\Gamma_{\bar c _Z c _A}(p^2=0) = \Gamma_{\bar c_ Ac _A}(p^2 = 0) = 0 
\end{equation}
can be fulfilled by adjusting the parameters $\hat \zeta ,\hat G$ and
the BRS-transformation matrix $\delta \hat g_{\a b}$

\newsubsubsection{The classical approximation}

First we solve the ghost equations in the classical
approximation, taking into account that we impose 
 normalization
conditions on the ghost 2-point functions as specified in 
 (\ref{ghostmass}),
(\ref{ghostdemix}) and (\ref{ghostres}). The 
 bilinear part of
the ghost action is  therefore fixed (see (\ref{gagenbil}))
 and all parameters have to
be determined as functions of $Z^g_{ab}$ and  ${\cal M}^g_{ab}$,
and of the free parameters of the vector and scalar  part
of the action.
We proceed therefore as in the vector sector and define on the
bilinear part bare ghosts by the following field redefinitions
\begin{equation}
\label{bareghosts}
c_a^o = z^g_{ab} c_b \qquad \bar c_a^o = \bar z ^g_{ab} \bar c_b
\end{equation}
Inserting these bare fields in the bilinear part of the ghost
action and requiring the bare action to have
a standard form determines $z^g$ and $\bar z^g$ up to a diagonal
matrix:
\begin{eqnarray}
\Ga ^{gen}_{bil, ghost}
 & = & - \int \bigl(\bar c_a Z^g_{ab} \Box c_b + \bar c_a
{\cal M }_{ab}^g c_b \bigr) \nonumber \\
& = & - \int \bigl(\bar c^o _a (\bar z^g)^{-1 T}_{aa'}
Z^g_{a'b'} \Box (z^g)\inv_{b'b}c^o_b + \bar c^o_a (\bar z^g)^{-1 T}_{aa'}
{\cal M }_{a'b'}^g  (z^g)^{-1 }_{b'b}c^o_b \bigr) \nonumber \\ 
& \stackrel ! = & - \int \bigl(\bar c^o_a \tilde I_{ab} \Box c^o_b + \bar c^o_a
{\cal M }_{ab}^{og} c^o_b \bigr) 
\end{eqnarray}
 Explicitly one gets a relation between the field redefinitions
of ghosts and antighosts  on the kinetic part:
\begin{equation}
\tilde I_{ab'}  (z^g)_{b'b}  =
(\bar z^g)^{-1 T}_{aa'} 
Z^g_{a'b} 
\end{equation}
The remaining undetermined matrix can be used to fix the ghost mass 
matrix as being diagonal:
\begin{equation}
\label{bareghostmass}
{\cal M}^{og}_{ab} =
\left(\begin{array}{cccc}
       0 & \zeta_W^ o M_{W}^{o2} & 0 & 0 \\
       \zeta_W^ oM_{W}^{o2} & 0 & 0 & 0 \\
        0& 0& \zeta_Z^o M_{Z}^{o2} &  0 \\
       0 & 0 & 0 & M_{gA}^{o2} \end{array} \right)
\end{equation}
and 
\begin{equation}
\label{gmassdiag}
{\cal M}^{og}_{ab} = (\bar z^g)^{-1 T}_{aa'}
{\cal M }_{a'c}^g \tilde I_{cc'}(Z^g)\inv_{c'b'} (\bar z^g)^{T }_{b'b}
\end{equation}
We have parameterized the $W$ and $Z$-ghost mass by the masses
of the $W$- and $Z$-boson and  independent parameters $\zeta_W^o$ and
$\zeta_Z^o$.
In contrast to the vector mass matrix  the ghost mass matrix is not 
required to be symmetric 
in the neutral
components. Indeed considering the 1-loop diagrams it is seen, that
the 2-point function $\Ga_{\bar c_Z c_A}$ and $\Ga_{\bar c_A c_Z}$
get different loop corrections, because  interactions of scalars and
ghosts are unsymmetric in $Z$-ghosts and $A$-ghosts in the tree approximation
(cf.~(\ref{gaghost})). For diagonalizing an arbitrary matrix
the equivalence transformation has to be carried out by
 an  invertible matrix. Therefore 
diagonalization of the ghost mass matrix
 determines the wave functions renormalization of
anti-ghosts $\bar z^g_{ab}$ up to 
  a diagonal
matrix.  Taking for the BRS-transformation matrix $\delta \hat g_{\a b}$
the ansatz
\begin{equation}
\label{deltahatg}
\delta \hat g_{\a b} =
\left(\begin{array}{cccc}
       1 & 0 & 0 & 0 \\
       0 & 1 & 0 & 0 \\
        0& 0& \cos \ga_Z & -\sin \ga_A \\
       0 & 0 &  \sin \ga_Z & \cos \ga_A \end{array} \right)
\end{equation}
it is possible
  to diagonalize
the ghost matrix by adjusting the arbitrary angles $\ga_Z$ and $\ga_A $,
and at the same time the three undetermined parameters of the diagonal matrix
 are fixed.
We want to mention that in massless nonabelian gauge theories 
the antighost field redefinitions are not determined on the
bilinear part of the action and can therefore be   completely fixed by 
the ST identity. 

  In the classical approximation
 the external field vertices $\delta \Ga^{gen}_{cl}/ \delta \rho_\a^\mu$
and  $\delta \Ga^{gen}_{cl}/ \delta Y_a $ are local field polynomials.
They have been determined in the last section as functions of vector
field redefinitions $z^V_{ab}$, of scalar field redefinitions $z^S_a$ and
of the matrix  $r^g_{\a b}$ and $\hat z^g_W$ (\ref{zgdef}).
 One can either take the explicit form of
$\Ga^{gen}_{ext.f.}$ or better one goes back to the bare form of
the action as given in (\ref{gagenextf}).
For proceeding with the bare form, we have also to transform
the scalars and vectors in the gauge fixing part to bare fields.
The parameters, which appear by carrying out this transformation,
are absorbed into a redefinition of $B$-fields, into the overall redefinition
of the external scalar fields and into a redefinition of the arbitrary
parameters into a  bare form. One has to note that in the classical
approximation the matrices $r^V_{\a b}$ and $r^S_a $ are determined
as functions of field redefinitions and bare masses (cf.~(\ref{rdetv}) and
(\ref{rdets})), especially one has in the 
classical approximation
\begin{eqnarray}
(r^V)_{a\a}\inv O_{\a b'}(\theta_W^o)(z^V)_{b' b} & = & z_W \delta_{ab} + z_W
\Bigl(\sqrt {\hfrac{\tan (\theta_W^o + \theta_Z)}
{\tan (\theta_W^o + \theta_A)}} -1 \Bigr) \delta_{a4} \delta_{b4} \\
 (r^S)_{a\a}\inv (z^S)_{\a b} & = & z_H \delta_{ab}  
\end{eqnarray}
The bare form of the gauge fixing is then given by
\begin{eqnarray}
\label{gagfbare}
\Ga^{gen}_{g.f.} & = & \int \Bigl( \frac 1 2 \xi^o
B_a^o  \tilde I _{a b}  B^o_b
+ 
\frac 1 2\hat \xi^o (\sin\theta_W ^o B^o_Z + \cos \theta_W^o B^o_A )^2
+ B^o_a\tilde I_{ab} \partial^\mu V^o_{\mu b} \\
& & \phantom{\int}
+ g_\xi \Bigl( \sum_{\a = \atop +,-,3}
 B^o_a  O^T_{a\a}(\theta_W^o) 
( \phi^o_b + \delta_{Hb} v^o)
\hat t_{ bc,\a}
( \hat \phi _c + \delta _{Hc} \hat \zeta^o v^o )  \nonumber \\
& & \phantom{\int + g_\xi}
+ \hat G^o B^o_a  O^T_{a4}(\theta_W^o)  ( \phi^o_b + \delta_{Hb} v^o)
\hat t_{ bc,4} ( \hat \phi^o _c + \delta _{Hc} \hat 
\zeta^o v^o )  \bigr) \Bigr)
\nonumber
\end{eqnarray}
with the bare fields
\begin{equation}
\label{bareb}
B_a^o = (z^V)^{-1T}_{ab} B_b \qquad
\hat \phi^o_a = r^S_a {z_W \over z_H  } \hat \phi_a
\end{equation}
and bare parameters
\begin{eqnarray}
\xi ^o & = & z^2_W \xi \\
\hat \xi^o &  = & z^2_W \Bigl( {\tan (\theta_W^o + \theta_Z) \over
\tan (\theta_W^o + \theta_A)} -1 \Bigr)  \xi  + z^2_W
 {\tan (\theta_W^o + \theta_Z) \over
\tan (\theta_W^o + \theta_A)}  \hat \xi   \nonumber \\
\hat \zeta^o & = & {z_W \over z_H^2} \hat \zeta \nonumber \\
\hat G^o & = & \sqrt {\tan (\theta_W^o + \theta_Z) \over
\tan (\theta_W^o + \theta_A)} \hat G   \nonumber
\end{eqnarray}
One has finally to transform the ghost fields and $B$-fields 
in the ST identity and in the external field part
(\ref{gagenextf}) to bare fields, absorbing
the field redefinition parameters into a redefinition
of the by now undetermined parameters $r^g_{\a b}$ and
$\hat z_W^g$ (\ref{zgdef})
and $\ga_Z, \ga_A$.
The bare transformation
matrix $\delta \hat g _{\a b}$, which depends on the bare
angles $\ga^o_Z$ and $\ga^o_A$ as defined in (\ref{deltahatg}),
is computed  via 
\begin{equation} 
O^T_{a \a}(\theta_W^o) \delta \hat g^o_{\a b}  = 
\Bigl( (z^V)\,O(\theta_W^o)\,(r^V)\inv\Bigr)_{a \beta} \delta \hat g
_{\b b'} (\bar z^g)^T_{b' b}
\end{equation}
This equation determines also those three parameters of the antighost field
redefinition matrix $(\bar z^g)_{ab}$, which are not specified on
the bilinear ghost part of the action.

Having transformed the general bilinear ghost action into its
standard form
 the ghost equations are solved quite simply. 
On the kinetic parts the matrix $a^{og}_{\a b}$ is related to
the angles of antighost transformations:
\[
\hat z_W^{og} = 1
\]
\begin{equation}
\label{agdet}
\begin{array}{lcl}
r^{og}_{3Z}& = & \cos \ga_Z^o \\
r^{og}_{4Z} &  = & \sin \ga_Z^o 
\end{array}
\qquad
\begin{array}{lcl}
r^{og}_{3A} & = & -\sin \ga_A^o \\
r^{og}_{4A} &  = & \cos \ga_Z^o 
\end{array}
\end{equation}
  The parameter
$\hat \zeta^o$ of the gauge fixing part
is determined from the mass ratio of $W$-boson and $W$-ghost:
\begin{eqnarray}
{g_\xi \over g_2 ^o} \hat \zeta^o  & = & \zeta_W^ o
\end{eqnarray}
Inserting this result yields on the neutral part the following equations: 
\begin{eqnarray}
\zeta_W^ o M_{W}^{o2}
\cos ( \ga_Z^o - \theta_W^0) \is \cos \theta_W ^o \cos \ga_Z^o
\zeta_Z^ o M_{Z}^{o2}\nonumber \\
\zeta_W^ o M_{W}^{o2}
\sin ( \ga_A^o - \theta_W^0) \is \cos \theta_W ^o \sin \ga_A^o
M_{g_A}^{o2} \nonumber \\
- \hat G^o 
\zeta_W^ o M_{W}^{o2}
\cos ( \ga_Z^o - \theta_W^0) \is \cos \theta_W ^o \sin \ga_Z^o
\zeta_Z^ o M_{Z}^{o2} \nonumber \\
 \hat G^o 
\zeta_W^ o M_{W}^{o2}
\sin ( \ga_A^o - \theta_W^0) \is \cos \theta_W ^o \cos \ga_A^o
M_{g_A}^{o2}  
\end{eqnarray}
They
determine the BRS-transformations of
antighosts, i.e.~$\ga^o_Z ,\ga^o_A$, and  the abelian parameter
of the gauge fixing part $\hat G^o$ as functions of
the vector boson mass ratio and the ghost mass ratio.
 The mass of the photon ghost
is  seen to be not a free parameter of the model but has to
vanish ($\cos \theta_W^o \equiv M_W^o / M_Z^o$ ):
\begin{equation}
\label{geqsol}
{\zeta_W^ o M_{W}^{o2} \over \zeta_Z^ o M_{Z}^{o2}}
  =  {\cos \ga^o_Z \cos \theta_W^o
\over \cos (\ga^o_Z - \theta_W^o)} \qquad
{M^o_{gA}}  =   0 
\end{equation}
\[
\hat G ^o = - \tan  \ga^o_Z \qquad \tan \ga_A^o =  \tan \theta_W^ o
\]

The whole point in this calculation is the adjustment of the
abelian coupling $\hat G$ via the mass of the Z-ghost. For arbitrary
$\hat G $ indeed one has to introduce the angle $\ga_Z$ into
the BRS-transformations of ghosts, as otherwise one is not
able to keep the normalization condition
\begin{equation}
\label{infrag}
\Ga_{\bar c_A c_Z}\Big|_{p^2  = 0} = 0
\end{equation}
which is crucial for infrared finite computations
for off-shell Green functions. In the tree approximation, of course
it is possible to fix the ghost mass ratio equal to the vector mass
ratio by the normalization condition:
\begin{equation}
\label{normgmsym}
{\mathrm Re}
\Ga_{\bar c_+ c_-} (p^2) \big|_{p^2 = \zeta M_W^2} = 0 \quad
{\mathrm Re}
\Ga_{\bar c_Z c_Z} (p^2) \big|_{p^2 = \zeta M_Z^2} = 0 
\end{equation}
Then the expressions (\ref{geqsol}) simplify to the ansatz we have taken
in the classical approximation:
\begin{equation}
\cos \ga_Z = {M_W \over M_Z} + O(\hbar) 
\end{equation}
and the BRS-transformation of antighosts is diagonal.
For higher orders, however, the normalization conditions (\ref{normgmsym})
together with the infrared condition (\ref{infrag}) does not imply 
a diagonal transformation matrix for antighosts. Conversely
requiring (\ref{normgmsym}) and a diagonal ghost transformation
one has to introduce then counterterms $\bar c_A c_Z$ in order to fulfil
the ST identity. These counterterms produce in the next
order off-shell infrared divergencies. That the ghost mass ratio
and the parameter $\hat G$ are indeed independent parameters
of the standard model, is
 already indicated by the computations we have carried out in the
classical approximation: Starting with the general bilinear ghost
action it is seen, that there is no parameter left, which could
adjust the bare ghost mass ratio to the bare vector  mass ratio.
Likewise when we transformed the gauge fixing part to the
bare form, we had to treat the parameter $\hat G$ as an independent
parameter of the theory.
   The Callan-Symanzik equation, we derive in the  section 4,
unambiguously
allows to determine
 the independent parameters of the model. There it is finally proven, that
  the ghost mass ratio is a further independent
parameter of the theory. The coupling $\hat G$ and $\ga_Z$ are then
determined order by order by normalization conditions, which
fix  the mass of the Z-ghost and
diagonalize the neutral ghost mass matrix at $p^2 = 0 $. 
 Taking therefore
the ghost mass ratio as arbitrary also in the tree approximation
as specified in (\ref{ghostmass}) we find
\begin{eqnarray}
\label{hatG}
\hat G & = & \tan\theta_W {1- {\zeta_W \over \zeta_Z} \cos^2 \theta_W
                           \over {\zeta_W \over \zeta_Z}(1 - \cos^2 \theta_W)}
 + O(\hbar)\nonumber  \\
 & = & 
 { {\zeta_Z M_Z^2 - \zeta_W M_W^2}  \over \zeta_W M_W \sqrt{M_Z^2 - M_W^2}}
+ O(\hbar)
\end{eqnarray}
This independent parameter does not only enter the gauge fixing
part but also enters the ST identity and the 
ghost interactions, and has, wherever it appears, to be differently
treated from the vector mass ratio, even if we fix
the ghost mass ratio to be the same as the vector mass ratio
by the normalization conditions (\ref{normgmsym}).
The ghost interactions  in the classical approximation
are immediately read off from (\ref{geqint}):
\begin{eqnarray}
\label{geqintbare}
& &  \Ga^{gen}_{ext.f.} (\rho^o_\a , Y_a^o)
 + \Ga^{gen}_{ghost} (\bar c ^o_a)
  =   
\Ga_{ext.f}^{gen} (\rho'^o_\a , Y'^o_a) 
   - 
   \bar c^o _a (\delta \hat g^o _{a4})\inv  (\delta \hat g^o)_{4b} 
\Box c^o_b  
\nonumber \\
  &    & -  g_\xi\bar c^o_a
\Bigl(\sum_{\a=\atop +,-,3}
  (\delta \hat g^o)_{a \a}\inv\hat t_{bc,\a }
  -  \tan \theta_G^o ( \delta \hat
g^o)_{a  4  }\inv  \hat t_{bc,4} \Bigr) 
\bigl( \phi^o_b + \delta _{Hb} 2 \hfrac{M_W^o}{g_2 ^o}\bigr)  \hat q^o_c
\end{eqnarray}
with 
\begin{eqnarray}
\rho'^{o}_\a &= &
\rho^o_\a + \partial \bar c^o_a (\delta \hat g^o)\inv_{a \a} \\
Y'^o_b   & = & 
Y^o_b -  
\bar c^o_a g_\xi \tilde I_{bb'}\Bigl(\sum_{\a=\atop +,-,3}
(  \delta \hat g^o)_{a \a}\inv \hat t_{b'c,\a } 
  -  \tan \theta^o_G ( \delta \hat
g^o)_{a  4  }  \inv\hat t_{b'c,4} \Bigr) (\hat \phi^o_c + \delta_{Hc} \zeta 
2 \hfrac  {M_W^o}{g_2^o} ) \nonumber
\end{eqnarray}
The interactions of the external fields with propagating fields
are summarized in
$\Ga_{ext.f}^{gen}$ (\ref{gagenextf}). It is not modified
by solving the ghost equations, but for bare ghosts
the arbitrary parameters therein
are now specified to be related to $\delta \hat g ^o$:
\begin{equation}
\label{hatgghw}
\delta \hat g^o_{\a b} =
\left(\begin{array}{cccc}
       1 & 0 & 0 & 0 \\
       0 & 1 & 0 & 0 \\
        0& 0& \cos \theta^o_G & - \sin \theta^o_W \\
       0 & 0 &  \sin \theta^o_G & \cos \theta^o_W \end{array} \right)
\end{equation}
For having simple notations we have introduced 
 the ghost angle $\theta^o_G$ in analogy
to the weak mixing angle $\theta_W^o$,
which both are defined in the on-shell scheme  by the vector mass 
and the ghost mass ratio, respectively:
\begin{eqnarray}
\label{ghostangle}
\cos\theta_W^o & \equiv & {M_W^o \over M_Z^o}  \nonumber \\
\cos \theta_G^o & \equiv & {\zeta^o_W M^o_W \over \zeta^o _Z M^o_Z} 
{\sqrt {1 - {M_W^{o2} \over M^{o2}_Z }} \over \sqrt {1 - {\zeta^o_W M_W^{o2} 
\over 
\zeta^o_Z M_Z^{o2}}}}
\end{eqnarray}
Respective expressions are defined for the physical on-shell masses.
It is quite instructive to consider the ghost transformation
matrix in the context of the classical field transformations:
When we introduced the ghosts in the classical approach by changing
the infinitesimal parameters of gauge transformations $\epsilon_\a (x)$
to anticommuting
parameters $c_a$ (\ref{onshellghosts}), we have already
mentioned that there is an arbitrariness
in defining them. This arbitrariness has now been exploited to
construct a diagonal ghost mass matrix by the transformation
\begin{equation}
\epsilon _\a \longrightarrow c_\a = \delta \hat g_{\a b} c_b
\end{equation}

 Finally it remains  to solve the ST identity 
for the interactions, which depend on 
 the external scalar field we have suppressed up to now.
Quite generally the dependence of the generating functional
of 1PI Green functions on the external scalars is governed by
the following equation, one derives from the ST identity:
\begin{equation}
\label{extfbrs}
\brs_\Ga 
\left({\delta \Ga\over \delta q_a}\right)
 = - {\delta \Ga \over \delta \hat \phi_a}
\end{equation}
Solving it in the classical approximation 
it turns out that one further parameter appears,  
 due to a field redefinition of the propagating
scalars into propagating and external scalars. The transformation
\begin{equation}
\label{extscred}
{\phi_a^o \choose   \hat \phi_a^o}  \longrightarrow
{\phi_a^o - x^o \hat \phi^ o_a\choose   \hat \phi_a^o}  
\end{equation}
is compatible with the ST identity and rigid symmetry
if we redefine the external field
part by
\begin{equation}
Y^o_a \hat t_{abc} ( \phi^o_b+ v^o \delta_{bH}) c^o_c  
\longrightarrow 
Y^o_a \hat t_{abc}  (\phi^o_b - x^o \hat \phi^o_b+ v^o \delta_{bH}) c^o_c  
- x^o Y^o_a \tilde I_{ab} q^ o_b
\end{equation}
As long as we do not want to interpret $\hat \phi_a$ as a background
field, the normalization of $x$ is irrelevant, because finally
Green functions are considered at $\hat \phi _a= 0 $. One
can fix $x$ by the following normalization condition,
\begin{equation}
\partial_{p^2}\Ga_{H \hat H} (p^2) \Big|_{p^2 = \mu^2_H} = x
\end{equation}
For the purpose of this paper we choose $x  = 0$, but nevertheless
one gets nonlocal higher order contributions between propagating
and external scalars.

\newsubsubsection{The solution of the ghost equations in higher orders}

The purpose of this section is to prove, that the ghost equations
can be indeed established to all orders in accordance with
  the normalization conditions on the ghost 2-point functions
(\ref{ghostmass}),
(\ref{ghostdemix}) and (\ref{ghostres}). 
If one is
able to  implement the ghost normalization conditions by
adjusting the free parameters in the external field part, the
gauge fixing and the ST identity, then in the construction of 
higher orders one has only to consider the non-local functional 
$\Ga^{nl}$ as defined in (\ref{geqint}).

First we  give the ghost  equations in momentum space
and test them with respect to the mass normalization conditions.
Introducing 
\begin{equation}
\Ga_{\rho^\mu_\a c_b }(p,-p) = - i p^\mu \Ga_{\rho_\a c_b} (p^2) =
-i p^\mu + O(\hbar)
\end{equation}
the ghost equation of the charged ghost tested at $p^2 =  \zeta_W M_W^2 $
reads:
\begin{equation}
\zeta _W M_W^2 {\mathrm Re}
\Ga_{\rho_+ c_-} (\zeta_W M_W^2)
+ i \hat \zeta M_W {\mathrm Re}
\Ga_{Y_+ c_ -}(\zeta_W M_W^2) = 0
\end{equation}
The $SU(2)$-components of the ghost equation are tested 
at $p^2 = \zeta _Z M_Z^2$ and $p^2 = 0$:
\begin{eqnarray}
\label{geqnaho}
 {\mathrm Re} \bigl(
\zeta _Z M_Z^2 \Ga_{\rho_3 c_Z} (\zeta_Z M_Z^2)
 - \hat \zeta M_W \Ga_{Y_\chi c_ Z}(\zeta_Z M_Z^2)\bigr) & = & 0 \\
 {\mathrm Re}\bigl(
\zeta _Z M_Z^2 \Ga_{\rho_3 c_A} (\zeta_Z M_Z^2)
 - \hat \zeta M_W \Ga_{Y_\chi c_A}(\zeta_Z M_Z^2) 
\bigr)& = & - \sin \ga_A{\mathrm Re} \Ga_{\bar c_A
c_A} (\zeta_Z M_Z^2) \nonumber \\
- \hat  \zeta M_W \Ga_{Y_\chi c_Z}(0) 
  & = &  \cos \ga_Z \Ga_{\bar c_Z
c_Z} (0) \nonumber \\
\hat   \zeta M_W  \Ga_{Y_\chi c_A}(0) 
\bigr)& = &  0  \nonumber
\end{eqnarray}
The same test is carried out on the abelian component of the 
ghost equations:
\begin{eqnarray}
\label{geqabho}
\zeta_Z M_Z^2 r^g_{4Z}
 + \hat G 
\hat \zeta M_W {\mathrm Re} \Ga_{Y_\chi c_ Z}(\zeta_Z M_Z^2)\bigr) & = & 0 \\
\zeta_Z M_Z^2 r^ g_{4A} + \hat G 
 \hat \zeta M_W {\mathrm Re} \Ga_{Y_\chi c_A}(\zeta_Z M_Z^2) 
& = &  \cos \ga_A {\mathrm Re}\Ga_{\bar c_A
c_A} (\zeta_Z M_Z^2) \nonumber \\
\hat G  \hat \zeta M_W \Ga_{Y_\chi c_Z}(0) 
& = &  \sin \ga _Z \Ga_{\bar c_Z
c_Z} (0)  \nonumber \\
\hat G  \hat \zeta M_W \Ga_{Y_\chi c_A}(0) 
\bigr) & = &  0 \nonumber
\end{eqnarray}

In order to evaluate these equations
 one has to take into account that the vertex functions
$\Ga_{\rho_a c_b} (p^2) $ and $\Ga_{Y_a c_b}(p^2 )$ are not independent
from each other but are related by the ST identity.
We assume now, that we had already  established the ST identity
and the Ward identities of rigid symmetry to order $n-1$ 
for all Green  functions and to order $n$ for   all tests with
respect to vectors and scalars. Having also applied the normalization
conditions on the 2-point functions and the one for fixing
the nonabelian coupling 
 the  external field part
is determined up to local contributions of order $n$.
These local contributions  are read off from the classical approximation
(\ref{gagenextf}).
If the  vertex functions $\Ga_{\rho_\a c_b} (p^2)$
and $\Ga_{Y_a c_b} (p^2) $  solve the ST identity then also
the vertex functions $\Ga'_{\rho_\a c_b} (p^2)$
and $\Ga'_{Y_a c_b}(p^2) $  solve
the ST identity and they are related by
\begin{eqnarray}
\Ga'^{(n)}_{\rho_ \a c_b} (p^2) \is
 \Ga^{(n)}_{\rho_\a c_b}(p^2) + a^{(n)}_{\a b} 
\nonumber \\
\Ga'^{(n)}_{Y_\pm c_\mp } (p^2) \is
\Ga ^{(n)}_{Y_\pm c_\mp } (p^2 ) \pm i M_W a^{(n)}_{++}
\nonumber \\
\Ga'^{(n)}_{Y_\chi c_a } (p^2) \is
\Ga ^{(n)}_{Y_\chi c_a } (p^2 ) +  M_Z 
\bigl(\cos \theta_W  a^{(n)}_{3 a}  + \sin \theta_W a^{(n)}_{4 a}
\bigr) c_a
\end{eqnarray}
and $a^{(n)}_{++} = a^{(n)}_{--}$ due to CP-invariance.
Inserting in the ghost equations and taking advantage of the quantum
action principle, which restricts the breakings of order $n$ to
local expressions, it is seen that the ghost equations can be 
fulfilled, if we adjust the
arbitrary parameters  $a^{(n)}_{\a b}$,
 the parameters of the gauge fixing part $\hat \zeta$ and $\hat G$
and the linear transformation parameters of the vectors $r^{g(n)}_{4a}$.
In fact it is the same  calculation as in the classical approximation.
In particular the  diagonalization of the ghost mass matrix at 
$p^2 = 0$ (\ref{infraghost}) implies
\begin{equation}
\Ga_{Y_\chi c_A } (0) = 0  \qquad \hbox{and} \qquad \Ga_{Y_\chi c_Z} (0) =
\cos \ga_Z {\Ga_{\bar c_Z c_Z} (0) \over \hat \zeta M_W}
\end{equation}
and
\begin{equation}
\hat G = - \tan \ga_Z  \qquad 
\end{equation}
Inserting the last relation into the solution of the
ghost equation  (\ref{geqint}) we find, that the condition 
\begin{equation}
\label{infraantig}
\Ga_{\bar c_A c_Z}(p^2 = 0) = 0 
\end{equation}
is now fulfilled by construction, whereas 
\begin{equation}
\label{infrag0}
\Ga_{\bar c_Z c_A } (p^2 = 0) = \Ga_{\bar c_A c_A } (p^2 = 0) = 0
\end{equation}
has to be established by requiring 
\begin{equation}
\Ga_{Y_\chi c_A} (p^2 = 0) = 0
\end{equation}
We want to mention, that in the BPHZL-scheme this condition is 
implemented due to the infrared degrees of $c_A$ and
$Y_\chi $. Otherwise this condition has to be introduced in addition
to the usual ones in order to be able to protect internal ghost loops from
off-shell infrared divergencies.

Explicitly it is seen that the normalization conditions 
at $p^2 = 0$ fix the counterterms
\begin{eqnarray}
\cos \theta _W a^{(n)}_{3 Z} & +  & \sin \theta _W r^{g(n)}_{4 Z} \nonumber \\
\cos \theta _W a^{(n)}_{3_A} & +  & \sin \theta _W r^{g(n)}_{4 A}
\end{eqnarray}
On-shell conditions for the charged ghosts and Z-ghost 
and  the separation of massive and massless ghosts at $p^2 = \zeta_Z M_Z^2$
determine finally the parameters
$ \hat \zeta,   \ga_Z ,  \ga_A   $
whereas  $a_{3Z} , a_{3A}$ and $a_{++}$ are fixed by
the normalization conditions on the residua of ghosts.

The construction of higher orders can be therefore indeed restricted 
to $\Ga ^{nl}$ as defined in (\ref{geqint}), but
we have to take into account, that we are not able to
dispose of the counterterms $Y_\chi c_Z$ and $Y_\chi c_A$,
because these counterterms have to be adjusted for establishing
the ghost equations without introducing infrared divergencies.
Thanks to the fact, that the
antighost transformations can be modified by introducing
the angles $\ga_Z$  and $\ga_A $,  the coefficients appearing
in the ST identity $r^ g_{4Z}$ and $r^ g_{4A}$ are at our disposal
for absorbing local breakings of the ST identity into corrections
of the ST operator, even if the ghost 2-point functions are
constructed with  on-shell  conditions.

\newsubsection{Summary of the classical approximation}

Because the approach we have chosen here for determining the
invariant local counter\-terms is somewhat unconventional, we want
to summarize the results of the last sections. 

The general invariant
action has been determined by requiring invariance with
respect to the ST identity and with respect to rigid $SU(2)$-symmetry,
\begin{equation}
\label{STsum}
{\cal S} (\Ga_{cl}^{gen}) = 0 \quad {\cal W}_\a \Ga_{cl}^{gen} = 0 
\end{equation}
The analysis is unconventional in so far as we did not prescribe 
the symmetry operators explicitly as e.g.~in the tree form, but we
specified them by field content and  algebraic properties, i.e.~nilpotency of
the ST operator, algebra of rigid operators and the consistency
relation. This is the only form appropriate for the treatment
of the standard model, because the
 parameters of the tree approximation get higher
order corrections as indicated by  the classical approximation, if
one separates the massless and massive particles
at $p^2 = 0$. This result can be read off from the explicit expressions
for the vector 2-point functions, which have been calculated in the
on-shell scheme in the literature (see e.g.~\cite{Denner} for
a complete list). Actually, in the abstract approach 
one notices
that one has to modify the symmetry operators of the
tree approximation, only if one classifies the higher order
breakings of the
ST identity according to their infrared and ultraviolet degree,
as we do it in section 5. 
Indeed it will be seen there, that we have already solved the infrared
part of the problem by solving 
 the classical approximation with the general symmetry operators.

A further important point in the treatment
is the observation, that the ST identity does not uniquely determine
all parameters of the standard model. For gauging the electromagnetic
current rather than the currents of lepton and quark number conservation
the Gell-Mann Nishijima relation has to be extended for off-shell
Green functions to higher orders. For doing this one has to derive
a $U(1)$ Ward identity and specify therein the free parameters as
the ones of electromagnetic current conservation:
\begin{equation}
\label{wqsum}
{\mathbf w}^Q = {\mathbf w}_{em} - {\mathbf w}_{3}
\end{equation}

The general invariant action as solution (\ref{STsum}) can be
decomposed as in the tree approximation into
\begin{equation}
\Ga^{gen}_{cl} =
\Ga^{gen}_{GSW}+ \Ga^{gen}_{ext.f.}+ \Ga^{gen}_{ghost}+ \Ga^{gen}_{g.f.}
\end{equation}
Apart from the mass of the photon and the photon ghosts, the parameters
of the bilinear action, i.e.~masses, residua and nondiagonal
mass matrix elements, are free parameters of the theory, and one
can  dispose of them by the normalization conditions we have specified
in section 4.1. Masslessness of the photon and the photon ghost has
to be proven to be in agreement with the ST identity in higher orders.
If we furthermore use the local $U(1)$-Ward identity for determining
the lepton and quark family couplings, then we remain with one free
parameter, the nonabelian coupling,
  which
can be chosen to the electromagnetic fine structure constant in the
Thompson limit by the normalization condition on the electron-photon 
vertex (\ref{coupl}).

The general action can be written in the bare form by eliminating
the non-diagonal mass matrix elements of the general bilinear part
and the arbitrary residua into a field redefinition, which
transforms the original fields to bare fields. The general expression
is obtained by undoing this transformations. Expressed in bare
fields the general action
depends  on the bare vector boson masses, the bare ghost  masses,
the bare fermion masses and the bare Higgs mass  and the
  the  nonabelian  coupling $g_2^o$ or likewise $e^o$.
\begin{equation}
\begin{array}{lcl}
M_W^{o2} \is M_W^2 + \delta M_W^2 \\ 
M_Z^{o2} \is M_Z^2 + \delta M_Z^2
\end{array}  \qquad 
\begin{array}{lcl}
m_H^{o2} \is m_H^2 + \delta m_H^2 \\
m_{f_i}^{o2} \is m_{f_i}^2 + \delta m_{f_i}^2
\end{array}
\end{equation}
\begin{equation} 
\begin{array}{lcl}
\zeta _W^ o M_W^{o2} \is \zeta_W M_W^2 + \delta \zeta_W M_W^2 + \zeta_W \delta
M_W^ 2 \\ 
\zeta _Z ^o M_Z^{o2} \is \zeta_Z M_Z^2 + \delta \zeta_Z  M_Z^2 + \zeta_Z \delta
M_Z^ 2
\end{array} 
\end{equation}
We want to point out that $\delta \zeta_Z$ and $\delta \zeta_W$
are independent higher order corrections even if we choose $\zeta_W =
\zeta_Z $
 in the tree approximation.
In a QED-like parameterization one has
\begin{equation}
g^o_2 = e {M_Z \over \sqrt{ M_Z^2 -  {M_W^2}  }} + \delta g_2
\end{equation}

$\Ga^{gen}_{GSW}$ has been given in (\ref{gagengsw}), where we have
to replace the couplings of lepton and quark currents according to
(\ref{lqfcouplch}) in order ot fulfil the abelian Ward identity of  
electromagnetic and weak current conservation. 
Taking for the gauge fixing function the most general ansatz, which is
 compatible
with rigid symmetry  and is linear in propagating fields 
(cf.~(\ref{gab}) and (\ref{gagfbare})),
the ghost equations relate $\Ga_{ghost}^{gen}$ to the external
field part according to (\ref{geqintbare}). Using 
the notations for the ghost angle and weak mixing angle as
introduced in  
(\ref{hatgghw}) and (\ref{ghostangle}) 
 the external field part as function of the bare vector masses
and bare ghost masses  is given by
\begin{eqnarray}
\Ga^{gen}_{ext.f.}  =   \int &\hspace{-3mm} \biggl(&\hspace{-3mm}
 -\frac 1{2} \sigma^ o _a \hat \ve_{\a \b \ga }   
\delta \hat g^o_{\b b} c^o_b  \delta \hat g^o_{\ga c}c^o_c  \\
&\hspace{-3mm} &\hspace{-3mm}
                     + \rho^o_{\mu\a} ( \partial^\mu \tilde I_{\a \b}
\delta \hat g^o_{\b b}
  c^o_b + \hat \ve _{\a \b \ga} 
O_{\b b} (\theta _W^o) 
V^{o\mu} _b \delta \hat g^o_{\ga c} c^o_c \bigr) \nonumber \\
&\hspace{-3mm} &\hspace{-3mm}
            + \Bigr(Y^{o\dagger } \Bigl( i   g^o_2 \bigl(
 \frac{{\tau}_\a}{2} \delta \hat g^o_{\a a}
  -\frac {\mathbf 1} 2 \tan \theta_W^o \delta \hat g^o_{4 a}\bigr)
   ( \Phi^o + \frac {\sqrt 2}{g^2_o}{0\choose  M_W^o} )  c^o_a - x 
\hat {\mathbf q}^ o \Bigr) + \hbox{h.c.} \Bigr)\nonumber\\
&\hspace{-3mm} &\hspace{-3mm}
                    + \sum_{i=1}^{N_F}\Bigl( \sum_{\delta = l,q}
 \bigl( \bar \Psi^{oR}_{\delta_i} i 
g^o_2 \bigl(
 \frac{{\tau}_\a}{2}
\delta g^o_{\a a}
  +   \frac {G^{\delta_i}} 2 \delta \hat g^o_{4 a} \bigr) c^o_a
   F^{oL}_{\delta_i} \nonumber \\
&\hspace{-3mm} &\hspace{-3mm} 
                \phantom{+ \sum_{i=1}^{N_F}}+ \sum_{f = e,d}
                   \bar \psi^{oL}_{f_i} 
i g_2^o \frac 12 (\tan \theta_W^o + G^{\delta_i} ) \bigr)  f_i^{oR} 
 \delta \hat g^o_{4a} c^o_a  \nonumber \\
&\hspace{-3mm} &\hspace{-3mm} 
                 \phantom{+ \sum_{i=1}^{N_F}} +
                   \bar \psi^{oL}_{u_i} 
 i g_2^o\frac 12 (-\tan \theta_W^o + G^{q_i}  \bigr) u_i^{oR} 
 \delta \hat g^o_{4a} c^o_{a}+ \hbox{h.c.}\Bigr)  \biggr)  \nonumber 
\end{eqnarray}
Here we have again rewritten   the scalars
 into complex doublets. 

The  ST operator and the Ward operators depend in the bare form
also on the vector mass ratio and ghost mass ratio:
\begin{eqnarray}
\label{STbare}
{\cal S}
(\Ga_{cl}^{gen} ) &\hspace{-3mm}= &\hspace{-3mm} \int \biggl(  
\bigl(\sin \theta_G^o \partial _\mu c _Z + \cos \theta_W^o
 \partial_\mu c^o_A\bigr)
             \Bigl( \sin
\theta_W^o {\delta   \over \delta Z^o_\mu } +  \cos \Theta^o_W
       {\delta   \over \delta A^o_\mu} \Bigr) \\  
 &\hspace{-3mm}
 & +     {\delta  \Ga_{cl}^{gen} 
\over \delta \rho^{o\mu}_3 } 
              \Bigl( \cos \theta_W^o 
       {\delta 
  \over \delta Z^o _\mu } -   \sin \theta _W^o
       {\delta   \over \delta A^o _\mu} \Bigr) 
       + {\delta \Ga_{cl}^{gen} \over \delta \sigma^o _3 } \hfrac 1 { \cos 
            (\theta_W^o - \theta _G^o)}
              \Bigl(\cos \theta_W^o {\delta   \over \delta c^o_Z } - 
          \sin \theta_G^o
       {\delta  \over \delta c^o_A} \Bigr) \nonumber \\ 
&\hspace{-3mm} &      + {\delta \Ga_{cl}^{gen}  \over \delta  \rho^{o\mu} _+ }
               {\delta    \over \delta W^o _{\mu - } }
      + {\delta \Ga_{cl}^{gen}   \over \delta \rho^{o\mu} _- }
               {\delta  \over \delta W^o _{\mu + } }
+  {\delta \Ga_{cl}^{gen}   \over \delta \sigma^o _+ }
               {\delta   \over \delta c^o_{-} }
+  {\delta \Ga_{cl}^{gen}   \over \delta \sigma^o _- }
               {\delta   \over \delta c^o_{+} } +
{\delta  \Ga_{cl}^{gen}  \over \delta Y^o _a}\tilde I_{aa'}
{ \delta   \over \delta \phi^o _{a'} } 
 \nonumber \\
&\hspace{-3mm}& + \sum_{i=1}^{N_F} 
\Bigl({\delta  \Ga_{cl}^{gen}  \over \delta \overline{\psi^{oL}
_{f_i}}}
{   \delta \over \delta f^{oR}_{i} }
+ {\delta  \Ga_{cl}^{gen}  \over \delta \overline{\psi^{oR}_{f_i}}}
{   \delta \over \delta f^{oL}_{i} } + 
  \hbox{h.c.} \Bigr)  \nonumber \\
& \hspace{-3mm}& 
+ B^o_a \left(\begin{array}{cccc}
       1 & 0 & 0 & 0 \\
       0 & 1 & 0 & 0 \\
        0& 0& \cos (\theta_G^o - \theta_W^o) & 0 \\
       0 & 0 &  \sin (\theta_G^o - \theta_W^o) & 1 \end{array}\right)_{ab}
 {\delta   \over \delta \bar c_b }  
   + q^o_a{\delta  
 \over \delta \hat \phi^o_a  }  \biggr) \Ga_{cl}^{gen} = 0
 \nonumber
\end{eqnarray}
The Ward operators of rigid symmetry are determined to ($\a = +,-,3$)
\begin{eqnarray}
\label{wardbare}
{\cal W}_\a = \tilde I_{\a\a'}\int dx &\hspace{-3mm}\biggl( & \hspace{-3mm}
V^{o\mu}_b  O^T_{b\b}(\theta_W^o)
 \hat \ve_{\b\ga \a'} O_{\ga c} (\theta_W^o)
\tilde I_{c c'} {\delta \over \delta V^{o\mu}_{c'} }
+ \{ B_a^o \} \\
&\hspace{-3mm}+& \hspace{-3mm}
c^o_b (\delta \hat g^o)^{T}_{b\b}  \ve_{\b\ga\a'}
(\delta \hat g^o)^{-1T}_{\ga c}  \tilde I_{cc'} {\delta \over \delta c^ o
_{c'} }
+\bar c^o_b  (\delta \hat g^o)^{-1}_{b\b}  \hat \ve_{\b \ga \a'}
(\delta \hat  g^o)_{\ga c}  \tilde I_{cc'}
 {\delta \over \delta \bar 
c^o_{c'} } \nonumber \\
&\hspace{-3mm}+& \hspace{-3mm}
(\phi^o_b + \delta_{Hb} 2\hfrac {M_W^o}{g_2^o})
\hat t_{bc,\a'} 
 \tilde I_{cc'} {\delta \over \delta \phi^o_{c'} }
+ \{Y^o_b  , \hat \phi^o_b + \hat \zeta^o 2\hfrac {M_W^o}{g_2^o}, \hat q_a^o\}
\nonumber \\
&\hspace{-3mm}+&\hspace{-3mm}
\rho^o_\b   \ve_{\b\ga,\a'} \tilde I_{\ga \ga'} 
{\delta \over \delta \rho^o_{\ga'} }
+\sigma^o_\b   \ve_{\b\ga,\a'} \tilde I_{\ga \ga'} 
{\delta \over \delta \sigma^o_{\ga'} } \nonumber \\
&\hspace{-3mm}+&\hspace{-3mm}
\sum_{i=1}^{N_F} \sum_{\delta = l,q} \Bigl(
\overline {F^{oL}_{\delta_i}} 
  i \frac {\tau_{\a'} } 2 
{\delta \over \delta \overline {F^{oL}_{\delta_i} }} -
{\delta \over \delta  F^{oL}_{\delta_i} }   \frac {\tau_{\a'}} 2 
F^{oL}_{\delta_i} \nonumber \\
&\hspace{-3mm} & \hspace{-3mm} \phantom{\sum \sum} 
+\overline {\Psi^{oR}_{f_i}}  \frac { i \tau_{\a' }} 2
{\delta \over \delta \overline {\Psi^{oR}_{f_i}} }  -
{\delta \over \delta  \Psi_{\delta_i}^{oR} }
  \frac { i \tau_{\a' }} 2 \Psi_{\delta_i}^R 
\Bigr)\biggr) \nonumber 
\end{eqnarray}
Comparing the ST identity and the Ward operators of rigid symmetries
expressed in terms of bare fields,
with the ones of the tree approximation, it is seen that
 they differ due to the appearance of the ghost angle,
which signals -- apart from field redefinitions -- different corrections
of the vector mass ratio and the ghost mass ratio in higher orders.

The local $U(1)$ Ward identity 
\begin{equation}
\label{wqlocbare}
\biggl( 
g_2^o \tan \theta_W^o {\mathbf w}_4^Q 
-  \partial{\delta \over \delta Z^o} \sin \theta _W^o
-  \partial{\delta \over \delta A^o} \cos \theta _W^o 
\biggr)\Ga^{gen}_{cl} =  \sin \theta_W^o \Box B^o_Z + \cos \theta_W^o
\Box B^o_A
\end{equation}
has been derived in the matter part of the action. It is broken
only by  gauge fixing the longitudinal parts
of  the abelian vector field combination.
It is valid in the ghost part of the action as it is, if we take care
to maintain rigid symmetry.

Transforming the bare fields to original fields yields the general
form of the action  and the general form of the ST identity and
Ward identities.  It allows order by order to determine the
invariant local contributions by expanding the parameters in
perturbation  theory. There it has to be noted that some of the
parameters appear explicitly in the symmetry operators. According
to our conventions (cf.~(\ref{defrv}),\ref{defrs} and (\ref{rghost}))
these parameters are 
\begin{equation}
\label{parfin}
r^V_Z, r^V_A , \Theta^V , r^S_a , r_{l_i}, r_{q_i}, r^g_{\a a} , \ga_Z , \ga_A
\end{equation}
As long as they are not needed for removing infrared divergent
contributions, they could be fixed in the symmetry transformations
as it is e.g.~for $r^S$ and $r_{q_i}, r_{l_i}$.
Furthermore one has explicit dependence on the shift of
the Higgs field $v$ and the shift of the external Higgs $\hat \zeta v $
in the Ward operators of rigid symmetry. These parameters are fixed by
on-shell conditions on the mass of the $W$-boson
and on the mass of charged ghosts, 
\begin{equation}
M_W , \zeta_W M_W 
\end{equation}
The remaining parameters which are only specified on the
general classical action are the field redefinitions
\begin{equation}
\label{fielddiv}
z_W , {\tan{(\theta_W^o + \theta_A)}\over \tan{(\theta_W^o + \theta_A)}},
Z^g_{+-}, z_H , z_{\nu_i} , z_{u_i}, \tilde z_{f_i} \, \hbox{and} \, x^o
\end{equation}
the nonabelian coupling, which is fixed to the fine structure constant
in QED-like schemes,
and  the remaining masses of the standard model:
\begin{equation}
\label{coupdiv}
e^ o \quad \hbox{and} \quad M^ o_Z ,  m^ o_H ,\zeta^ o_{Z} M^ o_Z , m^ o_{f_i}
\end{equation}
These parameters, in any case, have to be fixed by normalization
conditions on the finite 1PI Green functions.

The Callan-Symanzik equation, we derive in the next section,
enables one to calculate the asymptotic logarithmic corrections
to the invariants. There it is
seen, that the invariants, which are connected
with the parameters (\ref{fielddiv}) and (\ref{coupdiv})
get independent logarithmic corrections to the next order.

\newsection{The Callan-Symanzik equation}

The Callan-Symanzik (CS) equation describes the response of the Green
functions to the scaling of all momenta by an infinitesimal factor.
The dilatational operator
\begin{equation}
{\cal W}^D = \sum_{\hbox{\scriptsize all fields}}
\int (d_{\varphi_k} + x \partial_x)\varphi_k(x) {\delta
\over \delta \varphi _k(x)} \qquad d_{\varphi_k} = \dim ^{UV}{\varphi_k}
\end{equation}
acts on the 1PI-Green  functions in the same way as the differentiation
with respect to all the mass parameters of the theory:
\begin{eqnarray}
& &{\cal W}^{D} \Gamma = - m\partial _m \Gamma \quad\hbox{with} \\
& & m\partial_m \equiv {M_W}\partial_{M_W} +
{M_Z}\partial_{M_Z} +
{m_H}\partial_{m_H} + \sum_{i=1}^{N_F}\sum_f
{m_{f_{i}}}\partial_{m_{f_i}} +
\sum_{\kappa_a} \kappa_a \partial_{\kappa_a} \nonumber
\end{eqnarray}
$\kappa_a$ are the normalization points, which we have introduced to
fix the residua of fields. In concrete calculations one will introduce
only one $\kappa $ for fixing all the 
 infrared divergent residua of charged particles off-shell.

The CS equation \cite{CAL70} is of utmost interest in the abstract
approach as well as for concrete calculations. In the abstract approach
the $\beta$-functions and anomalous dimensions, which describe
the breaking of dilatations by anomalies in higher orders, allow
to determine the independent parameters of the theory in a scheme
independent way. When we solved the ST identity and 
the Ward identities
of rigid symmetries for the local contributions, we gave a  
list of free parameters, which were not fixed by the symmetries
and can be adjusted by normalization conditions. The CS equation
 singles out from these  parameters  the ones
which get independent logarithmic corrections in the asymptotic region,
where all momenta are much larger than the masses of the theory.
To be specific we consider as an
 example  the ghost mass ratio: It can, of course, be fixed
by a normalization condition to the vector boson mass ratio. However,
from the CS equation it is immediately derived, that the vertices,
which depend on the ghost mass ratio, get different logarithmic 
corrections in higher orders from the ones, which depend on the
vector mass ratio. Similar statements are true for other
mass ratios, one would like to relate to the vector boson mass ratio
in lowest order, as it is required for example 
in the context of noncommutative
geometry \cite{NG}. With the help of the CS equation it is
at least in massless theories possible to find relations
between independent parameters, which are compatible
 with renormalizability,
 using the principle of reduction of
couplings \cite{OEZI85, KUSI85}.
 It is, however, not obvious, how reduction has to be applied
to spontaneously broken theories, due to the fact that the $\beta$-functions
do not only depend   on the perturbative expansion parameters but also on
the mass ratios. This property has also the consequence,
that the $\beta$-functions
of the CS equation  depend on mass logarithms from 2-loop order onwards
and that they differ from the ones of the symmetric
theory \cite{INVCH}.

For the purpose of the present paper
concerning the renormalization of the standard model, 
the CS equation, especially the dilatational
operator,  is also of special interest, because it allows
to simplify the  algebraic cohomology as carried out in the next section
considerably.
 One can derive that all algebraic anomalies of the ST identity
and the Ward identities are
restricted to 4-dimensional field polynomials, 3-dimensional breakings
are immediately seen to be variations of terms with quantum numbers
of the action \cite{BABE78}. In this context we also want to mention
that the CS equation plays an important role
if one wants to prove  the  nonrenormalization theorems 
for the Adler-Bardeen anomaly as it appears in the ST identity
(see \cite{BABE80} and references therein).

For concrete calculations the most important outcome
of the CS equation is the determination of higher
orders leading logarithms. If there are large asymptotic
logarithms in 1-loop order,
the CS equation allows consistently to determine 
those large leading logarithms of higher orders, which 
are induced by the 1-loop logarithms of the lowest order. 

In this section we construct the symmetric operators of the CS equation,
 which define
the $\beta$-functions and anomalous dimensions
 of 1-loop order. It is important to note, that we are able to classify
the dilatational anomalies also if we have not proved that
the ST identity and Ward identities of rigid symmetries are 
established in 1-loop. The only ingredients are the symmetries of
lowest order  
 and the linear gauge fixing. The soft breakings, however,
can be only consistently classified when we have established rigid
symmetry of 1-loop order. Then the classification  works as
we present here in the tree approximation.

\newsubsection{The soft breaking of dilatations}

In the standard model dilatations are already broken in the tree 
approximation by all terms with mass dimensions less than 4,
 especially by   the mass terms of the fields.
 Due to the spontaneous symmetry breaking
all the masses of the physical fields are generated by the shift
of the Higgs field. According to the construction of the gauge
fixing sector using rigid symmetry 
the ghost masses and the masses of would-be-Goldstones
are generated by the shift of 
the external Higgs (cf.~(\ref{gfrig}),(\ref{gaugefix}) and (\ref{gaghost})).
In the tree approximation one  reads off the breakings
of dilatations by  applying $m\partial_m$ on the classical
action.  For proceeding to higher orders it is unavoidable to
characterize also the soft breakings by their symmetries. For this reason
 we do the same
in the tree approximation and solve thereby also the problem
of  constructing
the soft breakings of higher orders, if the ST identity
and rigid symmetry are established.
 We have in the tree approximation
\begin{equation}
m \partial_m \Ga_{cl} = \Delta_m
\end{equation}
where $ \Delta_m $ is an integrated
 field polynomial with mass dimension less than
4, CP-even and neutral with respect to electric and $\phi\pi$ charge.
Commuting the operator $m\partial _m$ with the ST operator 
it is seen that $\Delta_m$ is $\brs _{\Ga_{cl}}$-invariant.
\begin{equation}
 \brs _\Ga m\partial_m \Ga - m\partial _m {\cal S} (\Ga)
= 0 \Longrightarrow 
 \brs_{\Ga_{cl}} \Delta _m = 0
\end{equation}
The dilatational operator does not commute with the Ward identities
of rigid symmety:
\begin{equation}
\Bigl[{\cal W }_\a,m \partial _m\Bigr] = 
\Bigl[{\cal W }_\a,\int v\Bigl( {\delta \over \delta H } +
\hat \zeta {\delta \over \delta \hat H } \Bigr) \Bigr] 
\end{equation}
This implies
\begin{equation}
{\cal W_\a } \Delta_m = {\cal W}_\a \int v \Bigl( {\delta \over \delta H } +
\hat\zeta {\delta \over \delta \hat H } \Bigr) \Ga_{cl}
\end{equation}
Noting that the differentiation with respect to the Higgs
and the external Higgs is a BRS-variation quite generally
(\ref{extfbrs})
\begin{equation}
\brs_{\Ga } Y_H = {\delta \Ga \over \delta H } 
\qquad \hbox{and} \qquad \brs_{\Ga }\left( {\delta \Ga \over \delta \hat q_H} 
\right)
= {\delta \Ga \over \delta \hat H} 
\end{equation}
the polynomial $\Delta_m $ can be decomposed into 
\begin{equation}
\Delta_m = \int \Bigl( v {\delta \Ga_{cl}\over \delta H } + \hat \zeta 
v{\delta \Ga_{cl} \over \hat \delta H }\Bigr) + \Delta_{inv}
\end{equation}
and 
\begin{equation}
\brs_{\Ga_{cl}} \Delta_{inv} = 0 \qquad {\cal W}_\a \Delta_{inv} = 0
\end{equation}
Inspecting all 2- and 3-dimensional field polynomial  it is seen
that there is only one rigid and $\brs_{\Ga_{cl}}$- invariant 
polynomial:
\begin{equation}
\label{deltainv}
\Delta_{inv} \equiv \int( 2 \phi_+ \phi_- + \chi ^2 + H^2 + 2 v H) 
\end{equation}
This invariant field polynomial we couple to an external scalar
field $\hat \varphi_o$ with UV dimension 2 and IR-dimension 2
and add it  to the classical action
\begin{equation}
\label{gaclhso}
\Ga_{cl} \to \Ga_{cl} + \int \hat \varphi_o 
( 2 \phi_+ \phi_- + \chi ^2 + H^2 + 2 v H) \, .
\end{equation}
The enlarged classical action is ST-invariant and rigid invariant
if we assign:
\begin{equation}
\brs \hat \varphi_o = 0 \qquad \delta_\a \hat \varphi_o = 0
\end{equation}
Finally we are able to write $\Delta_m$ as a field operator 
acting on the classical action:
\begin{equation}
\label{cstree}
m\partial_m \Gamma_{cl} = \int v \Bigl( {\delta \over \delta H } +
\hat\zeta {\delta \over \delta \hat H }  +  
\frac {m_H^2}{2 v}  
{\delta \over \delta \hat \varphi _o} \Bigr) \, \Gamma_{cl}
\end{equation}
and $v$ is determined  in the QED-like parameterization
to
\begin{equation}
v=  2 {M_Z \over e} \sin \theta_W \cos \theta_W + O(\hbar)
\qquad \quad \cos \theta_W \equiv {M_W \over M_Z}
\end{equation}
The soft breaking of dilatations, and in particular the
breaking of the tree approximation, is therefore completely characterized
by its properties under the symmetry transformations and one
can proceed in higher orders as in the tree approximation,
if the ST identity and rigid symmetries are established. We want to
mention, that also at this stage rigid symmetry plays an important
role for classifying the soft breakings of dilatations.
If one breaks rigid symmetry in the gauge fixing and ghost sector
it is not possible to
derive an unambiguous expression for the soft breakings in higher orders.

Because we have   rewritten the  dilatational breaking
of lowest order  into
a sum of field differentiations we are able to proceed immediately
to the next order. When one applies 
  the symmetric
operator
\begin{equation}
\label{wdsym}
 {\cal W}^D_{sym} \equiv
 {\cal W}^D - \int v \Bigl( {\delta \over \delta H } +
\hat \zeta {\delta \over \delta \hat H }  +  
\frac {m_H^2}{2 v}  
{\delta \over \delta \hat \varphi _o} \Bigr) 
\end{equation} 
on the functional of 1PI Green functions,
all  breakings of 1-loop order are now known
to be local due to the action principle.
But  in the construction of the ST identity and rigid Ward identities
we have
 also to consider Green functions which include the invariant
external field $\hat \varphi_o$.
 For the invariant
counterterms as derived in the last sections the only linear dependence
on $\hat \varphi _o $ enters via the
interaction with the 2-dimensional scalar invariant (\ref{gaclhso}),
which reads for the bare scalar fields $\tilde\phi^o_a =
 \phi^o_a + x ^o \hat \phi^o_a$ (cf.~(\ref{extscred})):
\begin{equation}
\Ga ^{gen}_{\hat \varphi_o} = 
\int \hat \varphi^o_o (2 \tilde\phi^o_+ \tilde\phi^o_- 
+ \tilde\chi ^{o2} + \tilde H^{o2} + 2 v^o \tilde H^o) 
\end{equation}
The field renormalization of the external scalar field
$\hat \varphi_o$ can be fixed by setting the coefficient of
the field differentiation with respect to $\hat\varphi_o $
in the CS equation.

\newsubsection{The dilatational anomalies -- hard breakings}

In higher orders the dilatations are not only broken by the
soft mass terms but also by hard terms, the dilatational anomalies.
The importance of the CS equation is founded in the fact, that
these anomalies can be  
absorbed into differential operators, which are connected
with the anomalous 
dimensions and $\beta$-functions. 

The dilatational
anomalies of one-loop order are normalization point independent,
but the differential operators introduced depend on the special
parameterization and the specific form of the breaking mechanism.
They are essentially characterized  by the symmetries of the
tree approximation. Because we want finally to use the CS operator
for classifying the breakings of the ST identity we only
assume that the ST identity is established in lowest order
and so for the rigid symmetries.
However, for deriving
the CS equation we cannot completely stick to the tree approximation
as given in the section~2, but we have to treat the ghost mass ratio
as an independent parameter. The ST identity and
Ward identities of the tree approximation
have then the form as given in (\ref{STbare}) and (\ref{wardbare}), replacing
all fields and parameters by ordinary fields and parameters.
Therefore the symmetry operators depend  in the tree approximation
on the vector mass ratio, i.e.~$\cos \theta_W$, and ghost
mass ration, i.e.~$\cos \theta_G$ (\ref{ghostangle}).

Using the  action principle in its quantized version
one derives that all symmetries
of the tree approximation are 
broken in 1-loop order by integrated field polynomials
\begin{equation}
 \Bigl({\cal S } (\Ga)\Bigr)^{(\le 1)}  = \Delta^{(1)}_{brs} \qquad
\Bigl( {\cal W}_\a \Ga \Bigr)^{(\le 1)} = \Delta^{(1)}_\a 
\end{equation}
\[
 \Bigl( {\cal W}^D_{sym} \Ga \Bigr)^{(\le 1)} = \Delta^{(1)}_m
\]
These field polynomials are restricted according to the UV-degree
and IR-degree of the classical operators:
\begin{equation}
\begin{array}{lcl}
\dim^{UV} \Delta_{brs}^{(1)}& \le &4 \\
\dim^{UV} \Delta_\a^{(1)}& \le &4 \\
\dim^{UV} \Delta_{m}^{(1)}&  \le&  4
\end{array}
\qquad
\begin{array}{lcl}
\dim^{IR} \Delta_{brs}^{(1)} & \ge & 3 \\
\dim^{IR} \Delta_\a^{(1)} & \ge & 2 \\
\dim^{IR} \Delta_m^{(1)} & \ge & 2  
\end{array}
\end{equation}
We want to mention already here that breakings with
infrared dimension 2 are alarming, because they do in general not
exist as integrated insertions in higher orders. They potentially
contain mass insertions for massless particles, which are seen
to be infrared divergent off-shell.
 In particular,
absence of mass insertions for massless particles, i.e. $A^\mu A_\mu$ and
$\bar c_A c_A $, has to be verified in the CS equation
by a test with respect to
the respective normalization conditions to all orders. 

Because the functional of 1PI
Green functions is invariant under the global charge
symmetries (\ref{chargecons}) and (\ref{qlfcons}) and CP-transformations,
 all breakings have a well-defined transformation
behaviour under these symmetries:
 $\Delta_{brs}$ is CP-even, has $\phi\pi$ charge 1 and is neutral
with respect to electric charge,
whereas $\Delta_{\a}$ is CP-odd, neutral
with respect to $\phi\pi$ charge and has electric charge $\pm 1$
for $\a = +,-$ and is neutral for $\a = 3$.
The important point is that $\Delta_m$ in fact has the same quantum
numbers as the general classical action: It is neutral with respect
to electric and $\phi\pi$-charge and CP-even. 
Up to linear field polynomials the most general basis
for the integrated insertion $\Delta_m$ is equivalent to the general
renormalizable action, from which one has constructed the general
invariant solution in section 5. 
For this reason we are able to make use of the symmetric dilatational
operator, when we consider the algebraic cohomology problem, for
classifying the breakings of the ST identity.

For proceeding we take advantage from the property that
 the operator ${\cal W}^D_{sym}$ (\ref{wdsym}) commutes with
the ST operator and the Ward operators of rigid symmetry.
Therefrom one derives the following consistency relations:
\begin{eqnarray}
\label{wdcomm}
{\cal W}^D_{sym} \Delta^{(1)}_\a =  - {\cal W _\a } \Delta^{(1)}_m \qquad
{\cal W}^D_{sym} \Delta^{(1)}_{brs} =  - {\brs}_{\Ga} \Delta^{(1)}_m
\end{eqnarray}
We decompose now  the local insertions $\Delta_\a ,
\Delta_m$ and  $\Delta_{brs}$  according to their
transformation under ${\cal   W}^D_{sym}$ 
\begin{eqnarray}
\label{dimequ}
\Delta_{op}^{(1)} & = & \sum_{k=1}^4\Delta_{op}^k
\qquad \hbox{with}\qquad {\cal W}^D_{sym}   \Delta^k_{op} = (k-4) \Delta^k_{op}
\end{eqnarray}
It is proven immediately that this decomposition is unique, because
it is nothing else but classifying field polynomials according to their mass
dimension and taking the Higgs, the external Higgs and the
neutral scalar $\hat \varphi_o$ in a shifted version
\begin{equation}
H+ v\, , \quad \hat H + \hat \zeta v\, , \quad 
\hat \varphi_o + {m_H^2 \over 4 }
\end{equation}
The consistency relations split up into equations for
any of the $\Delta^{k}_{op}$ and give different informations
as long as we did not establish the ST operator and rigid symmetry
in 1-loop order. The 4-dimensional breakings $\Delta^4_m$ are seen
to be $\brs_{\Ga_{cl}}$-symmetric and rigid symmetric
\begin{eqnarray}
\label{P4inv}
 0 = -{\cal W_\a} \Delta_m^4  \qquad
 0  = -{\brs}_{\Ga_{cl}} \Delta_m^4
\end{eqnarray}
These equations are used to construct the hard anomalies of the
CS equation, which are related to
$\beta$-functions and
anomalous dimensions. The further equations for lower dimensional
field polynomials 
state that the breakings $\Delta_{brs}^{\le 3}$
and  $\Delta_{\a}^{\le 3}$ 
 can be immediately written as $\brs_{\Gacl}$ and ${\cal W}_\a$
variations, respectively, of integrated
field polynomials with quantum numbers of the action.

For constructing the hard breakings of the CS equation
we have therefore the task to find all independent field polynomials
satisfying the above constraints and to express them in form of
symmetric differential operators. The first problem has been already
solved in the last section, because $\Delta_m$ has the same quantum
numbers as the general renormalizable action. We have only
to fix the parameters, which appear explicitly in the symmetry
operators (\ref{parfin})
to their tree values and have to expand the remaining
ones   (\ref{fielddiv}) and (\ref{coupdiv})
to the next order, singling out the polynomials
which contain only lower dimensional field polynomials.
\begin{eqnarray}
{\cal S}(\Ga_{cl}^{gen})(\theta_W , \theta_G) & = &
{\cal S}(\Ga_{cl}+ \sum_{n=1}^\infty\Ga^{(n)}_{inv})(\theta_W , \theta_G)
\nonumber \\
& = &
\Bigl({\cal S}(\Ga_{cl}) + \brs_{\Ga_{cl}}\sum_{n=1}^\infty\Ga^{(n)}_{inv} 
\Bigl) (\theta_W , \theta_G) = 0 
\end{eqnarray}

For finding the invariant operators corresponding to the invariant
field polynomials we construct in the usual way symmetric operators,
which commute with the lowest order rigid  Ward operators
and the lowest order ST operator and this all done we
identify the operators with invariant field polynomials.
It is well-known that
the invariant field polynomials are separated into two  classes: The first
one contains all invariant field polynomials,
which are  $\brs _{\Ga_{cl}}$-variations. These invariants
are generated by acting with symmetric field differentiation operators
on the classical action and are connected with field redefinitions
and anomalous dimensions.
The second class comprises the invariants
which are generated by differentiation with respect to independent
parameters of the theory.
 These field polynomials
 are $\brs _{\Ga_{cl}}$-invariants without being variations.

First we give a list of all symmetric field differentiation
operators. Because the vectors, $B$-fields  and ghosts are
rotated from the $SU(2)$ and $U(1)$-fields to mass eigenstates
by the matrix $O_{\a a}(\theta_W) $ and $\delta g_{\a b}$,
the field differentiation operators are not purely leg counting
operators, but mix massless and massive neutral fields.
In the vector ghost sector we find the following invariant field
differentiation operators
\begin{eqnarray}
{\cal N}_V &=& \int \Bigl( V_a {\delta \over \delta V_a} -
                        \rho_\alpha    {\delta \over \delta \rho_\alpha }+
                      \hbox{$ \frac 1 \cwg $} (\sg c_Z + \cw c_A ) 
                       \bigl( \sw {\delta \over \delta c_Z   } 
                              +\cg{\delta \over \delta c_A   } \bigr)\Bigr)
\nonumber\\  
\hat {\cal N}_V & = & \int \Bigl( (\sw Z + \cw A ) \bigl( \sw
                       {\delta \over \delta Z } + 
                       \cw {\delta \over \delta A    } \bigr) \nonumber \\
  & & +                      \hbox{$ \frac 1 \cwg$} (\sg c_Z + \cw c_A )
                     \bigl( \sw {\delta \over \delta c_Z   }                 
                       +\cg{\delta \over \delta c_A   } \bigr)\Bigr)\nonumber
\\
\label{Nvec}
{\cal N}_B & = & \int \Bigl( B_a {\delta \over \delta B_a} +
\bar c_a {\delta \over \delta \bar c_a}    
                           \Bigr) \\
\hat{\cal N}_B & = & \int \Bigl(    (\sw B_Z + \cw B_ A ) \bigl( \sw
                       {\delta \over \delta B_Z } + 
                       \cw {\delta \over \delta B_A    } \bigr)\nonumber
  \\
  & & +                      \hbox{$ \frac 1 \cwg$} (\sw \bar c_Z + \cg \bar 
                           c_A )
                     \bigl( \sg {\delta \over \delta \bar c_Z   }
                       +\cw{\delta \over \delta \bar c_A   } \bigr)\Bigr)
 \nonumber\\   
{\cal N}_c & = & \int \Bigl(c _+ {\delta \over \delta c_+   }      +
                           c _- {\delta \over \delta c_-   }      
                    +        \hbox{$ \frac 1 \cwg$} (\cg c_Z - \sw c_A )
                     \bigl( \cw {\delta \over \delta c_Z   }                 
                       -\sg{\delta \over \delta c_A   } \bigr)\nonumber\\   
             & & -           \sigma _+ {\delta \over \delta \sigma_+   }      
                 -          \sigma _- {\delta \over \delta \sigma_-   }      
                    -         \sigma _3
                          {\delta \over \delta \sigma _3   }      
\Bigr) \nonumber
\end{eqnarray}
The symmetric field differentiation  operators in the fermion sector 
have to commute also with the operators of lepton
and quark family conservation. They are also not leg-counting
operators for massive fermions, but involve the $\ga_5$.
It is convenient to split these operators into the ones for
left-handed and right handed fields, which are both invariant operators:
\begin{eqnarray}
\label{Nferm}
{\cal N}^L_{F_{\delta_i}}& =& \int \Bigl( \overline {F^L_{\delta_i}}
 {\delta \over \delta 
                         \overline {F^L_{\delta_i}}} 
                      -\overline {\Psi^R_{\delta_i}} {\delta \over \delta 
                         \overline {\Psi^R_{d,i}}} +
                     {\delta \over \delta    {F^L_{\delta_i}}}     {F^L_{
\delta_i}} 
-                     {\delta \over \delta    {\Psi^R_{\delta_i}}}  
{\Psi^R_{\delta_i}} 
\Bigr) \quad \delta= l,q   \\
{\cal N }^R_{f_i} & = & \int \Bigl(  \overline {f^R_{i}} {\delta \over \delta 
                         \overline {f^R_{i}}} 
                      -\overline {\psi^L_{i}} {\delta \over \delta 
                         \overline {\psi^L_{i}}} +
                     {\delta \over \delta    {f^R_{i}}}     {f^R_{i}} 
-                     {\delta \over \delta    {\psi^L_{i}}}  {\psi^L_{i}} 
\Bigr)  \qquad f_i = e_i, d_i , u_i \nonumber 
\end{eqnarray}

The invariant scalar field differentiation
operators comprise  the ones of the propagating scalars and of
external scalars.
 They are symmetric with respect to the rigid
operators, if one includes the shift of the Higgs and external 
Higgs in the transformation.
\begin{eqnarray}
\label{Nscal}
{\cal N}_S + v \int {\delta \over \delta H} & = &  \int \Bigl( 
         \phi_a         {\delta \over \delta \phi_a} + v
                      {\delta \over \delta H}
            -  Y_a {\delta \over \delta Y_a}
                     \Bigr)  \\
{\cal N}_{\hat  S} +\hat\zeta v \int {\delta \over \delta \hat H} & = &
  \int \Bigl( 
         \hat \phi_a {\delta \over \delta \hat \phi_A} +  
              \hat \zeta v \int {\delta \over \delta \hat H} 
    +  \hat q_a {\delta \over \delta q_a}  \Bigr)\nonumber 
\end{eqnarray}
When 
acting on $\Ga_{cl}$ the invariant differentiation operators summarized
in (\ref{Nvec}), (\ref{Nferm}) and (\ref{Nscal}) 
are in one  to one correspondence
with the field redefinition parameters listed in (\ref{coupdiv}):
\begin{equation}
\begin{array} {rcl}
\bigl({\cal N}_V - {\cal N}_B \bigl)\, \, \Gacl & \longleftrightarrow & z_W \\
\bigl(\hat {\cal N}_V - \hat {\cal N}_B \bigl)\, \, \Gacl & \longleftrightarrow &
{ \tan{\theta_W + \theta_Z} \over \tan (\theta_W + \theta_A) } \\
 {\cal N}_c \, \, \Gacl & \longleftrightarrow &  Z^g_{+-} \end{array}
\qquad  \begin{array} {rcl}
{\cal N}_{S} \, \, \Gacl &  \longleftrightarrow &  z_H \\
{\cal N}^L_{F_{l_i}} \, \, \Gacl & \longleftrightarrow & z_{\nu_i} \\
{\cal N}^L_{F_{q_i}} \, \, \Gacl & \longleftrightarrow & z_{u_i} \\
{\cal N}^R_{f_i} \, \, \Gacl &  \longleftrightarrow & \tilde z_{f_i} 
\end{array}
\end{equation}
The operators ${\cal N}_B, \hat {\cal N} _B $ and ${\cal N}_{\hat S} $
correspond to field redefinitions of $B$-fields and external scalars,
which are, however, fixed in the gauge fixing part  to the ones
of vectors, propagating scalars and coupling redefinitions 
(cf.~(\ref{bareb})). Taking them as independent operators in the CS equation
their coefficients are determined quite simply by a test on the
local gauge fixing polynomial.

The invariant field polynomial, which corresponds to the field
redefinition of the pro\-pa\-ga\-ting into the external scalar, i.e.~to
 the parameter  $x^o$ (\ref{extscred}),
 is generated by the mixed field differentiation 
operator:
 \begin{eqnarray}
\label{Nextprop}
 {\tilde {\cal N}}_S +\hat
 \zeta v \int {\delta \over \delta H} & = &  \int \Bigl( 
         \hat \phi_a   {\delta \over \delta \phi_a} + 
         \hat  \zeta v \int {\delta \over \delta H}   \Bigr)
\end{eqnarray}
It is symmetric with respect to rigid symmetry but not with respect
to the ST operator. The $\brs_{\Gacl}$-invariant  insertion is then
given by
\begin{equation}
\label{extscal}
 \bigl(  {\tilde {\cal N}}_S +\hat \zeta v \int {\delta \over \delta H}\bigr)
              \Gamma_{cl} +
          \int   q_a \tilde I_{ab} Y_b 
\end{equation}

Now we turn to the non-variations among invariant field
polynomials. From the general classical symmetric solution they
are read off
by expanding the independent parameters  (\ref{coupdiv})
 in perturbation 
theory. Equivalently they are  generated by differentiating
the classical action with respect to the independent parameters:
 These are
the  coupling $e$, which is the perturbative expansion parameter,
and furthermore the mass ratios, $\frac {M_W}{M_Z}$, for the weak
interactions, $\frac{m_H}{M_Z}$ for the scalar interaction and
$\frac {m_{f_i}}{M_Z}$ for the Yukawa interactions. At this stage it
is unavoidable to treat $\theta _G$ i.e.~the ghost mass ratio as
an independent parameter, because its differentiation corresponds
to an independent insertion in the gauge fixing and ghost sector.
Similarly it turns out that also the differentiation with respect
to the both gauge parameters, $\xi $ and $\hat \xi$, has to be included. 
The differentiation with respect to parameters, which do not
appear in the ST identity and the rigid Ward operators 
of the tree approximation immediately correspond to 
respective invariant field polynomials:
\begin{equation}
\label{symcoup}
  {m_{H}} \partial _{ m_{H}} , \,
 {m_{f_i}} \partial _ {m_{f_i}} , \, \xi \partial_\xi,
\, \hat\xi \partial _{\hat \xi}
\end{equation}
The differentiation with respect to the coupling $e$ is not a rigid
invariant, but has to be symmetrized by including the shift:
the operator
\begin{equation}
\label{coup}
e \partial_e - e \partial _e v \int \Bigl( {\delta \over \delta H } +
\hat \zeta {\delta \over \delta \hat H} \Bigr)  =
e \partial_e + \frac 2 e M_Z \sw \cw \int \Bigl( {\delta \over \delta H } 
+ \hat \zeta {\delta \over \delta \hat H} \Bigr) 
\end{equation}
is $\brs_{\Gacl}$- and rigid symmetric.
Without using the local $U(1)$-Ward identity there would
be seemingly invariant field polynomials corresponding to
the lepton and  quark family coupling $G_{\delta_i}$.
They are however singled out by deriving
quite in analogy to (\ref{wdcomm}) and (\ref{P4inv}),
 that the field polynomials $\Delta_m^4$
are $U(1)$-gauge invariant. This result finally relates
also the $\beta$-functions of the coupling $e$ and the mass ratio
$M_W \over M_Z$ to the anomalous dimensions (see (\ref{abrel})).

In order to find the $\brs_{\Gacl} $-invariants of the mass ratios
$M_W \over M_Z $ and $\zeta_W M_W \over \zeta_Z M_Z$ it is not sufficient
to expand the mass ratios  only in the 
general symmetric classical action,
 but one has to take into account, that such a mass expansion concerns
also the ST operator and Ward operators of rigid symmetry.
 This subtlety comes in, because these mass ratios
take a twofold role: They appear in the
field transformation matrices, which are introduced for constructing
mass eigenstates (cf.~(\ref{othetaw}) and (\ref{hatgghw})) , and they
take at the same time the role of the
abelian gauge coupling and the
abelian coupling  $\hat G$, respectively
(cf.~(\ref{coupthetawrel}) and (\ref{hatG})).
Expanding the bare  mass ratios  in perturbation theory
\begin{equation}
\cos \theta_W^o = \cos (\theta_W + \delta \theta_W)
\qquad
\cos \theta_G^o = \cos (\theta_G + \delta \theta_G)
\end{equation}
 one finds from the
general symmetric solution:
\begin{equation}
{\brs^{(0)} _{\Gacl}} \Ga^{(1)}_{inv} (\delta \theta_W, \delta \theta_G )
+ (\delta S) ^{(1)} (\Ga_{cl})(\delta \theta_W , \delta \theta_G)
= 0 + O (\hbar ^2 )
\end{equation}
with
\begin{equation}
{\cal S}({\Ga}) = \bigr({\cal S}^{(0)} + {\delta S}^{(1)}\bigl) (\Ga) + 
                  O(\hbar^2)
\end{equation}
Corresponding to these expressions
 differentiation with respect to $\theta_W $ as well as $\theta_G$
are not  $\brs_{\Gacl}$-invariant operators, because their action
on $\Gacl$ produces only $\Ga^{(1)}_{inv}$. In addition one has to enlarge
them both with mixed massless -- massive
field differentiation operators for
 being $\brs_{\Gacl}$-invariant operators:
\begin{eqnarray}
\label{thetaw}
\tilde \partial _{\theta_W} & \equiv&
\partial_{\theta_W} +  \int\Bigl( A {\delta \over \delta Z} - 
                            Z{\delta \over \delta Z}  
               +         B_ A {\delta \over \delta B_Z} - 
                          B_  Z{\delta \over \delta B_A} \Bigr) \nonumber \\   
               & &       + \hbox{$\frac 1 \cwg $} \int 
 c_ A \Bigl( {\delta \over \delta c_Z} + \swg 
                            {\delta \over \delta c_Z} \Bigr)  \nonumber \\
 & &-  \hbox{$\frac 1 \cwg $} \int \Bigl(  \bar
 c_ Z + \swg \bar c_A \Bigr) {\delta \over \delta\bar c_A } \\ 
\tilde \partial_{\theta_G} &\equiv &
\partial_{\theta_G}   -  \hbox{$\frac 1 \cwg $} \int 
 c_ Z \Bigl( {\delta \over \delta c_A} + \swg 
                            {\delta \over \delta c_Z} \Bigr)  \nonumber \\
\label{thetag}
& & +  \hbox{$\frac 1 \cwg $} \int \Bigl( \swg \bar
 c_ Z + \bar c_A \Bigr) {\delta \over \delta\bar c_Z} 
\end{eqnarray}
 The operator
$\tilde \partial_{\theta_G}$ is a symmetric operator with respect
to rigid symmetry, whereas
$\tilde \partial_ {\theta_W}$ has to be enlarged by
the contributions from the shift, because it acts on $v$ already in the
lowest order.
The operator
\begin{equation}
\label{weinsym}
\tilde \partial _{\theta_W} - \partial _{\theta_W} v \int \Bigl(
{\delta \over \delta H} + \hat \zeta {\delta \over \delta \hat H} \Bigr) = 
\tilde \partial _{\theta_W} - \frac 2e M_Z cos 2\theta_W  \int \Bigl(
{\delta \over \delta H} + \hat \zeta {\delta \over \delta \hat H} \Bigr) 
\end{equation}
is then also rigid symmetric.

Acting with the symmetric operators 
(\ref{Nvec}), (\ref{Nferm}), (\ref{Nscal}),
(\ref{symcoup}), (\ref{coup}), (\ref{thetag}) and
(\ref{weinsym})  on the classical action one produces
together with the polynomial (\ref{extscal})
a complete basis for the hard breakings of the symmetric dilatational operator
(\ref{wdsym}) in 1-loop order. Therefore it is possible to give
the dilatational anomalies
 in the form of a CS equation, i.e.~as a linear combination
of differential operators. Writing all the soft breakings produced 
by symmetrization with respect to the shift on the r.h.s we get
 the CS equation of the standard model in 1-loop order:
\begin{eqnarray}
\label{cs}
\! &\! \Bigl( \! & \! 
m\partial _m + \beta_e e\partial_e - \beta_{M_W} \tilde \partial _{\theta_W}
+ \beta_{m_H} m_H \partial _{m_H} + 
\sum_{i=1}^{N_F}\sum_f\beta _{m_{f_i}}  {m_{f_i}} \partial
_{m_{f_i}} 
   \\
\! & \! -\! & \! 
\gamma_V \bigl( {\cal N}_V  
- {\cal N}_B + 2 \xi  \partial_\xi + 2 \hat \xi \partial_{\hat \xi}
+ \sin \theta_G \cos \theta_G \tilde\partial_{\theta_G} \bigr) 
-\gamma_c {\cal N}_c 
\nonumber \\
\! & \! -\! & \! 
 \hat \gamma_V \bigl(\hat {\cal N}_V
-\hat \gamma_B \hat {\cal N}_B  + 2 (\xi + \hat \xi) \partial_{\hat \xi} 
\bigr)  -
 \gamma_S  {\cal N}_S 
-  \gamma_{\hat S}{\cal N}_{\hat S}   
 - \tilde \gamma_S \tilde {\cal N}_S  \nonumber \\
\! & \!-\! & \! 
\sum_{i=1}^{N_F} \bigl( \gamma_{F_{l_i}}  {\cal N}^L_{F_{l_i}}
+\gamma_{F_{q_i}}  {\cal N}^L_{F_{q_i}}
+ \gamma_{e_{i}}  {\cal N}^R_{e_{i}}
+ \gamma_{u_{i}}  {\cal N}^R_{u_{i}}
+ \gamma_{d_{i}}  {\cal N}^R_{d_{i}} \bigr)  \,
\Bigr) \, \Gamma ^{(\le 1)}\Big|_{\hat \varphi_o = 0}  \nonumber  \\
\! & \! = \! & \! \int \Bigl( (1 + 
 \beta_e e\partial_e  - \beta_{M_W}\partial_{\theta_W}) v
\Bigl( {\delta \Gamma \over \delta H } +
\hat \zeta {\delta \Gamma \over \delta \hat H } \Bigr)  
 +  v (\gamma_S +\tilde \gamma_S ){\delta \Gamma \over \delta H } +
 \hat \zeta v \gamma_{\hat  S}{\delta \Gamma \over \delta \hat H } +
\frac {m_H^2}2
{\delta \Gamma \over \delta \hat \varphi_o }  \Bigr) \nonumber \\
\!& \! + \! & \! \int \tilde \gamma_S  
 \hat q_a \tilde I_{ab}Y_b + \Delta_m^{\le 3}
\nonumber 
\end{eqnarray}
On the right hand side we have also collected all local lower
dimensional 1-loop
breakings, which are not classified by the lowest order
symmetries, into $\Delta_m^{\le 3}$. When  finally 
the ST identity and Ward identities are established,
 we are able to prove that $\Delta_m^{\le 3 }$ is vanishing.
The CS equation takes then essentially this
form to all orders, in particular the number of independent
operators we have introduced is exhausted by the list giving
above. It is only the explicit form of higher order
operators, which is changed due to the symmetrization with
respect to the general Ward operators and the
generalized ST identity.
 
Calculating the commutator of the CS operator and the local 
$U(1)$-Ward operator yields the abelian relation between the
$\beta$-functions and anomalous dimensions.
\begin{equation}
\label{abrel}
\beta_e = \frac {\sw}{\cw} \beta_{M_W} + \gamma_V + \hat \gamma_V
\end{equation}

In the CS equation we have already inserted the result, which comes
out from the test on the  local $B$-dependent part of the action
(\ref{gab}) and (\ref{geqint}). In addition one derives
for the anomalous dimension of the external scalars $\hat \ga_S$
in the QED-like parameterization (\ref{gxi})
\begin{equation}
\label{garel}
\gamma_{\hat S} =  \beta_e + \frac {\cw}{\sw} \beta_{M_W} + 
\gamma_V - \gamma_S 
\end{equation}
whereas $\tilde \ga_S$ is an independent anomalous dimension and can
 be determined on the mixed external -- propagating scalar 2-point
functions. This function is only important, when one
wants to interpret $\hat \phi_a$  as a
background field.

 We want to give here the results for the $\beta$-functions
and anomalous dimensions of the vector-scalar sector. A complete list
of the $\beta$-functions can be found in \cite{wekr96}.
The $\beta$-functions of the electromagnetic coupling and
the vector mass ratio are determined to
\begin{eqnarray}
\label{beta}
\beta_e & = & -{e^2 \over 24\cdot4 \pi^2} 
\Bigl( 42 - \hbox{$\frac{64}3$} N_F \Bigr) \nonumber \\
\beta_{M _W} & = & - {e^2 \over 4\cdot 24 \pi^2\sw \cw} 
                     \Bigl( (43 - 8 N_F) - (42 - 
                     \hbox{$\frac {64}3$} N_F ) \sin^2 \theta_W \Bigr) .
\end{eqnarray}
The remarkable point is, that the $\beta$-function of the
electromagnetic coupling is indeed QED-like in the sense, that it
does not involve mass ratios, as does $\beta_{M_W}$.
In contrast to QED it involves the nonabelian contributions
of the charged vectors. This has as a consequence that the sign
of $\beta_e$ is negative, if one considers the standard model
without fermions or includes only one family.
The anomalous dimensions of vectors in general gauges compatible
with rigid symmetry are given by
\begin{eqnarray}
\label{anodim}
\gamma_V &=& {e^2 \over 4 \pi^2 \sin^2 \theta_W}
\Bigl( {6 \xi -25 \over 24 } 
                         + {1 \over 3 } N_F \Bigr) \\
\hat \gamma_V &=&  {e^2 \over 4 \pi^2 }
 \Bigl( - {6 \xi -25 \over 24  \sin^2 \theta_W} 
                             + {1 \over 24  \cos^2  \theta_W} 
                             +{-3 + 8 \sin^2 \theta_W \over 9  
                                \sin^2 \theta_W \cos^2 \theta_W}
 N_F \Bigr) ,   
\end{eqnarray}
The CS-functions  (\ref{beta}) and (\ref{anodim})
are  seen to fulfil the abelian relation (\ref{abrel}).

Because the anomalous dimension $\ga_V$ is nonvanishing, the coefficient
of the
ghost mass differentiation $\partial _{\theta_G}$  is also
nonvanishing and the ghost angle
has therefore to be treated as an independent parameter.
 This proves finally that the ghost
mass ratio gets independent higher order corrections. 
A choice compatible with renormalizability is to set
 all ghost masses  equal, i.e.~$\theta_G = 0$.
 Such a choice, however, is connected with
 nondiagonal vector-scalar propagators and not adequate for
concrete calculations. Similarly it is seen that also the abelian
gauge parameter $\hat \xi$ is an independent parameter of the theory.
 
Before we turn to the higher order breakings of the ST identity
and Ward identities we want to consider the off-shell infrared
problem as it appears in the CS equation. Because $\Delta_m$
has infrared dimension 2, it has to be proven explicitly
that the insertions
\begin{equation}
\int A_\mu A^\mu \, , \bar c_A c_A ,\,  H
\end{equation}
do not appear on the r.h.s. The proof is carried out best by
using the Zimmermann algebraic identities, which relate
insertions of infrared dimension 2 to insertions with infrared
dimension 3. (The technique has been presented and applied
 for deriving the CS equation in the 
spontaneously broken Higgs-Yukawa model \cite{CalSym} and works here
in the same way.) Then the r.h.s. contains all the 2-dimensional field
polynomials explicitly and one is able to
test the CS equation with respect to these field polynomials
at $p^2 = 0$.  Due to the existence of
the field mixing operators appearing e.g.~in $\hat {\cal N}_V $,
and in the operator $\tilde \partial_{\theta_W}$ the l.h.s. will
only vanish, 
if the mixed 2-point functions of massless and massive fields
vanish at $p^2 = 0$:
\begin{equation}
\Ga_{ZA} \Big|_{p^2= 0} = 0 \qquad \Ga_{\bar c_Z c_A }\Big|_{p^2 = 0} = 
\Ga_{\bar c_A c_Z }\Big|_{p^2 = 0} = 0 
\end{equation}
Otherwise, nonintegrable infrared divergencies appear to the next order
and make it impossible to derive the CS equation of higher orders.

\newsection{Higher orders}
\newsubsection{The quantum numbers of higher orders breakings}

We complete now the analysis of the renormalization of the
standard model by proving, that the ST identity
(\ref{STgen2}), Ward identities
of rigid symmetry (\ref{wardopequivclass})
and local $U(1)$-symmetry 
 (\ref{wardidho})
 can be established in the general form
as given in section 5.2 to all
orders and lead to unique  expressions for finite renormalized Green functions.
The basic ingredient of this proof is
 the action principle in its
quantized version, as it is valid in presence of massless particles
in the framework of the BPHZL scheme \cite{LOW76, CLLO76}.
If the symmetries are established to a definite order $n$ in perturbation
theory, the breakings of the next order are restricted to be
  local field polynomials with definite ultraviolet and infrared degree:
\begin{equation}
\label{indn}
\begin{array}{lcr}
\bigl({\cal S} (\Gamma )\bigr)^{(\le n-1)} & = &0  \\
\bigl({\cal W}_\a (\Gamma )\bigr)^{(\le n-1)} & = &0\end{array}
\quad \Longrightarrow \quad \begin{array}{lcl}
\bigl({\cal S} (\Gamma )\bigr)^{(\le n)} & = &\Delta_{brs}^{(n)} \\
\bigl({\cal W}_\a (\Gamma )\bigr)^{(\le n)} & = &\Delta^{(n)}_\a 
\end{array}
\end{equation}
with 
\begin{equation}
\label{uvir}
\begin{array}{lcl}
\dim^{UV} \Delta_{brs}^{(n)}& \le &4 \\
\dim^{UV} \Delta_\a^{(n)}& \le &4 \\
\end{array}
\qquad
\begin{array}{lcl}
\dim^{IR} \Delta_{brs}^{(n)} & \ge & 3 \\
\dim^{IR} \Delta_\a^{(n)} & \ge & 2 \\
\end{array}
\end{equation}
The ultraviolet degree of the breakings is deduced from a pure
power counting analysis of renormalizable quantum field theory,
but the infrared degree is assigned due to the BPHZL scheme.
The BPHZL scheme  implements those normalization conditions in the scheme,
which  have to be
fulfilled for being able to carry out infrared finite computations
for off-shell Green functions in presence of massless particles. 
Having the Green functions constructed
in a different scheme as dimensional regularization these
normalization conditions have to be established finally
by adjusting local counterterms. The conditions, which ensure
infrared finitiness for off-shell Green functions are read off
form the BPHZL-scheme:
\begin{equation}
\label{BPHZLn}
\begin{array}{lcl}
\Ga_{Z^\mu A^\nu}(p^2= 0) \is 0 \\
\Ga_{Y_\chi c_A } (p^2  = 0) \is  0 \\
\Ga_{\bar c_A c_Z}(p^2= 0) \is 0
\end{array}
\qquad
\begin{array}{lcl}
\Ga_{A^\mu A^\nu}(p^2= 0) \is 0 \\
\Ga_{\bar c_Z c_A}(p^2= 0) \is 0  \\
\Ga_{\bar c_A c_A}(p^2 = 0 )\is 0 
\end{array}
\end{equation}
The breakings are furthermore restricted by the
global symmetries, i.e.~by electromagnetic charge conservation
(\ref{chargecons})
\begin{equation}
\label{chconsbreak}
\begin{array}{lcl}
{\cal W}_{em} \Delta^{(n)}_{brs} \is 0 
\end{array}
\qquad
\begin{array}{lcl}
{\cal W}_{em} \Delta^{(n)}_{\pm} \is \mp i  \Delta^{(n)}_{\pm} \\
{\cal W}_{em} \Delta^{(n)}_{3} \is 0
\end{array}
\end{equation}
by Faddeev-Popov charge conservation
\[
\begin{array}{lcl}
{\cal W}_{\phi\pi} \Delta^{(n)}_{brs} \is \Delta^{(n)}_{brs}
\end{array}
\qquad
\begin{array}{lcl}
{\cal W}_{\phi\pi} \Delta^{(n)}_{\a} \is 0
\end{array}
\]
and by conservation of lepton and quark family number (\ref{qlfcons})
\[
\begin{array}{lcl}
{\cal W}_{\delta_i} \Delta^{(n)}_{brs} \is 0
\end{array}
\qquad
\begin{array}{lcl}
{\cal W}_{\delta_i} \Delta^{(n)}_{\a} \is 0
\end{array}
\]
The algebraic restrictions on the breakings derived from
nilpotency of the  ST operator (\ref{brsnil2}), algebra of Ward operators
(\ref{wardalg2})
and the consistency equation (\ref{cons}) read:
\begin{eqnarray}
\label{algbreak}
\brs_{\Gacl} \Delta^{(n)}_{brs} & = & 0 + O (\hbar ^{n+1})\\
{\cal W}_{\a } \Delta^{(n)}_{\b}-
{\cal W}_{\b } \Delta^{(n)}_{\a} & = & \ve _{\a \b \ga}
\tilde I_{\ga \ga'} \Delta^{(n)}_{\ga'} +  O (\hbar ^{n+1})\\
\brs_{\Gacl} \Delta^{(n)} _\a - {\cal W}_\a \Delta^{(n)}_{brs} &
= & 0 + O (\hbar ^{n+1})
\end{eqnarray}
The symmetry operators involved are the ones of the tree approximation,
because higher order contributions are not effective when acting
 in perturbation theory on a polynomial of order $\hbar^n$.

In the higher order analysis one has 
to find all the breakings compatible with quantum numbers
and symmetries, and one has to prove, that they can be absorbed into
a redefinition of the ST operator, the Ward operators of
rigid symmetry and by adjusting finite counterterms, without
destroying the on-shell normalization conditions on the 2-point functions.
This computation is well-defined and can be carried out
straightforwardly. However,
there are a lot of terms which have to be considered in the standard
model, even if CP-invariance is assumed.
The analysis is simplified enormously, when we take advantage of the
fact, that
 the breakings are also classified under the symmetric dilatational operator
${\cal W}^{D}_{sym}$ (\ref{wdsym}),
and when we finally use the knowledge about local symmetry
invariants. These local invariants
have been completely characterized, when we
 solved the ST identiy and Ward identities in the
classical approximation in generality without  using a perturbative
expansion and explicit  form of the operators.

We proceed now as follows:  First we consider the breakings of the
ST identity and classify them in
variations and non-variations, the anomalies (section 7.2):
\begin{equation}
\Delta^{(n)}_{brs} = \brs_{\Gacl} \Ga^{(n)}_{gen} + r^{(n)} \Delta^{anom}_{brs}
\end{equation}
Here $\Ga ^{(n)}_{gen}$ is a general local field polynomial with
quantum numbers of the classical action and UV dimension less than four.
One is not able to exclude at this stage
field polynomials of infrared dimension three. 
It is well-known, that there are anomalies in the standard model,
which are given in the next section. 
The anomalies have to be shown to vanish -- in the 1-loop order
by inspection of diagrams, in higher orders by applying the
non-renormalization theorems. These theorems state, that the anomalies
of the ST identity
 vanish in higher orders, if they vanish in lowest  order 
(\cite{BABE80} and references therein).
Application of the non-renormalization theorems to the standard
model will be considered elsewhere and we take its validity as granted
for the purpose of the paper. 
The variations we  absorb  as far as possible  into finite
counterterms to the action. In particular,
 breakings of IR dimension three cannot be absorbed into counterterms
for reasons of infrared definiteness,
although they are variations.
Then we establish Ward identities of
rigid symmetry and are finally able to define a unique ST identity
(section 7.3).
In this analysis  we do not have to consider
tests with respect to $B$-field and
with respect to antighosts, because this part has been already
constructed in agreement with the symmetries by solving the ghost equation
(see section 5.4). In particular we can carry out the variable
transformation $\rho_\a \to \rho '_\a $ and $Y _a \to Y' _a$ as
given in (\ref{geqint}) and establish the symmetries on
$\Ga^{nl} (\rho' , Y_a' )$, as usually done (cf.~e.g~\cite{PISO95}).
When we have established both, the rigid Ward identities  and the ST identity,
the abelian local Ward identity is identified. Its
breakings are known to be total divergencies. The variations
 can be absorbed into a redefinition of the 
lepton and quark couplings of the abelian subgroup and the anomalous
currents vanish, if the anomalies in the ST identity vanish.
At the very end one has  one single unspecified parameter, which is not fixed
on the 2-point functions. This parameter  can
be finally adjusted to be the electromagnetic fine structure constant in
the Thompson limit (\ref{coupl}), as it is done in the QED-like on shell
schemes (see for a review \cite{hollik} and references therein).

\newsubsection{The cohomolgy and the Adler-Bardeen anomaly}

In the first step we concentrate completely in finding the
non-variations of the breakings under the ST identity, the Adler-Bardeen
anomalies \cite{ADL69, BEJA69, BAR69}.
 In the construction of the Callan-Symanzik equation
we have shown that anomalies of the ST identity 
can appear only in the 4-dimensional
field polynomials (\ref{wdcomm}). This analysis is valid to all orders,
once  the CS equation  is established to all orders (section 7.4).
\begin{equation}
\label{dilbr}
{\cal W}^D_{sym} \Delta^{anom}_{brs} = 0
\end{equation}
all lower dimensional polynomials have been already seen to be variations.
 The consistency equation with the rigid transformation
operators furthermore tells that only rigid invariant field polynomials
can contribute to the anomaly:
\begin{equation}
{\cal W}_\a \Delta^{anom}_{brs} = 0
\end{equation}
Therefore the algebraic problem can be  indeed formulated in symmetric
variables $\a = +,-,3,4$:
\begin{equation}
\begin{array}{lcl}
c_\a & = & \delta \hat g_{\a b} c_b \\ 
V_\a  & = & O_{\a a} (\theta_W ) V_b 
\end{array}\qquad 
\begin{array}{lcl}
       H' & = & H + v \\
  \hat  H'&  = & \hat H + \hat \zeta v 
\end{array}
\end{equation}
and the analysis is  the same as one has to
carry out in  the symmetric theory.
Here we see in the abstract approach that ultraviolet divergencies
of the spontaneously broken theories are not worse than the ones
of the symmetric theory \cite{HO71, SY72}.

Since we have split off the transformation of the abelian component
  in the ST identity, the external field part is essentially treated
as in a $SU(2)$  gauge theory.
Therefore we remain finally with polynomials depending on
vectors,  scalars and fermions and arrive at the well-known
Wess-Zumino consistency condition \cite{WEZU71}: 
\begin{equation}
{\mathbf w}_\a P_\b - {\mathbf w }_\b P_\a = \ve _ {\a \b \ga}
\tilde I_{\ga \ga '} P_{\ga '}
\end{equation}
where ${\mathbf w}_\a $ are the gauge transformations of the tree
approximation given in (\ref{gaugeward}) 
and $P_\a$ is a 4-dimensional polynomial depending only on
the propagating fields of the standard model:
\begin{equation}
\Delta^{anom}_{brs} = \int c_a O^T _{a \a} (\theta_W) P_\a (V_a, \phi_a, 
f^L_i, f^R_i)
\end{equation}
The solution of the  consistency equation has been analyzed quite
generally in \cite{BABE78} 
and can be evaluated in the standard model
without further complications. Having CP-invariance there are 
even no abelian contributions, which escape the algebraic
treatment of consistency, because those terms are
CP odd. We end up with the following explicit 
expression for the anomalies ($a = +,-,Z,A$ are physical field indices,
$O(\theta_W)$ is defined in (\ref{othetaw}):
\begin{eqnarray}
\label{anom}
\Delta^{anom}_{brs} & = & r_1 \int \ve_{\mu \nu \rho \sigma}
                            O_{4a}(\theta)   c_a 
\partial^\mu \Bigl( O_{4b}(\theta_W) V_b^\nu \partial^\rho O_{4c}
(\theta_W ) V_c^\sigma \Bigr) \\
& + &  r_2 \int \ve_{\mu \nu \rho \sigma}
 O_{4a} (\theta_W) c_a \partial^\mu 
\Bigl(  V_b^\nu \tilde I_{bc} \partial^\rho  V_c^\sigma 
       - \frac 1 3 \hat \ve_{bcd} (\theta_W ) V^\nu_b V^\rho_c  V^\sigma_d
\Bigr) \nonumber
\end{eqnarray}
with
\begin{equation}
\hat \varepsilon _{abc} = \varepsilon_{\a \b \ga} O _{\a a}(\theta_W)
O _{\b b}(\theta_W) O _{\ga c}(\theta_W)
\end{equation}
The form of the anomaly is unique up to the addition of BRS-variations. 
The general classical action contains two rigid
invariant field polynomials in vectors which are odd under
parity transformations:
\begin{equation}
\Ga_{cl}^{P} (V_a) = 
\int \ve_{\mu \nu \rho \sigma} O_{4a} ( \theta_W)V^\mu_a 
\bigl( k_1 V^\nu_{b}
\tilde I_{bc} \partial ^{\rho} V^\sigma _c  + k_2
\hat \ve _{a b c} V^\mu_{a} V^\nu_{b} V^\rho_{c} V^\sigma_4  \bigr)
\end{equation}
 We have used the $\brs_{\Gacl}$-variations
of these field polynomials to bring the anomaly in the form given
above, where it only depends
on the abelian ghost combination.

In 1-loop order the coefficients of the anomaly vanish. One has
to note that the purely abelian part vanishes due to electromagnetic
current conservation and therefore depends crucially on
establishment of a local Ward identity in connection with electromagnetic
current conservation. 

In the following, we assume that the non-renormalization theorems
on the Adler-Bardeen is valid, if we are able to prove a local abelian
Ward identity and establish the Callan-Symanzik equation order
by order in perturbation theory. These two equations are the
necessary prerequisite  for proving the non-renormalization theorems
in higher orders \cite{BABE80}.

\newsubsection{The establishment of symmetries}

We start the consideration in 1-loop order and take for the lowest
order the usual standard model Lagrangian as given in section 2.
 If one calculates
the finite Green functions $\Ga^{ren}$
with the Feynman rules of the tree approximation
in a specific scheme to 1-loop order, the ST identity
is in general broken by  the local field polynomial  $\Delta_{brs}$:
\begin{equation}
\label{stbr}
({\cal S} \Ga^{ren} )^{(\le 1)} = \Delta_{brs}
\end{equation}
Because the coefficient of the anomaly vanishes in 1-loop order, 
the breaking can be rewritten
as a variation of integrated field polynomials:
\begin{equation}
\label{wabr}
\Delta_{brs} = \brs_{\Gacl} \Ga^{(1)}_{gen} 
\end{equation}
and
\begin{equation}
\dim^{UV} \Ga^{(1)}_{gen}  \le 4  \qquad
\dim^{IR} \Ga ^{(1)}_{gen}  \ge  3
\end{equation}
In the BPHZL scheme it is obvious that we do not have
to introduce counterterms with respect to a photon mass term,
because the variation of the
photon mass term has infrared dimension 2.
\begin{equation}
\brs_{\Gacl} \int A^\mu A_\mu = \int 2 \partial^\mu c_A A_\mu + ...
\end{equation}
But all further field polynomials appear in principle in
$\Ga^{(1)}_{gen}$.  In order to be able to establish the on-shell
conditions and conditions on the residua of all propagating
fields, we have to show that we have not to dispose of those
field polynomials which are fixed by the normalization conditions.
 These field polynomials
are listed in (\ref{gagenbil}). Due to the fact that we have eliminated
the antighost contributions by using the ghost equations
(\ref{geqint}), the normalization
conditions specified  on the ghost 2-point functions, are now
translated into normalization conditions on the external field
part. Explicitly we are
not able to dispose of the terms $Y_{\chi} c_Z$ and $Y_{\chi} c_A$
 for establishing on-shell conditions without
introducing infrared divergencies (cf.~eqs.~(\ref{geqnaho}),
(\ref{geqabho}) and also (\ref{chconsbreak})).
The terms 
$\rho_\a a^g_{\a b} \partial c_b$ are kept arbitrary and are
finally adjusted on the residua of ghost
propagators. Therefore we find the following list of field polynomials,
which are not available for adjusting finite counterterm contributions
\begin{eqnarray}
\label{gagenbil2}
\Ga^{gen}_{bil} & = &
\int \biggl(-\hfrac 14 
\bigr(\partial^\mu V^\nu_a - \partial^\nu V^\mu_a \bigl)  Z^V_{ab}
\bigr(\partial_\mu V_{\nu b} - \partial_\nu V_{\mu b} \bigl)  
+ \hfrac 12 V^\mu_a  {\cal M}^V_{ab} V_{\mu b} \nonumber \\
& & \phantom{\int } \:
+ \hfrac 12 \partial^{\mu} \phi_a Z_{ab}^S \partial_\mu\phi_b 
- \hfrac 12 M_H^2 H^2(x) \nonumber \\
& & \phantom{\int } \:
+ i Z_{f_i}^R\bar f_{i}^R \ds f_{i}^R +
i Z_{f_i}^L\bar f_{i}^L \ds f_{i}^L
- M_{f_i} (\bar f_{i}^R  f_{i}^L +\bar f_{i}^R  f_{i}^L) \nonumber \\
& & \phantom{\int } \:
+  \rho^{\mu}_\a a_{\a b}^g \partial_{\mu} c_b 
+  Y_{\chi} {m}^g_{\chi b} c_b
\biggr)
\end{eqnarray}

These terms can be eliminated if they are in one to one correspondence
with $\brs_{\Gacl}$-
invariants. 
From the detailed considerations of the classical approximation
it is seen, that there are left three polynomials, namely
 $\int Z^{\mu} A_{\mu}$, $\int Y_{\chi } c_Z$ and
$\int Y_{\chi } c_A$ 
which do not
correspond to $\brs_{\Gacl}$-invariants.
Therefore we are able to write
\begin{eqnarray}
\Delta_{brs} & = & \brs_{\Gacl} \Ga'_{break} + u_1 \brs_{\Gacl}\int M_Z^2
 Z^{\mu} A_{\mu}+ u'_2 M_Z \brs _{\Gacl} \int Y_{\chi } c_Z + u'_3 M_Z
 \brs_{\Gacl} \int Y_{\chi} c_A  \nonumber \\
& = & \brs_{\Gacl} \Ga _{break} + u_1 \brs_{\Gacl} \int \Bigl(
A^{\mu}  {\delta \over \delta Z^\mu} - 
Z^{\mu}  {\delta \over \delta A^\mu}  \Bigr)\Gacl \nonumber \\
& & + u_2 \brs _{\Gacl} \int  (  \sin \theta_W c_Z +  \cos \theta _W c_ A )
                \Bigl(\sin \theta_W {\delta \over \delta c_Z} +
\cos \theta_W {\delta \over \delta c_A}\Bigr) \Gacl \nonumber \\
& &+ u_3 \brs _{\Gacl} \int  (  \cos \theta_W c_Z -  \sin \theta _W c_ A )
                \Bigl(\sin \theta_W {\delta \over \delta c_Z} +
\cos \theta_W {\delta \over \delta c_A} \Bigr) \Gacl
\end{eqnarray}
There we have rewritten the field polynomials with
infrared dimension 3 into field operators
acting on the classical action, and have the remaining terms shifted
into $\Ga_{break}$. $\Ga_{break} $ consists of all integrated
CP-even field polynomials except the ones listed in (\ref{gagenbil2}),
i.e.~it has especially infrared dimension 4.
Applying the consistency equation between the Ward operators of
rigid symmetry and the ST identity (\ref{algbreak})
we find that the breakings
of the Ward operators take the following form:
\begin{eqnarray}
\Bigr({\cal W}_\a \Ga^{ren}\Bigl)^ {(\le 1)} & = & \Delta_\a^{inv}
+ {\cal W}_\a \Ga_{break} +
u_1 {\cal W}_\a \int \Bigl(
A^{\mu}  {\delta \over \delta Z^\mu} - 
Z^{\mu}  {\delta \over \delta A^\mu}  \Bigr)\Gacl  \\
& & +  u_3 {\cal W}_\a \int  (  \cos \theta_W c_Z -  \sin \theta _W c_ A )
                \Bigl(\sin \theta_W {\delta \over \delta c_Z } +
\cos \theta_W {\delta \over \delta c_A }
\Bigr)  \Gacl \nonumber 
\end{eqnarray}
$\Delta_\a^{inv}$ comprises all field polynomials, which are
$\brs_{\Gacl}$-invariants.
\begin{equation}
\brs_{\Gacl} \Delta_\a^{inv} = 0
\end{equation}
Considering the list of all possible breakings compatible
with the algebra of rigid symmetry and discrete and global symmetries
it is seen, that $\Delta_\a^{inv}$ itself can be written as
a $\brs_{\Gacl}$ and ${\cal W}_\a $-variation. Explicitly we find
the following list of contributions:
\begin{equation}
\Delta_\a = {\cal W}_\a \biggl(\sum_k u_k {\cal N}_k \Gacl + 
\int v ^{(1)}
{\delta  \over \delta H} \Gacl \biggr)
\end{equation}
and ${\cal N}_k$ comprises the following field operators
\begin{eqnarray}
{\cal N}_{ZA} & = & \int
 \Bigl( Z {\delta \over \delta Z} + A {\delta \over \delta A} +
 c_Z {\delta \over \delta c_Z} + c_A {\delta \over \delta c_A} -
                        \rho_3   {\delta \over \delta \rho_3 } -
 \sigma_3 {\delta \over \delta \sigma_3} \Bigr) \\
{\cal N}_{\widetilde {ZA}} & = & \int\! \biggl(
 \Bigl( \sin \theta _W Z  + \cos \theta_W A \Bigl)
\Bigr( \cos \theta_W {\delta \over \delta Z} 
 - \sin \theta_W  {\delta \over \delta A} \Bigr)
- \rho_3 \partial
(\sin \theta_W c_Z + \cos \theta_W c_A) 
\nonumber \\
& & \phantom{\int} +
 \Bigl( \sin \theta _W c_Z  + \cos \theta_W c_A \Bigr)
\Bigl( \cos \theta_W {\delta \over \delta c_ Z} 
 - \sin \theta_W  {\delta \over \delta c_A} \Bigr) \biggr) \nonumber \\
{\cal N}_{c_Z c_Z} &  = & \int \Bigl(
 \Bigl( \cos \theta _W c_Z  - \sin \theta_W c_A \Bigl)
\Bigr( \cos \theta_W {\delta \over \delta c_ Z} 
 - \sin \theta_W  {\delta \over \delta c_A} \Bigl) -
\sigma_3 {\delta \over \delta \sigma_3}\Bigr)\nonumber \\
{\cal N}_{\widetilde {c_Z c_A}} &  = & \int
  \Bigl( \sin \theta _W c_Z  + \cos \theta_W c_A \Bigl)
\Bigr( \cos \theta_W {\delta \over \delta c_ Z} 
 - \sin \theta_W  {\delta \over \delta c_A} \Bigl) \nonumber \\
{\cal N}_{\phi_+ } &= & \int \phi_+{ \delta \over \delta \phi_+ } 
+ \phi_- {\delta \over \delta \phi_-} - Y_+ {\delta \over \delta
Y_+} - Y_- {\delta \over \delta Y_-} \nonumber \\
{\cal N}_{\chi } & = &\int \Bigl( \chi { \delta \over \delta \chi } 
 - Y_{\chi} {\delta \over \delta Y_{\chi}} \Bigr) \nonumber \\
{\cal N }_{e_i} & = & \int \Bigl(  \overline {e^L_{i}} {\delta \over \delta 
                         \overline {e^L_{i}}} 
+ {\delta \over \delta 
                          {e^L_{i}}} e^ L_i 
                      -\overline {\psi^R_{e_i}} {\delta \over \delta 
                         \overline {\psi^R_{e_i}}} -
{\delta \over \delta  {\psi^R_{e_i}}} \psi^ R_{e_i} \Bigr) \nonumber \\
{\cal N }_{d_i} & = & \int \Bigl(  \overline {d^L_{i}} {\delta \over \delta 
                         \overline {d^L_{i}}}
+ {\delta \over \delta 
                          {d^L_{i}}} d^ L_i  
                      -\overline {\psi^R_{d_i}} {\delta \over \delta 
                         \overline {\psi^R_{d_i}}} -  
{\delta \over \delta  {\psi^R_{d_i}}} \psi^ R_{d_i} \Bigr) 
\nonumber
\end{eqnarray}
For absorbing these polynomials into
the Ward operators (\ref{wardopequivclass}) 
we have finally to note that the expansion
of the coefficients to 1-loop order can be also rewritten
into a field differentiation acting on $\Gacl$. Denoting
with ${\cal W}^{(0)}_\a $ the Ward operator of the tree approximation
(\ref{wardna})
then we write
\begin{equation}
{\cal W }_\a = {\cal W_\a }^{(0)} + \delta {\cal W}_\a^{(1)} + O(\hbar^2)
\end{equation}
with
\begin{eqnarray}
\label{wfd}
 \delta {\cal W}_\a^{(1)} \Gacl &= & 
{\cal W}^ {(0)}_\a \int 
\biggl( \delta r_Z  Z{\delta   \over \delta Z}
 + \delta r_A A{\delta   \over \delta A} 
 + \delta \theta^V \bigl(
Z{\delta \over \delta A} - A {\delta \over \delta Z} \bigr) \\
& & \phantom{{\cal W}_\a \int }
 + \delta r^g_{33}   \Bigl( \cos \theta _W c_Z  - \sin \theta_W c_A \Bigl)
\Bigr( \cos \theta_W {\delta \over \delta c_ Z}- 
 \sin \theta_W  {\delta \over \delta c_A} \Bigl) \nonumber\\
& & \phantom{{\cal W}_\a \int }
+ \delta r^g_{34}
 \Bigl( \cos \theta _W c_Z  - \sin \theta_W c_A \Bigl)
\Bigr( \sin \theta_W {\delta \over \delta c_ Z} 
 + \cos \theta_W  {\delta \over \delta c_A} \Bigl) \nonumber \\
& & \phantom{{\cal W}_\a \int }
+ \delta r^g_{43}
 \Bigl( \sin \theta _W c_Z  + \cos \theta_W c_A \Bigl)
\Bigr( \cos \theta_W {\delta \over \delta c_ Z} 
 - \sin \theta_W  {\delta \over \delta c_A} \Bigl) \nonumber \\
& & \phantom{{\cal W}_\a \int }
 + \delta r^S _+ {\cal N}_{\phi_+} + \delta r^S_\chi {\cal N}_{\chi}
 + \sum_{i=1}^{N_F} \bigl( \delta r_{l_i} {\cal N}_{e_i} +
\delta r_{q_i} {\cal N}_{d_i} \bigr) + \delta v {\delta \over \delta
H} \biggr)\Gacl \nonumber
\end{eqnarray}
Therefrom it is seen that the scalar and fermion contributions are
immediately absorbed into a redefinition of the tree Ward operators
compatible with the algebra.
For the ghosts and vectors only parts of the invariants are
absorbed, but a straigthforward calculation shows that all
the remaining contributions can be shifted into a $\hat \Ga_{break}$,
which again includes only interaction terms.
Therefore we remain with
\begin{equation}
\Bigr({\cal W} \Ga^{ren} \Bigl)^{(\le 1)} =
- \delta {\cal W}_\a^{(1)} \Gacl + {\cal W}_\a (\Ga_{break} + \hat
\Ga_{break}) 
\end{equation}
If one goes back with these expressions to the ST identity
we have now for reasons of consistency to split off therein
the corresponding contributions $\delta {\cal S}^{(1)}$, because
otherwise the consistency relations are not valid to the next order.
\begin{equation}
\label{sfd}
{\cal S} (\Ga ^{ren})^ {(\le 1)} = - \delta S^{(1)}\Ga_{cl} + \brs_{\Gacl}
(\Ga_{break} + \hat \Ga_{break})
\end{equation}
where 
\begin{eqnarray}
\delta S^{(1)}\Gacl & = &\brs_{\Gacl}\int 
\biggl( \delta r_Z  Z{\delta   \over \delta Z}
 + \delta r_A A{\delta   \over \delta A} 
 + \delta \theta^V \bigl(
Z{\delta \over \delta A} - A {\delta \over \delta Z} \bigr) \\
& & \phantom{{\brs_{\Gacl}} \int }
 + \delta r^g_{33}   \Bigl( \cos \theta _W c_Z  - \sin \theta_W c_A \Bigl)
\Bigr( \cos \theta_W {\delta \over \delta c_ Z} - 
 \sin \theta_W  {\delta \over \delta c_A} \Bigl) \nonumber\\
& & \phantom{\brs_{\Gacl} \int }
+ \delta r^g_{34}
 \Bigl( \cos \theta _W c_Z  - \sin \theta_W c_A \Bigl)
\Bigr( \sin \theta_W {\delta \over \delta c_ Z} 
 + \cos \theta_W  {\delta \over \delta c_A} \Bigl) \nonumber \\
& & \phantom{\brs_{\Gacl} \int}
+ \delta r^g_{44}
 \Bigl( \sin \theta _W c_Z  + \cos \theta_W c_A \Bigl)
\Bigr( \sin \theta_W {\delta \over \delta c_ Z} 
 + \cos \theta_W  {\delta \over \delta c_A} \Bigl) \biggr) \Gacl 
\nonumber
\end{eqnarray}
Defining the generating functional of Green functions of the
standard model by
\begin{equation}
\Ga = \Ga_{cl} +\Ga^{ren} - \Ga_{break} - \hat \Ga_{break}
+ O(\hbar^ 2)
\end{equation}
and the Ward operators and ST operator to 1-loop order by
\begin{equation}
{\cal W}_\a = {\cal W}_\a^{(o)} + \delta {\cal W}_\a^{(1)} \qquad
{\cal S}(\Ga) = {\cal S}^{(0)}(\Ga) + \delta {\cal S}^{(1)} (\Ga)
\end{equation}
we have proceeded to absorb all breakings into 
counterterms of the action and a redefinition of
the symmetry operators compatible with the algebra.
\begin{equation}
{\cal S}( \Ga ) = 0 + O(\hbar^ 2)
\qquad
{\cal W}_\a \Ga = 0 + O(\hbar^ 2) 
\end{equation}
 The field polynomials,
on which the normalization
conditions on the 2-point functions are established, are not touched
in the construction.
It is worth to note that at higher orders as it was in the
classical approximation the ST identity is only completely specified,
if we construct simultanouesly the Ward identities of rigid symmetry.
By now we have suppressed those contributions which depend on the
external field $\hat \phi_a$. They do not contribute to anomalies
and the absorption of their breakings proceeds as in \cite{KS},
 where we have carried out the same analysis
in the abelian Higgs model.

Since we have  only determined the normalization conditions
on the 2-point functions in the above construction, the finite
 Green functions are not unique  by now.
 First we have to fix the remaining coupling constant by a normalization
condition on an interaction vertex as given e.g~in (\ref{coupl}).
 From the construction of the general classical invariant action it is
seen, that furthermore the abelian couplings of fermions are
not specified. 
 The contributions
which remain arbitrary can be read off from the fermionic
 part $\Ga^{gen}_{matter}$  (\ref{gagenym})
and the external field part $\Ga^{gen}_{ext.f.}$ (\ref{gagenextf})
 of the general invariant classical action. 
They are  obtained in their explicit form by expanding
 $G_{\delta_i}$ to 1-loop order and setting all other coefficients
to their tree value. We denote with $\Ga_{\delta_i}$ the corresponding
$\brs_{\Gacl}$-invariant field polynomials.
\begin{equation}
\label{undet}
\Ga' = \Ga + \sum_{i=1}^{N_F}
\sum_{\delta = l,q} G_{\delta_i}^{(1)} \Ga_{\delta_i}
\end{equation}
and 
$\Ga$ and $\Ga'$ satisfy both the  ST identity and rigid Ward identities
in the same form.

For fixing these undetermined parameters
 one has to use the local $U(1) $ Ward identity (\ref{wardidho}).
Having constructed $\Ga$ in accordance with ST identity and
rigid symmetry, it is obvious 
that the abelian Ward identity is only broken by total divergencies
in 1-loop order.
Consistency (\ref{wqcons}) furthermore restricts the breakings
to be again $\brs_{\Gacl}$-invariants and rigid invariants.
One has
\begin{eqnarray} & &
\biggl( {e \over \cos \theta_W}
{\mathbf w}^Q - \partial \Bigl(\frac  1 {r_Z}
\sin \Theta {\delta \over \delta Z}
+ \frac 1 {r_A} \cos \Theta {\delta \over \delta A} \Bigr)\biggr)\Ga \\
 & & = \Box \Bigl( \frac 1 {r_Z}\sin \Theta  B_Z + 
\frac 1  {r_A}\cos \Theta B_A \Bigr) 
\nonumber \\
&  & +
 \delta g_1 \partial _\mu j^{ matter}_\mu + 
\delta g_{\delta_i} \partial _\mu j^{\delta_i} + r_i
\partial _\mu J_i ^{anom}  + O(\hbar^2) \nonumber
\end{eqnarray}
The breakings are given by
\begin{equation}
\partial ^ \mu j^{matter}_\mu = {\mathbf w}^Q \Gacl \qquad
\partial ^ \mu j^{\delta_i}_\mu = {\mathbf w}_{\delta_i} \Gacl
\end{equation}
${\mathbf w}_{\delta_i}$ denotes the non-integrated version of
the operators of lepton and quark familiy conservation (\ref{qlfcons}).
 The anomalous contribution is determined to
\begin{eqnarray}
\partial^\mu J^{anom}_\mu &  = &  
r_1 \int \ve_{\mu \nu \rho \sigma}
          \partial^\mu \Bigl( O_{4b}(\theta_W) V_b^\nu \partial^\rho O_{4c}
(\theta_W ) V_c^\sigma \Bigr) \\
& + &  r_2 \int \ve_{\mu \nu \rho \sigma}
\partial^\mu 
\Bigl(  V_b^\nu \tilde I_{bc} \partial^\rho  V_c^\sigma 
       - \frac 1 3 \hat \ve_{bcd} (\theta_W ) V^\nu_b V^\rho_c  V^\sigma_d
\Bigr) \nonumber
\end{eqnarray}
The coefficients of the anomalous currents in the Ward identity
are related to the ones of the ST identity. (This can be seen
by establishing the abelian Ward identity as a $\brs_{\Gacl}$-variation
of a ghost equation.)  In particular, they
vanish in 1-loop order, and
they vanish to all orders, if the non-renormalization
theorems are valid in the standard model. Vanishing of
the purely abelian current anomaly to all orders
can be proved only by means of the local abelian Ward identity
 \cite{BABE80}.

The absorption of non-anomalous currents   proceeds
as in the classical approximation
(cf.~(\ref{wqlocgen}) -- (\ref{coupl})): The lepton and quark family
currents are absorbed by fixing the by now undetermined constants
$G_{\delta_i}^{(1)}$ in (\ref{undet}) and
the matter current 
$\partial j^{matter}$ is absorbed
into the overall  normalization
of the Ward identity.

 The finite renormalized Green functions are constructed in
1-loop order uniquely: They satisfy the ST identity, the Ward identities
of rigid symmetry and the local abelian Ward identity. The
2-point functions of physical fields and of Faddeev-Popov ghosts
have one particle properties, and especially the mass matrices
of massiv massless particles are diagonalized at $p^2 = 0 $.
This property ensures that the Green functions of the next order
exist in renormalized perturbation theory.

\newsubsection{Induction to all orders}

Having constructed the Green functions of 1-loop order in accordance
with the symmetries and in accordance with off-shell infrared
existence (\ref{BPHZLn}), the action principle can be applied
to the renormalized Green functions of the next order in the same way, as
it applied, when we proceeded from lowest order to 1-loop 
(\ref{stbr}) and (\ref{wabr}).
For this reason we are able to carry out the proof to all orders
by induction (\ref{indn}).
 Assuming that the ST identity (\ref{STgen2}), 
 the Ward identities 
of rigid symmetry 
(\ref{wardopequivclass}) and the local $U(1)$ Ward identity  (\ref{wardidho})
are
established for the 
 Green functions to order $n -1$, then one can make
the induction step to order $n$. The important point is the fact that
the UV dimension and IR dimension of the breakings is not changed,
because the counterterms, we had to add for establishing the
symmetries, are compatible 
with
UV  and IR dimension 4. In particular, the
$\Ga_{eff}$, which governs the perturbative expansion of 
Green functions in the BPHZL scheme, is a 4-4 insertion
 (see \cite{PISO95} for
details).

Because the one-loop breakings have been absorbed in accordance with the
algebraic properties and the consistency equation
\begin{eqnarray}
\label{consf}
\brs _\Ga {\cal S} (\Ga )  & = & 0 \quad \hbox{ for any} \quad \Ga \\
\brs_\Ga \brs _\Ga  & = & 0 \quad \hbox{if} \quad 
{\cal S}(\Ga ) = 0 \nonumber \\
{\cal W}_\a {\cal S} (\Ga ) - \brs _\Ga {\cal W} _\a \Ga  & = & 0 
\quad \hbox{ for any} \quad \Ga \nonumber\\
\Bigl[ {\cal W}_\a , {\cal W} _\b \Bigr]  &= &
\ve_{\a \b \ga} \tilde I_{\ga \ga'}
{\cal W}_{\ga'} \nonumber
\end{eqnarray}
 the breakings of order $n$ are
algebraically characterized by (\ref{algbreak}).
Therefore we are able to proceed from order $n-1$ to order $n$
in the same way as from lowest order to 1-loop order,
 since we did not
use explicit expressions of 1-loop order, but only algebraic and
power counting properties. 

The only ingredient of 1-loop order has been the characterization
of the anomaly candidates by the  CS equation (\ref{dilbr}).
In order to close the arguments 
 we have
finally to derive the Callan-Symanzik  to order $n-1$. The
CS equation of 1-loop is given in (\ref{cs}). Since the
symmetries are established to order $n$ the
the unsymmetric soft field polynomial $\Delta_m^{\le 3} $ vanishes.
 The construction of the higher order
CS equation proceeds for the hard breakings as given in section 6.2,
especially there are the same number of independent
parameters and $\brs_{\Gacl}$-invariants. All differentiations
with respect to couplings and mass parameters 
 act on the parameters, which appear in higher
orders as corrections in the ST identity and the Ward identities
of rigid symmetry.   As it is for the  differentiation  with respect
to $M_W$ (\ref{thetaw}) in 1-loop order,
 they have all supplemented by field operators
 in order
to commute with the ST operator and Ward operators of rigid symmetry.
The explicit expressions can be read off from 
eqs.~(\ref{wfd}) and (\ref{sfd}) and will be given in detail elsewhere.
We denote with $\tilde \partial_\l ,
\l = e , m_H, m_{f_i}, \theta_W, \theta_G $
the rigid and $\brs_{\Gacl}$
symmetric operators of higher orders.
 (The weak mixing angle and the ghost angle are given by
the on-shell definition (\ref{ghostangle})). 
Also the 
higher order field differentiation
operators (\ref{Nvec}),   (\ref{Nferm}) and 
(\ref{Nscal}), which correspond to the anomalous dimensions, are
modified in an obvious way. 
The soft breakings of the CS equation 
are constructed as in the tree approximation, because they are
completely characterized by their algebraic properties.
The 
CS equation is then finally given by
\begin{eqnarray}
\label{cs2}
\! &\! \Bigl( \! & \! 
m\partial _m + \beta_e e\tilde
\partial_e - \beta_{M_W} \tilde \partial _{\theta_W}
+ \beta_{m_H} m_H \tilde \partial _{m_H} + 
\sum_{i=1}^{N_F}\sum_f\beta _{m_{f_i}}  {m_{f_i}} \tilde \partial
_{m_{f_i}} 
   \\
\! & \! -\! & \! 
\gamma_V \bigl( {\cal N}_V  
- {\cal N}_B + 2 \xi  \partial_\xi + 2 \hat \xi \partial_{\hat \xi} \bigr)
   - \beta _{\theta_G} \tilde\partial_{\theta_G} 
-\gamma_c {\cal N}_c 
\nonumber \\
\! & \! -\! & \! 
 \hat \gamma_V \bigl(\hat {\cal N}_V
-\hat \gamma_B \hat {\cal N}_B  + 2 (\xi + \hat \xi) \partial_{\hat \xi} 
\bigr)  -
 \gamma_S  {\cal N}_S 
-  \gamma_{\hat S}{\cal N}_{\hat S}   
 - \tilde \gamma_S \tilde {\cal N}_S  \nonumber \\
\! & \!-\! & \! 
\sum_{i=1}^{N_F} \bigl( \gamma_{F_{l_i}}  {\cal N}^L_{F_{l_i}}
+\gamma_{F_{q_i}}  {\cal N}^L_{F_{q_i}}
+ \gamma_{e_{i}}  {\cal N}^R_{e_{i}}
+ \gamma_{u_{i}}  {\cal N}^R_{u_{i}}
+ \gamma_{d_{i}}  {\cal N}^R_{d_{i}} \bigr)  \,
\Bigr) \, \Gamma \Big|_{\hat \varphi_o = 0}  \nonumber  \\
\! & \! = \! & \! \int \biggl(\Bigl( (1 +  \sum_{\l} \beta_\l \partial_\l ) v 
\Bigr)
 {\delta \Gamma \over \delta H } +
\Bigl( (1 +  \sum_{\l} \beta_\l \partial_\l ) \hat \zeta v 
\Bigr)
 {\delta \Gamma \over \delta \hat H }  \nonumber \\
\!& \!  \! & \! + 
   v (\gamma_S +\tilde \gamma_S ){\delta \Gamma \over \delta H } +
\hat \zeta v \gamma_{\hat  S}{\delta \Gamma \over \delta \hat H } +
\frac {m_H^2}2
{\delta \Gamma \over \delta \hat \varphi_o }    \biggr)
+
\int \tilde \gamma_S  
 \hat q_a \tilde I_{ab}Y_b \nonumber
\end{eqnarray}
 Infrared existence of the CS equation can be proved as
in 1-loop order,
 since the conditions for off-shell
infrared existence (\ref{BPHZLn}) have been maintained in the construction
of symmetry operators.

With the establishment of the CS equation the construction of
standard model Green functions to all orders is completed.
The ST identity and Ward identities of rigid symmetry
\begin{equation}
{\cal S} (\Ga ) = 0  \qquad {\cal W} _\a \Ga = 0
\end{equation}
the local abelian Ward identity
\begin{equation}
\biggl( g_1 {\mathbf w}^Q - 
 \partial \Bigl(\frac  1 {r_Z}
\sin \Theta {\delta \over \delta Z}
+ \frac 1 {r_A} \cos \Theta {\delta \over \delta A} \Bigr)\biggr)\Ga 
  = \Box \Bigl( \frac 1 {r_Z}\sin \Theta  B_Z + 
\frac 1  {r_A}\cos \Theta B_A \Bigr) 
\end{equation}
with
\begin{equation}
{\mathbf w}^Q \equiv {\mathbf w}_{em} - {\mathbf w}_3
\end{equation}
define uniquely the Green functions of the standard
model of electroweak interactions to all orders of perturbation
theory in the on-shell scheme.

\pagebreak
\newsection{Conclusions and outlook}

In this article we have constructed the finite renormalized Green 
functions of the standard model of electroweak interactions to
all orders of perturbation theory. Special attention has been paid
to the construction of 2-point functions in the on-shell scheme.
Only if the Green functions have one-particle properties in the
LSZ-limit (apart from the problem of unstable particles), can
one proceed to construct the S-matrix and finally prove unitarity
of the physical S-matrix. These properties are the main requirements
for being able to interpret a quantum field theory as a physical
theory of fundamental interactions. Since the standard model
contains massless particles, mass diagonalization of massless
and massive fields is connected with off-shell infrared existence
of finite renormalized Green functions.

The analysis  has  been carried out using the method of algebraic
renormalization, which until now was applied  mainly to
theories with semisimple gauge groups. In order to apply the
method to the standard model with the non-semisimple 
$SU(2) \times U(1)$ group
we had to generalize the method of algebraic renormalization at
some points. In particular, we had to obtain the
symmetry operators by means of their algebraic properties instead
of postulating them in an explicit form a priori. The parameters, which appear
in the general solution of the algebra, correspond to field redefinitions
of individual fields and, in particular, to non-diagonal
field redefinitions of neutral massive/massless fields. The adjustment
of the latter parameters is essential for diagonalizing the 
mass matrix of neutral vectors at $p^2 = 0$. 
Due to the non-semisimple group structure,  the
 abelian component
of the action  contains additional free
parameters, which are not specified by the Slavnov-Taylor identity.
These are interpreted as the couplings of the currents
of lepton and quark  family
conservation. Classically, these currents are conserved in the
standard model, if one neglects mixing of quarks due to the CKM
matrix.  In the general case
 there are  classically   the conserved currents  of fermion family
number conservation and of baryon number conservation.
 These currents are not gauged, but are not distinguished
from the electromagnetic current
 in the theoretical prescription,
 since they have  the
same quantum numbers.
In order to characterize the interactions prescribed by the standard model as
the ones of  weak and electromagnetic interaction, the
electromagnetic and the lepton and quark number  currents
have to be identified and fixed by a Ward identity. 
Because the electromagnetic Ward identity
of current conservation cannot be derived for off-shell Green functions,
we have to use a specific form of the abelian local Ward identity.
This identity
is the functional generalization of the Gell-Mann Nishijima relation.
We want to point out that in the general case, with  quark family mixing,
the  local Ward identity becomes even  more important for correct
adjustment of the electromagnetic current.

An abelian Ward identity cannot be derived  a priori, but has to
be characterized in the group structure as being abelian. For this
reason, we have to require invariance under the nonabelian rigid 
symmetries. 
The construction of Green functions in agreement with rigid symmetry
restricts the gauge fixing sector and the number of independent parameters
appearing therein. In order to be able to diagonalize the  mass matrix
of  neutral ghosts at $p^2 = 0$, one has to introduce an additional
ghost angle into the BRS-transformations of antighosts. In the
on-shell scheme, this angle is related to the ghost mass ratio in a
way similar
to that in which the vector mass ratio is related to the weak mixing angle.
From the Callan-Symanzik equation, one can see that the
ghost mass ratio indeed
has to be introduced  as an independent parameter
of the theory, since it has independent higher order corrections.

The most remarkable consequence of the higher order 
 construction is the observation,
that  the standard model provides exactly the right number of
 parameters 
to bring the propagators to a form in which they have one-particle properties
in the LSZ-limit.
As we have pointed out, one has to adjust all of these parameters and 
one has to take
into  account {\it all} deformations allowed by the algebra. If we had not
succeeded with the analysis as prescribed in the paper, then
we would have had to prove that one-particle properties are the consequence of 
a symmetry. 
Such a procedure has to be carried out finally in the
unphysical sector proving mass degeneracy for all unphysical
fields by means of the Slavnov-Taylor identity \cite{BRS76, KUOJ78}.
From this point of view
 the vector and the
unphysical sector will be analysed carefully, when
the renormalization is extended
to   CP-violating  interactions.

Having constructed the symmetry operators, we can  apply them
 immediately to explicit one-loop and higher loop calculations.
 The parameters appearing therein are mainly determined
on 2-point functions, which are 
 listed for  one-loop order in the literature. 
One is then able to prove, if finite Green functions satisfy 
the Slavnov-Taylor identity. In dimensional regularization one
 has  to pay most attention to such breakings,
which are absorbed into parity violating counterterms of the effective
action. Due to parity non-conservation, it is not evident that
the finite Green functions satisfy the Slavnov-Taylor identity, if 
poles
are subtracted by means of d-dimensional  symmetric counterterms.

A first insight into higher order non-local contributions can be
gained by considering the Callan-Symanzik equation. The Callan-Symanzik
equation of the standard model of electroweak interaction has
a completely different form from that of the corresponding symmetric
theory. It contains mixed field operators between massless/massive neutral
fields and, in particular,  $\beta$-functions with respect to
the independent mass parameters of the theory. It is, however, not
a matter of taste, whether one wants to derive the Callan-Symanzik
 equation in terms
of physical
fields
or symmetric fields, since it  does not even exist, if one does not
include the mass diagonalization conditions of massless/massive
particles at $p^2 = 0 $ in the construction. For this reason,
the considerations which concern the renormalization group analysis
of the unbroken $SU(2) \times U(1) $-theory are not applicable to the standard
model. It was one of
the main intentions of the present work, to make the
differences between unbroken and broken theories apparent.
In particular, what one has to consider  in the spontaneously
broken case,
are the large mass logarithms, which are induced from the lowest
order $\beta$-functions to higher orders. These 
large-mass logarithms are specific for the model
in its spontaneously broken form and have been analysed in a much simpler
broken theory in 
 \cite{INVCH}. The corresponding systematic investigation is now also
feasible in the electroweak standard model.
\vspace{15mm}

{\it Acknowledgements}
I want to thank K.~Sibold for initial common work, many helpful discussions
and a critical reading of the manuscript. I am grateful to
G.~Weiglein, B.A.~Kniehl and A.~Denner for comments on 
 explicit one-loop
renormalization.

\clearpage
\begin{appendix}

\section{The quantum numbers of fields}

In this appendix we list the quantum numbers of fields. We give
the electromagnetic charge $Q_{em}$, the Faddeev-Popov charge
$Q_{\phi\pi}$ and the properties under charge conjugation C and parity
transformation P. Parity transformation to 
 massive fermions is assigned  in accordance with parity
conservation in electromagnetic interactions. The infrared ($\dim^ {IR}$)
and
ultraviolet ($\dim^ {UV}$) dimensions of fields is adjusted in agreement
with the BPHZL-scheme \cite{ZIM69, LOW76}.

\begin{table}[!h]
\begin{center}
\begin{tabular}{l|c|c|c|c|c|c}
   & $dim^{UV}$ & $dim^{IR}$ & $Q_{em}$ & $Q_{\phi \pi}$ & $C$ & 
          $P(x^{\mu} \rightarrow x_{\mu})$ \\ \hline
$e^L$ & $\frac 3 2$ & 2 & -1 & 0 & $-i \gamma^2 e^{R*}$ & 
          $\gamma^0 e^R$  \\ \hline 
$e^R$ & $\frac 3 2$ & 2 & -1 & 0 & $-i \gamma^2 e^{L*}$ & 
          $\gamma^0 e^ L$  \\ \hline 
$u^L$ & $\frac 3 2$ & 2 & $+\frac 2 3$ & 0 & $-i \gamma^2 u^{L*}$ & 
          $\gamma^0 u^R$  \\ \hline 
$u^R$ & $\frac 3 2$ & 2 & $+\frac 2 3$ & 0 & $-i \gamma^2 u^{R*}$ & 
          $\gamma^0 u^ L$  \\ \hline 
$d^L$ & $\frac 3 2$ & 2 & $-\frac 1 3$ & 0 & $-i \gamma^2 d^{L*}$ & 
          $\gamma^0 d^R$  \\ \hline 
$d^R$ & $\frac 3 2$ & 2 & $-\frac 1 3$ & 0 & $-i \gamma^2 d^{R*}$ & 
          $\gamma^0 d^ L$  \\ \hline 
$\nu^L$ & $\frac 3 2$ & $\frac 3 2$ & 0 & 0 & $CP: -i \gamma^2 
\ga^0 \nu^{L*}$ & 
          $ $  \\ \hline 
$\psi^R_e$ & $\frac 5 2$ & 2 & -1 & 1 & $-i \gamma^ 2 \psi^{L *}_e$ & 
          $\gamma^0 \psi^L_e$  \\ \hline 
$\psi^L_e$ & $\frac 5 2$ & 2 & -1 & 1 & $-i \gamma^2 \psi^{R *}_e$ & 
          $\gamma^0 \psi^R_e$  \\ \hline 
$\psi^R_u$ & $\frac 5 2$ & 2 & $+\frac 2 3$ & 1 & $-i \gamma^2 \psi^{L *}_u$ & 
          $\gamma^0 \psi^L_u$  \\ \hline 
$\psi^L_u$ & $\frac 5 2$ & 2 & $+\frac 2 3$ & 1 & $-i \gamma^2 \psi^{R *}_u$ & 
          $\gamma^0 \psi^R_u$  \\ \hline 
$\psi^R_d$ & $\frac 5 2$ & 2 & $-\frac 1 3$ & 1 & $-i \gamma^2 \psi^{L *}_d$ & 
          $\gamma^0 \psi^L_d$  \\ \hline 
$\psi^L_d$ & $\frac 5 2$ & 2 & $-\frac 1 3$ & 1 & $-i \gamma^2 \psi^{R *}_d$ & 
          $\gamma^0 \psi^R_d$  \\ \hline 
$\psi^R_\nu$ & $\frac 5 2$ & $\frac 5 2$ & 0 & 1 & $
                            CP: -i \gamma^2 \ga^0\psi_\nu^{R*}$ &   
          $$  \\ \hline 
\end{tabular}
\end{center}
{{\sl Table 1}: Quantum numbers of the fermion fields}
\end{table}  

\clearpage

\newpage
\begin{table}
\begin{center}
\begin{tabular}{l|c|c|c|c|c|c}
   & $dim^{UV}$ & $dim^{IR}$ & $Q_{em}$ & $Q_{\phi \pi}$ & $C$ & 
          $P(x^{\mu} \rightarrow x_{\mu})$ \\ \hline
$W^{\mu}_{\pm}$ & 1 & 2 & $\pm 1$ & 0 & $-W^{\mu}_{\mp}$ & 
          $W_{\mu \pm}$ \\ \hline
$Z^\mu$ & 1 & 2 & 0 & 0 & $-Z^\mu$ & 
          $Z_\mu$  \\ \hline 
$A^\mu$ & 1 & 1 & 0 & 0 & $-A^\mu$ & 
          $A_\mu$  \\ \hline 
$\rho^{\mu}_{\pm}$ & 3 & 3 & $\pm 1$ & -1 & $-\rho^{\mu}_{\mp}$ & 
          $\rho_{\mu \pm}$ \\ \hline
$\rho^{\mu}_3$ & 3 & 3 & 0 & -1 & $-\rho^{\mu}_{3}$ & 
          $\rho_{\mu 3}$ \\ \hline
$c_{\pm}$ & 0 & 1 & $\pm 1$ & 1 & $-c_{\mp}$ & 
          $c_{\pm}$ \\ \hline
$c_{Z}$ & 0 & 1 & 0 & 1 & $-c_{Z}$ & 
          $c_{Z}$ \\ \hline
$c_{A}$ & 0 & 0 & 0 & 1 & $-c_{A}$ & 
          $c_{A}$ \\ \hline
$\sigma_{\pm}$ & 4 & 4 & $\pm 1$ & -2 & $-\sigma_{\mp}$ & 
          $\sigma_{\pm}$ \\ \hline
$\sigma_{3}$ & 4 & 4 & 0 & -2 & $-\sigma_{3}$ & 
          $\sigma_{3}$ \\ \hline
$\bar{c}_{\pm}$ & 2 & 3 & $\pm 1$ & -1 & $-\bar{c}_{\mp}$ & 
          $\bar{c}_{\pm}$ \\ \hline
$\bar{c}_{Z}$ & 2 & 3 & 0 & -1 & $-\bar{c}_{Z}$ & 
          $\bar{c}_{Z}$ \\ \hline
$\bar{c}_{A}$ & 2 & 1 & 0 & -1 & $-\bar{c}_{A}$ & 
          $\bar{c}_{A}$ \\ \hline
$B_{\pm}$ & 2 & 3 & $\pm 1$ & 0 & $-B_{\mp}$ & 
          $B_{\pm}$ \\ \hline
$B_{Z}$ & 2 & 3 & 0 & 0 & $-B_{Z}$ & 
          $B_{Z}$ \\ \hline
$B_{A}$ & 2 & 2 & 0 & 0 & $-B_{A}$ & 
          $B_{A}$ \\ \hline
$\phi_{\pm}$ & 1 & 1 & $\pm 1$ & 0 & $\phi_{\mp}$ & 
          $\phi_{\pm}$ \\ \hline
$H$ & 1 & 2 & 0 & 0 & $H$ & 
          $H$ \\ \hline
$\chi$ & 1 & 2 & 0 & 0 & $-\chi$ & 
          $\chi$ \\ \hline
$Y_{\pm}$ & 3 & 3 & $\pm 1$ & -1 & $Y_{\mp}$ & 
          $Y_{\pm}$ \\ \hline
$Y_{H}$ & 3 & 3 & 0 & -1 & $Y_{H}$ & 
          $Y_{H}$ \\ \hline
$Y_{\chi}$ & 3 & 3 & 0 & -1 & $-Y_{\chi}$ & 
          $Y_{\chi}$ \\ \hline
$\hat{\phi}_{\pm}$ & 1 & 1 & $\pm 1$ & 0 & $\hat{\phi}_{\mp}$ & 
          $\hat{\phi}_{\pm}$ \\ \hline
$\hat{H}$ & 1 & 2 & 0 & 0 & $\hat{H}$ & 
          $\hat{H}$ \\ \hline
$\hat{\chi}$ & 1 & 2 & 0 & 0 & $-\hat{\chi}$ & 
          $\hat{\chi}$ \\ \hline
$\hat{q}_{\pm}$ & 1 & 1 & $\pm 1$ & 1 & $+\hat{q}_{\mp}$ & 
          $\hat{q}_{\pm}$ \\ \hline
$\hat{q}_{H}$ & 1 & 2 & 0 & 1 & $\hat{q}_{H}$ & 
          $\hat{q}_{H}$ \\ \hline
$\hat{q}_{\chi}$ & 1 & 2 & 0 & 1 & $-\hat{q}_{\chi}$ & 
          $\hat{q}_{\chi}$ \\ \hline
$\hat{\varphi}_{0}$ & 2 & 2 & 0 & 0 & $\hat{\varphi}_{0}$ & 
          $\hat{\varphi}_{0}$ \\ \hline
\end{tabular}
\end{center}
{{\sl Table 2}: Quantum numbers of boson fields}
\end{table}  

\end{appendix}

\end{document}